\documentstyle[12pt]{article}

\newdimen\SaveWidth \SaveWidth=\textwidth
\newdimen\SaveHeight \SaveHeight=\textheight
\advance\SaveWidth by -\textwidth
\advance\SaveHeight by -\textheight
\divide\SaveWidth by 2
\divide\SaveHeight by 2
\advance\hoffset by \SaveWidth
\advance\voffset by \SaveHeight

\def\sgn{\mathop{\rm sgn}}
\def\GeV{{\rm GeV}}
\def\etmiss{\slashchar{E}_T}

\def\slashchar#1{\setbox0=\hbox{$#1$}           
   \dimen0=\wd0                                 
   \setbox1=\hbox{/} \dimen1=\wd1               
   \ifdim\dimen0>\dimen1                        
      \rlap{\hbox to \dimen0{\hfil/\hfil}}      
      #1                                        
   \else                                        
      \rlap{\hbox to \dimen1{\hfil$#1$\hfil}}   
      /                                         
   \fi}                                         %

\def\simge{
    \mathrel{\rlap{\raise 0.511ex
        \hbox{$>$}}{\lower 0.511ex \hbox{$\sim$}}}}
\def\simle{
    \mathrel{\rlap{\raise 0.511ex 
        \hbox{$<$}}{\lower 0.511ex \hbox{$\sim$}}}}

\begin{document}


\rightline{BNL-HET-03/28}
\rightline{UH-511-1038-03}

\vskip.5in

\centerline{\Large\bf ISAJET 7.69}
\medskip
\centerline{\Large\bf A Monte Carlo Event Generator}
\medskip
\centerline{\Large\bf for $pp$, $\bar pp$, and $e^+e^-$ Reactions}
\bigskip\bigskip
\centerline{\bf Frank E. Paige and Serban D. Protopopescu}
\smallskip
\centerline{Physics Department}
\centerline{Brookhaven National Laboratory}
\centerline{Upton, NY 11973, USA}
\bigskip
\centerline{\bf Howard Baer}
\smallskip
\centerline{Department of Physics}
\centerline{Florida State University}
\centerline{Talahassee, FL 32306}
\bigskip
\centerline{\bf Xerxes Tata}
\centerline{Department of Physics and Astronomy}
\centerline{University of Hawaii}
\centerline{Honolulu, HI 96822}

\vskip1cm

\centerline{\bf Abstract}
\bigskip

ISAJET 7.69 is now available, replacing version 7.64 and several
intermediate versions.

The complete set of 1-loop radiative corrections to sparticle masses has
been included in the ISAJET SUSY spectrum calculation, according to the
formulae given by Pierce {\it et al.}, Nucl.\ Phys.\ {\bf B491}, 3
(1997).  Previous versions of ISAJET had included the logarithmic 1-loop
corrections by freezing out soft mass parameters at a scale equal to
their mass. The current calculation includes all finite corrections as
well. Crucial contributions to the encoding of these expressions were
made by Tadas Krupovnickas. We have also adjusted slightly the input
$\overline{DR}$ gauge couplings at $M_Z$ to coincide with measured
central values.  Logarithmic threshold corrections to gauge and Yukawa
couplings are accounted for as in previous ISAJET versions by changing
the beta functions in the RGEs as various soft mass thresholds are
passed. We have also enlarged the set of SUSY RGEs to include running
of the Higgs field vevs. The net effect of these changes is to modify
various sparticle masses by typically 1-5\% from the predictions of
ISAJET 7.64.

The size of the small mass splitting between $\tilde\chi_1^\pm$ and
$\tilde\chi_1^0$ is important for AMSB phenomenology. Rather than
extracting this splitting from the general result, for the AMSB model we
calculate it from the tree and the 1-loop vector boson loop graphs
[taken from Cheng, Dobrescu and Matchev, Nucl.\ Phys.\ {\bf B543}, 47,
(1999)] and use it to set the $\tilde\chi_1^\pm$ mass.

We have modified the RGE solution to use the previous SUSY masses to
compute $\mu$ before computing the new SUSY masses and from them the
loop corrections to $\mu$. This seems to improve the convergence, e.g.,
in the focus point, or hyperbolic branch region, of the mSUGRA model. 
The RGE solution still does not converge properly in isolated regions.

We have also corrected a sign mistake in the $A$-term boundary
conditions for the anomaly-mediated SUSY breaking model.

The new version can be obtained either by anonymous FTP from 
\begin{verbatim}
ftp://ftp.phy.bnl.gov/pub/isajet
\end{verbatim}
or by HTTP from
\begin{verbatim}
http://www.phy.bnl.gov/~isajet
\end{verbatim}
or via AFS from
\begin{verbatim}
/afs/cern.ch/user/p/paige/public/isajet
\end{verbatim}
All of these contain the same card image PAM file isajet.car and Unix
Makefile to build ISAJET from isajet.car on all supported Unix
platforms.

\newpage
\setcounter{page}{1}


\centerline{\Large\bf ISAJET 7.69}
\medskip
\centerline{\Large\bf A Monte Carlo Event Generator}
\medskip
\centerline{\Large\bf for $pp$, $\bar pp$, and $e^+e^-$ Reactions}
\bigskip\bigskip
\centerline{\bf Frank E. Paige and Serban D. Protopopescu}
\smallskip
\centerline{Physics Department}
\centerline{Brookhaven National Laboratory}
\centerline{Upton, NY 11973, USA}
\bigskip
\centerline{\bf Howard Baer}
\smallskip
\centerline{Department of Physics}
\centerline{Florida State University}
\centerline{Talahassee, FL 32306}
\bigskip
\centerline{\bf Xerxes Tata}
\centerline{Department of Physics and Astronomy}
\centerline{University of Hawaii}
\centerline{Honolulu, HI 96822}

\bigskip\bigskip
\tableofcontents

\newpage
\section{Introduction\label{INTRO}}

      ISAJET is a Monte Carlo program which simulates $pp$, 
$\bar pp$ and $e^+e^-$ interactions at high energies. 
ISAJET is based on
perturbative QCD plus phenomenological models for parton and beam jet
fragmentation. Events are generated in four distinct steps:
\begin{itemize}
\item A primary hard scattering is generated according to the
appropriate QCD cross section.
\item QCD radiative corrections are added for both the initial and the
final state.
\item Partons are fragmented into hadrons independently, and particles
with lifetimes less than about $10^{-12}$ seconds are decayed.
\item Beam jets are added assuming that these are identical to a
minimum bias event at the remaining energy.
\end{itemize}

      ISAJET incorporates ISASUSY, which evaluates branching ratios for
the minimal supersymmetric extension of the standard model. H.~Baer and
X.~Tata are coauthors of this package, and they have done the original
calculations with various collaborators. See the ISASUSY documentation
in the patch Section~\ref{SUSY}.

      ISAJET is supported for ANSI Fortran and for Cray, DEC Ultrix,
DEC VMS, HP/9000 7xx, IBM VM/CMS 370 and 30xx, IBM AIX RS/6000, Linux,
Silicon Graphics 4D, and Sun computers. The CDC 7600 and ETA 10
versions are obsolete and are no longer supported. It is written
mainly in ANSI standard FORTRAN 77, but it does contain some
extensions except in the ANSI version. The code is maintained with a
combination of RCS, the Revision Control System, and the Patchy code
management system, which is part of the CERN Library. The original
sources are kept on physgi01.phy.bnl.gov in
\verb|~isajet/isalibrary/RCS|; decks revised in release \verb|n.nn|
are kept in \verb|~isajet/isalibrary/nnn|. ISAJET is supplied to BNL,
CERN, Fermilab, and SLAC; it is also available from
\begin{verbatim}
http://www.phy.bnl.gov/~isajet
ftp://ftp.phy.bnl.gov/pub/isajet
\end{verbatim}
or by request from the authors.

      Patch ISAPLT contains the skeleton of an HBOOK histogramming
job, a trivial calorimeter simulation, and a jet-finding algorithm.
(The default is HBOOK4; HBOOK3 can be selected with a Patchy switch.)
These are provided for convenience only and are not supported.
\newpage
\section{Physics\label{PHYSICS}}

      ISAJET is a Monte Carlo program which simulates $pp$, $\bar pp$
and $e^+e^-$ interactions at high energy. 
The program incorporates
perturbative QCD cross sections, initial state and final state QCD
radiative corrections in the leading log approximation, independent
fragmentation of quarks and gluons into hadrons, and a
phenomenological model tuned to minimum bias and hard scattering data
for the beam jets.

\subsection{Hard Scattering\label{hard}}

      The first step in simulating an event is to generate a primary
hard scattering according to some QCD cross section. This has the
general form
$$
\sigma = \sigma_0  F(x_1,Q^2) F(x_2,Q^2)
$$
where $\sigma_0$ is a cross section calculated in QCD perturbation
theory, $F(x,Q^2)$ is a structure function incorporating QCD scaling
violations, $x_1$ and $x_2$ are the usual parton model momentum
fractions, and $Q^2$ is an appropriate momentum transfer scale.

      For each of the processes included in ISAJET, the basic cross
section $\sigma_0$ is a two-body one, and the user can set limits on
the kinematic variables and type for each of the two primary jets. For
DRELLYAN and WPAIR events the full matrix element for the decay of the
W's into leptons or quarks is also included.

      The following processes are available:

\subsubsection{Minbias} No hard scattering at all, so that the event
consists only of beam jets. Note that at high energy the jet cross
sections become large. To represent the total cross section it is
better to use a sample of TWOJET events with the lower limit on pt
chosen to give a cross section equal to the inelastic cross section or
to use a mixture of MINBIAS and TWOJET events.

\subsubsection{Twojet} All order $\alpha_s^2$ QCD processes, which
give rise in lowest order to two high-$p_t$ jets. Included are, e.g.
\begin{eqnarray*}
g + g &\to& g + g\\
g + q &\to& g + q \\
g + g &\to& q + \bar q
\end{eqnarray*}
Masses are neglected for $c$ and lighter quarks but are taken into
account for $b$ and $t$ quarks. The $Q^2$ scale is taken to be
$$
Q^2 = 2stu/(s^2+t^2+u^2)
$$
The default parton distributions are those of the CTEQ Collaboration,
fit CTEQ5L, using lowest order QCD evolution. A few older sets of parton
distributions are included. There is also an interface to the CERN
PDFLIB compilation of parton distributions. Note that structure
functions for heavy quarks are included, so that processes like
$$
g + t \to g + t
$$
can be generated. The Duke-Owens parton distributions do not contain b
or t quarks.

      Since the $t$ is so heavy, it decays before it can hadronize, so
instead of $t$ hadrons a $t$ quark appears in the particle list. It is
decayed using the $V-A$ matrix element including the $W$ propagator
with a nonzero width, so the same decays should be used for $m_t < m_W$
and $m_t > m_W$; the $W$ should {\it not} be listed as part of the decay
mode.  The partons are then evolved and fragmented as usual; see
below. The real or virtual $W$ and the final partons from the decay,
including any radiated gluons, are listed in the particle table,
followed by their fragmentation products.  Note that for semileptonic
decays the leptons appear twice: the lepton parton decays into a
single particle of the same type but in general somewhat different
momentum. In all cases only particles with $\verb|IDCAY| = 0$ should be
included in the final state.

      A fourth generation $x,y$ is also allowed. Fourth generation
quarks are produced only by gluon fusion. Decay modes are not included
in the decay table; for a sequential fourth generation they would be
very similar to the t decays. In decays involving quarks, it is
essential that the quarks appear last.

\subsubsection{Drellyan} Production of a $W$ in the standard model,
including a virtual $\gamma$, a $W^+$, a $W^-$, or a $Z^0$, and its
decay into quarks or leptons. If the transverse momentum QTW of the
$W$ is fixed equal to zero then the process simulated is
\begin{eqnarray*}
q + \bar q \to W &\to& q + \bar q \\
                 &\to& \ell + \bar\ell
\end{eqnarray*}
Thus the $W$ has zero transverse momentum until initial state QCD
corrections are taken into account. If non-zero limits on the
transverse momentum $q_t$ for the $W$ are set, then instead the
processes
\begin{eqnarray*}
q + \bar q &\to& W + g \\
g + q      &\to& W + q
\end{eqnarray*}
are simulated, including the full matrix element for the $W$ decay.
These are the dominant processes at high $q_t$, but they are of course
singular at $q_t=0$. A cutoff of the $1/q_t^2$ singularity is made by
the replacement
$$
1/q_t^2 \to 1/\sqrt{q_t^4+q_{t0}^4} \quad q_{t0}^2 =  (.2\,\GeV) M
$$
This cutoff is chosen to reproduce approximately the $q_t$ dependence
calculated by the summation of soft gluons and to give about the right
integrated cross section. Thus this option can be used for low as well
as high transverse momenta.

      The scale for QCD evolution is taken to be proportional to the
mass for lowest order Drell-Yan and to the transverse momentum for
high-$p_t$ Drell-Yan. The constant is adjusted to get reasonable
agreement with the $W + n\,{\rm jet}$ cross sections calculated from
the full QCD matrix elements by F.A. Berends, et al., Phys.\ 
Lett.\ B224, 237 (1989).

      For the processes $g + b \to W + t$ and $g + t \to Z + t$, cross
sections with a non-zero top mass are used for the production and the
$W/Z$ decay. These were calculated using FORM 1.1 by J.~Vermaseren. The
process $g + t \to W + b$ is {\it not} included. Both $g + b \to W^- +
t$ and $g + \bar t \to W^- + \bar b$ of course give the same $W^- + t
+ \bar b$ final state after QCD evolution. While the latter process is
needed to describe the $m_t = 0$(!) mass singularity for $q_t \gg
m_t$, it has a pole in the physical region at low $q_t$ from on-shell
$t \to W + b$ decays. There is no obvious way to avoid this without
introducing an arbitrary cutoff.  Hence, selecting only $W + b$ will
produce a zero cross section. The $Q^2$ scale for the parton
distributions in these processes is replaced by $Q^2 + m_t^2$; this
seems physically sensible and prevents the cross sections from
vanishing at small $q_t$.

\subsubsection{Photon} Single and double photon production through the
lowest order QCD processes
\begin{eqnarray*}
g + q &\to& \gamma + q \\
q + \bar q &\to& \gamma + g \\
q + \bar q &\to& \gamma + \gamma
\end{eqnarray*}
Higher order corrections are not included. But $\gamma$'s, $W$'s, and
$Z$'s are radiated from final state quarks in all processes, allowing
study of the bremsstrahlung contributions.

\subsubsection{Wpair} Production of pairs of W bosons in the standard
model through quark-antiquark annihilation,
\begin{eqnarray*}
q + \bar q &\to& W^+ + W^- \\
           &\to& Z^0 + Z^0 \\
           &\to& W^+ + Z^0, W^- + Z^0 \\
           &\to& W^+ + \gamma, W^- + \gamma \\
           &\to& Z^0 + \gamma
\end{eqnarray*}
The full matrix element for the W decays, calculated in the narrow
resonance approximation, is included. However, the higher order
processes, e.g.
$$
q + q \to q + q + W^+ + W^-
$$
are ignored, although they in fact dominate at high enough mass.
Specific decay modes can be selected using the WMODEi keywords.

\subsubsection{Higgs} Production and decay of the standard model Higgs
boson. The production processes are
\begin{eqnarray*}
g + g      &\to& H \quad\hbox{(through a quark loop)} \\
q + \bar q &\to& H \quad\hbox{(with $t + \bar t$ dominant)} \\
W^+ + W^-  &\to& H \quad\hbox{  (with longitudinally polarized $W$)} \\
Z^0 + Z^0  &\to& H \quad\hbox{ (with longitudinally polarized $Z$)}
\end{eqnarray*}
If the (Standard Model) Higgs is lighter than $2 M_W$, then it will
decay into pairs of fermions with branching ratios proportional to
$m_f^2$. If it is heavier than $2 M_W$, then it will decay primarily
into $W^+ W^-$ and $Z^0 Z^0$ pairs with widths given approximately by
\begin{eqnarray*}
\Gamma(H \to W^+ W^-) &=& {G_F M_H^3 \over 8 \pi \sqrt{2} } \\
\Gamma(H \to Z^0 Z^0) &=& {G_F M_H^3 \over 16 \pi \sqrt{2} }
\end{eqnarray*}
Numerically these give approximately
$$
\Gamma_H = 0.5\,{\rm TeV} \left({M_H \over 1\,{\rm TeV}}\right)^3
$$
The width proportional to $M_H^3$ arises from decays into longitudinal
gauge bosons, which like Higgs bosons have couplings proportional to
mass.

      Since a heavy Higgs is wide, the narrow resonance approximation is
not valid. To obtain a cross section with good high energy behavior, it
is necessary to include a complete gauge-invariant set of graphs for the
processes
\begin{eqnarray*}
W^+ W^- &\to& W^+ W^- \\
W^+ W^- &\to& Z^0 Z^0 \\
Z^0 Z^0 &\to& W^+ W^- \\
Z^0 Z^0 &\to& Z^0 Z^0
\end{eqnarray*}
with longitudinally polarized $W^+$, $W^-$, and $Z^0$ bosons in the
initial state. This set of graphs and the corresponding angular
distributions for the $W^+$, $W^-$, and $Z^0$ decays have been
calculated in the effective $W$ approximation and included in HIGGS.
The $W$ structure functions are obtained by integrating the EHLQ
parameterization of the quark ones term by term. The Cabibbo-allowed
branchings
\begin{eqnarray*}
q &\to& W^+ + q' \\
q &\to& W^- + q' \\
q &\to& Z^0 + q
\end{eqnarray*}
are generated by backwards evolution, and the standard QCD evolution is
performed. This correctly describes the $W$ collinear singularity and
so contains the same physics as the effective $W$ approximation.

      If the Higgs is lighter than $2M_W$, then its decay to
$\gamma\gamma$ through $W$ and $t$ loops may be important. This is
also included in the HIGGS process and may be selected by choosing
\verb|GM| as the jet type for the decay.

      If the Higgs has $M_Z < M_H < 2M_Z$, then decays into one real
and one virtual $Z^0$ are generated if the \verb|Z0 Z0| decay mode is
selected, using the calculation of Keung and Marciano, Phys.\ Rev.\
D30, 248 (1984). Since the calculation assumes that one $Z^0$ is
exactly on shell, it is not reliable within of order the $Z^0$ width
of $M_H = 2M_Z$; Higgs and and $Z^0 Z^0$ masses in this region should
be avoided. The analogous Higgs decays into one real and one virtual
charged W are not included.

      Note that while HIGGS contains the dominant graphs for Higgs
production and graphs for $W$ pair production related by gauge invariance,
it does not contain the processes
\begin{eqnarray*}
q + \bar q &\to& W^+ W^- \\
q + \bar q &\to& Z^0 Z^0
\end{eqnarray*}
which give primarily transverse gauge bosons. These must be generated
with WPAIR.

      If the \verb|MSSMi| or \verb|SUGRA| keywords are used with
HIGGS, then one of the three MSSM neutral Higgs is generated instead
using gluon-gluon and quark-antiquark fusion with the appropriate SUSY
couplings. Since heavy CP even SUSY Higgs are weakly coupled to W
pairs and CP odd ones are completely decoupled, $WW$ fusion and $WW
\to WW$ scattering are not included in the SUSY case. ($WW \to WW$ can
be generated using the Standard Model process with a light Higgs mass,
say 100 GeV.) The MSSM Higgs decays into both Standard Model and SUSY
modes as calculated by ISASUSY are included. For more discussion see
the SUSY subsection below and the writeup for ISASUSY. The user must
select which Higgs to generate using HTYPE; see Section 6 below. If a
mass range is not specified, then the range mass $M_H \pm 5\Gamma_H$
is used by default. (This cannot be done for the Standard Model Higgs
because it is so wide for large masses.) Decay modes may be selected
in the usual way.

\subsubsection{WHiggs} Generates associated production of gauge and
Higgs bosons, i.e.,
$$
q + \bar q \to H + W, H + Z\,,
$$
in the narrow resonance approximation. The desired subprocesses can be
selected with JETTYPEi, and specific decay modes of the $W$ and/or $Z$
can be selected using the WMODEi keywords. Standard Model couplings are
assumed unless SUSY parameters are specified, in which case the SUSY
couplings are used.

\subsubsection{SUSY} Generates pairs of supersymmetric particles from
gluon-quark or quark-antiquark fusion. If the MSSMi or SUGRA parameters
defined in Section 6 below are not specified, then only gluinos and
squarks are generated:
\begin{eqnarray*}
g + g      &\to& \tilde g + \tilde g \\
q + \bar q &\to& \tilde g + \tilde g \\
g + q      &\to& \tilde g + \tilde q \\
g + g      &\to& \tilde q + \tilde{\bar q} \\
q + \bar q &\to& \tilde q + \tilde{\bar q} \\
q + q      &\to& \tilde q + \tilde q
\end{eqnarray*}
Left and right squarks are distinguished but assumed to be degenerate.
Masses can be specified using the \verb|GAUGINO|, \verb|SQUARK|, and
\verb|SLEPTON| parameters described in Section 6. No decay modes are
specified, since these depend strongly on the masses. The user can
either add new modes to the decay table (see Section 9) or use the
\verb|FORCE| or \verb|FORCE1| commands (see Section 6).

      If \verb|MSSMA|, \verb|MSSMB|, and \verb|MSSMC| are specified,
then the ISASUSY package is used to calculate the masses and decay
modes in the minimal supersymmetric extension of the standard model
(MSSM), assuming SUSY grand unification constraints in the neutralino
and chargino mass matrix but allowing some additional flexibility in
the masses. The scalar particle soft masses are input via
\verb|MSSMi|, so that the physical masses will be somewhat different
due to $D$-term contributions and mixings for 3rd generation sparticles.
$\tilde t_1$ and $\tilde t_2$ production and decays are now included.
The lightest SUSY particle is assumed to be the lightest neutralino
$\tilde Z_1$. If the \verb|MSSMi| parameters are specified, then the
following additional processes are included using the MSSM couplings
for the production cross sections:
\begin{eqnarray*}
g + q    &\to& \tilde Z_i + \tilde q, \quad \tilde W_i + \tilde q \\
q + \bar q &\to& \tilde Z_i + \tilde g, \quad \tilde W_i + \tilde g \\
q + \bar q &\to& \tilde W_i + \tilde Z_j \\
q + \bar q &\to& \tilde W_i^+ + \tilde W_j^- \\
q + \bar q &\to& \tilde Z_i + \tilde Z_j \\
q + \bar q &\to& \tilde\ell^+ + \tilde\ell^-, \quad \tilde\nu + \tilde\nu
\end{eqnarray*}
Processes can be selected using the optional parameters described in
Section 6 below.

      Beginning with Version 7.42, matrix elements are taken into
account in the event generator as well as in the calculation of decay
widths for MSSM three-body decays of the form $\tilde A \to \tilde B f
\bar f$, where $\tilde A$ and $\tilde B$ are gluinos, charginos, or
neutralinos. This is implemented by having ISASUSY save the poles and
their couplings when calculating the decay width and then using these
to reconstruct the matrix element. Other three-body decays may be
included in the future. Decays selected with \verb|FORCE| use the
appropriate matrix elements.

      An optional keyword \verb|MSSMD| can be used to specify the second
generation masses, which otherwise are assumed degenerate with the first
generation. An optional keyword \verb|MSSME| can be used to specify
values of the $U(1)$ and $SU(2)$ gaugino masses at the weak scale rather
than using the default grand unification values. The chargino and
neutralino masses and mixings are then computed using these values.

\subsubsection{SUSY Models} The 24 MSSMi parameters describe the MSSM at
the weak scale with the additional assumptions of exact flavor and $CP$
conservation; the general MSSM has 105 parameters. These weak-scale
SUSY-breaking parameters presumably arise from spontaneous SUSY breaking
in a hidden sector that is communicated to the MSSM at some scale $M \gg
M_Z$. There are a number of plausible models in which this symmetry
breaking is simple, so that the MSSM at the high scale involves only a
small number of parameters. These are then related to those at the weak
scale by the renormalization group equations (RGE's).

      Isajet therefore implements in subroutine SUGRA the complete
2-loop RGE's for the gauge couplings, Yukawa couplings, and soft
breaking terms. Contributions from right-handed neutrinos are
optionally included. The RGE's are solved iteratively, running from
the weak scale to the high scale $M$ and back using Runge-Kutta
integration. After each iteration the SUSY masses are recalculated,
and the renormalization group improved 1-loop corrected 
Higgs potential is calculated and minimized. These results are
used to modify the RGE $\beta$-functions appropriately as each
threshold is crossed during the next iteration. Beginning with Version
7.65, the constant parts as well as the logarithms from these
thresholds are included using the results of Pierce {\it et al.},
Nucl.\ Phys.\ {\bf B491}, 3 (1997). The whole process is repeated,
increasing the number of Runge-Kutta steps by a factor of 1.2 for each
iteration, until all the RGE variables except $\mu$ and $B$ differ by
less than 0.3\%. Since $\mu$ and $B$ vary rapidly near the weak scale,
they are only required to differ by less than 5\%. The requirement of
good electroweak symmetry breaking, $\mu^2>0$, is only imposed after
the iterative solution has converged.

\medskip

      A number of different models for SUSY breaking at the high scale
are included in ISAJET. The \verb|SUGRA| parameters must be
specified for the
minimal supergravity framework. This assumes that the gauge couplings
unify at the GUT scale, $M \sim 10^{16}\,{\rm GeV}$, defined by
$\alpha_2=\alpha_1$. SUSY breaking occurs at that scale with universal
soft breaking terms produced by gravitational interactions with a hidden
sector. The parameters of the model are
\begin{itemize}
\item $m_0$: the common scalar mass at the GUT scale;
\item $m_{1/2}$: the common gaugino mass at the GUT scale;
\item $A_0$: the common soft trilinear SUSY breaking parameter at the
GUT scale;
\item $\tan\beta$: the ratio of Higgs vacuum expectation values at the
electroweak scale;
\item $\sgn\mu=\pm1$: the sign of the Higgsino mass term.
\end{itemize}
An attractive feature of this model is that the Higgs are unified with
the other scalars at the GUT scale but $m_{H_u}^2$ is driven negative by
the large top Yukawa coupling $f_t$. Isajet imposes this radiative
symmetry breaking for the SUGRA model but not other possible constraints
such as $b$-$\tau$ unification or limits on proton decay.

      The SUGRA model with exact compling constant unification produces
too large a value of $\alpha_s$ at the weak scale. The default is to use
the experimental value, assuming that threshold effects at the GUT scale
produce this. Exact unification can also be imposed.

      The assumption of universality at the GUT scale is rather
restrictive and may not be valid. A variety of non-universal SUGRA
(NUSUGRA) models can be generated using the \verb|NUSUG1|, \dots,
\verb|NUSUG5| keywords. These might be used to study how well one could
test the minimal SUGRA model. The keyword \verb|SSBCSC| can be used to
specify an alternative scale (i.e., not the coupling constant
unification scale) for the RGE boundary conditions.

\medskip

      An alternative to the SUGRA model is the Gauge Mediated SUSY
Breaking (GMSB) model of Dine, Nelson, and collaborators. In this model
SUSY breaking is communicated through gauge interactions with messenger
fields at a scale $M_m$ small compared to the Planck scale and are
proportional to gauge couplings times $\Lambda_m$. Since $M_m$ is small
and the masses at it are the same for each generation, there are no
flavor changing neutral currents. The messenger fields should form
complete $SU(5)$ representations to preserve the unification of the
coupling constants. The parameters of the GMSB model, which are
specified by the \verb|GMSB| keyword, are
\begin{itemize}
\item $\Lambda_m = F_m/M_m$: the scale of SUSY breaking, typically
10--$100\,{\rm TeV}$;
\item $M_m > \Lambda_m$: the messenger mass scale; 
\item $N_5$: the equivalent number of $5+\bar5$ messenger fields.
\item $\tan\beta$: the ratio of Higgs vacuum expectation values at the
electroweak scale;
\item $\sgn\mu=\pm1$: the sign of the Higgsino mass term;
\item $C_{\rm grav}\ge1$: the ratio of the gravitino mass to the value it
would have had if the only SUSY breaking scale were $F_m$.
\end{itemize}
In GMSB models the lightest SUSY particle is always the nearly massless
gravitino $\tilde G$. The parameter $C_{\rm grav}$ scales the gravitino
mass and hence the lifetime of the next lightest SUSY particle to decay
into it. The \verb|NOGRAV| keyword can be used to turn off gravitino
decays. 

      A variety of non-minimal GMSB models can be generated using
additional parameters set with the GMSB2 keyword. These additional
parameters are
\begin{itemize}
\item $\slashchar{R}$, an extra factor multiplying the gaugino masses
at the messenger scale. (Models with multiple spurions generally have
$\slashchar{R}<1$.)
\item $\delta M_{H_d}^2$, $\delta M_{H_u}^2$, Higgs mass-squared
shifts relative to the minimal model at the messenger scale. (These
might be expected in models which generate $\mu$ realistically.)
\item $D_Y(M)$, a $U(1)_Y$ messenger scale mass-squared term
($D$-term) proportional to the hypercharge $Y$.
\item $N_{5_1}$, $N_{5_2}$, and $N_{5_3}$, independent numbers of
gauge group messengers. They can be non-integer in general.
\end{itemize}
For discussions of these additional parameters, see S. Dimopoulos, S.
Thomas, and J.D. Wells, hep-ph/9609434, Nucl.\ Phys.\ {\bf B488}, 39
(1997), and S.P. Martin, hep-ph/9608224, Phys.\ Rev.\ {\bf D55}, 3177
(1997).

      Gravitino decays can be included in the general MSSM framework by
specifying a gravitino mass with \verb|MGVTNO|. The default is that such
decays do not occur.

\medskip

     Another alternative SUSY model choice allowed is anomaly-mediated
SUSY breaking, developed by Randall and Sundrum.  In this model, it is
assumed that SUSY breaking takes place in other dimensions, and SUSY
breaking is communicated to the visible sector via the superconformal
anomaly. In this model, the lightest SUSY particle is usually the
neutralino which is nearly pure wino-like. The chargino is nearly mass
degenerate with the lightest neutralino. It can be very long lived, or
decay into a very soft pion plus missing energy.  The model incorporated
in ISAJET, based on work by Ghergetta, Giudice and Wells
(hep-ph/9904378), and by Feng and Moroi (hep-ph/9907319) adds a
universal contribution $m_0^2$ to all scalar masses to avoid problems
with tachyonic scalars. The parameters of the model, which can be set
via the \verb|AMSB| keyword, are
\begin{itemize}
\item	$m_0$: Common scalar mass;
\item 	$m_{3/2}$: Gravitino mass, typically $\simge 10\,{\rm TeV}$ since 
$m_i = (\beta_i/g_i) m_{3/2}$.
\item	$\tan\beta$: Usual ratio of vev's at weak scale;
\item	$\sgn\mu$: Usual sign of $\mu$, $\pm1$.
\end{itemize}
Care should be taken with the chargino decay, since it may have
macroscopic decay lengths, or even decay outside the detector.

\medskip

     Since neutrinos seem to have mass, the effect of a massive
right-handed neutrino has been included in ISAJET, when calculating the
sparticle mass spectrum. If the keyword \verb|SUGRHN| is used, then the
user must input the 3rd generation neutrino mass (at scale $M_Z$) in
units of GeV, and the intermediate scale right handed neutrino Majorana
mass $M_N$, also in GeV. In addition, one must specify the soft
SUSY-breaking masses $A_n$ and $m_{\tilde\nu_R}$ valid at the GUT scale.
Then the neutrino Yukawa coupling is computed in the simple see-saw
model, and renormalization group evolution includes these effects
between $M_{GUT}$ and $M_N$. Finally, to facilitate modeling of $SO(10)$
SUSY-GUT models, loop corrections to 3rd generation fermion masses have
been included in the ISAJET SUSY models.

\bigskip\bigskip

      The ISASUSY program can also be used independently of the rest of
ISAJET, either to produce a listing of decays or in conjunction with
another event generator. Its physics assumptions are described in more
detail in Section~\ref{SUSY}. ISASUSY accepts soft SUSY breaking
parameters at the weak scale and calculates the masses and decay modes
from them. The ISASUGRA program can also be used independently to solve
the renormalization group equations with SUGRA, NUSUGRA, GMSB, or AMSB
boundary conditions and then to call ISASUSY to calculate the decay
modes. The main programs described in Section~\ref{sugrun} prompt for
interactive input and print the results to a file.

      Generally the MSSM, SUGRA, or GMSB option should be used to study
supersymmetry signatures; the SUGRA or GMSB parameter space is clearly
more manageable. The more general option may be useful to study
alternative SUSY models. It can also be used, e.g., to generate
pointlike color-3 leptoquarks in technicolor models by selecting squark
production and setting the gluino mass to be very large. The MSSM or
SUGRA option may also be used with top pair production to simulate top
decays to SUSY particles.

\subsubsection{$e^+e^-$} An $e^+e^-$ event generator is also included in
ISAJET. The
Standard Model processes included are $e^+e^-$ annihilation through
$\gamma$ and $Z$ to quarks and leptons, and production of $W^+W^-$ and
$Z^0Z^0$ pairs. In contrast to WPAIR and HIGGS for the hadronic
processes, the produced $W$'s and $Z$'s are treated as particles, so
their spins are not properly taken into account in their decays.
(Because the $W$'s and $Z$'s are treated as particles, their decay
modes can be selected using \verb|FORCE| or \verb|FORCE1|, not
\verb|WMODEi|. See Section [6] below.)  Other Standard Model
processes, including $e^+ e^- \to e^+ e^-$ ($t$-channel graph) and $e^+ e^-
\to \gamma \gamma$, are not included.  Once the primary reaction has been
generated, QCD radiation and hadronization are done as for hadronic
processes. 

The $e^+e^-$ generator can be run assuming no initial state
radiation (the default), or an initial state electron structure function
can be used for bremsstrahlung or the combination bremsstrahlung/beamstrahlung
effect. Bremsstrahlung is implemented using the Fadin-Kuraev
$e^-$ distribution function, and can be turned on using the \verb|EEBREM|
command while stipulating the minimal and maximal subprocess energy.
Beamstrahlung is implemented by invoking the \verb|EEBEAM| keyword.
In this case, in addition the beamstrahlung parameter $\Upsilon$ and
longitudinal beam size $\sigma_z$ (in mm) must be given.
The definition for $\Upsilon$ in terms of other beam parameters can be 
found in the article Phys. Rev. D49, 3209 (1994) by Chen, Barklow and Peskin.
The bremsstrahlung structure function is then convoluted with the 
beamstrahlung distribution (as calculated by P. Chen) and a spline fit
is created. Since the cross section can contain large spikes, event generation
can be slow if a huge range of subprocess energy is selected for light 
particles; in these scenarios, \verb|NTRIES| must be increased well beyond
the default value.

      $e^+e^-$ annihilation to SUSY particles is included as well with
complete lowest order diagrams, and cascade decays.  The processes
include
\begin{eqnarray*}
e^+ e^- &\to& \tilde q \tilde q \\
e^+ e^- &\to& \tilde\ell \tilde\ell \\
e^+ e^- &\to& \tilde W_i \tilde W_j \\
e^+ e^- &\to& \tilde Z_i \tilde Z_j \\
e^+ e^- &\to& H_L^0+Z^0,H_H^0+Z^0,H_A^0+H_L^0,H_A^0+H_H^0,H^++H^-
\end{eqnarray*}
Note that SUSY Higgs production via $WW$ and $ZZ$ fusion, which can
dominate Higgs production processes at $\sqrt{s} > 500\,\GeV$,
is not included. Spin correlations are neglected, although 
3-body sparticle decay matrix elements are included.

      $e^+e^-$ cross sections with polarized beams are included for
both Standard Model and SUSY processes. The keyword \verb|EPOL| is
used to set $P_L(e^-)$ and $P_L(e^+)$, where
$$
P_L(e) = (n_L-n_R)/(n_L+n_R)
$$
so that $-1 \le P_L \le +1$. Thus, setting \verb|EPOL| to $-.9,0$ will
yield a 95\% right polarized electron beam scattering on an unpolarized
positron beam.

\subsubsection{Technicolor} Production of a technirho of arbitrary
mass and width decaying into $W^\pm Z^0$ or $W^+ W^-$ pairs. The cross
section is based on an elastic resonance in the $WW$ cross section
with the effective $W$ approximation plus a $W$ mixing term taken from
EHLQ.  Additional technicolor processes may be added in the future.

\subsubsection{Extra Dimensions} The possibility that there might be
more than four space-time dimensions at a distance scale $R$ much larger
than $G_N^{1/2}$ has recently attracted interest. In these theories,
$$ 
G_N = {1 \over 8\pi R^\delta M_D^{2+\delta}}\,, 
$$
where $\delta$ is the number of extra dimensions and $M_D$ is the
$4+\delta$ Planck scale. Gravity deviates from the standard theory at a
distance $R \sim 10^{22/\delta-19}\,{\rm m}$, so $\delta\ge2$ is
required. If $M_D$ is of order $1\,{\rm TeV}$, then the usual heirarchy
problem is solved, although there is then a new heirarchy problem of why
$R$ is so large.

      In such models the graviton will have many Kaluza-Klein
excitations with a mass splitting of order $1/R$. While any individual
mode is suppressed by the four-dimensional Planck mass, the large number
of modes produces a cross section suppressed only by $1/M_D^2$. The
signature is an invisible massive graviton plus a jet, photon, or other
Standard Model particle. The \verb|EXTRADIM| process implements this
reaction using the cross sections of Giudice, Rattazzi, and Wells,
hep-ph/9811291. The number $\delta$ of extra dimensions, the mass scale
$M_D$, and the logical flag \verb|UVCUT| are specified using the keyword
\verb|EXTRAD|. If \verb|UVCUT| is \verb|TRUE|, the cross section is cut
off above the scale $M_D$; the model is not valid if the results depend
on this flag.

\subsection{Multiparton Hard Scattering}

      All the processes listed in Section~\ref{hard} are either $2\to2$
processes like \verb|TWOJET| or $2\to1$ $s$-channel resonance processes
followed by a 2-body decay like \verb|DRELLYAN|. The QCD parton shower
described in Section~\ref{qcdshower} below generates multi-parton final
states starting from these, but it relies on an approximation which is
valid only if the additional partons are collinear either with the
initial or with the final primary ones. Since the QCD shower uses exact
non-colliear kinematics, it in fact works pretty well in a larger region
of phase space, but it is not exact.

      Non-collinear multiparton final states are interesting both in
their own right and as backgrounds for other signatures. Both the matrix
elements and the phase space for multiparton processes are complicated;
they have been incorporated into ISAJET for the first time in
Version~7.45. To calculate the matrix elements we have used the MadGraph
package by Stelzer and Long, Comput.\ Phys.\ Commun.\ {\bf81}, 357
(1994), hep-ph/9401258. This automatically generates the amplitude using
\verb|HELAS|, a formalism by Murayama, Watanabe, and Hagiwarak
KEK-91-11, that calculates the amplitude for any Feynman diagram in
terms of spinnors, vertices, and propagators. The MadGraph code has been
edited to incorporate summations over quark flavors. To do the phase
space integration, we have used a simple recursive algorithm to generate
$n$-body phase space. We have included limits on the total mass of the
final state using the \verb|MTOT| keyword. Limits on the $p_T$ and
rapidity of each final parton can be set via the \verb|PT| and \verb|Y|
keyworks, while limits on the mass of any pair of final partons can be
set via the \verb|MIJTOT| keyword. These limits are sufficient to shield
the infrared and collinear singularities and to render the result
finite. However, the parton shower populates all regions of phase space,
so careful thought is needed to combine the parton-shower based and
multiparton based results.

      While the multiparton formalism is rather general, it still takes
a substantial amount of effort to implement any particular process. So
far only one process has been implemented.

\subsubsection{$Z + {\rm 2\ jets}$} The \verb|ZJJ| process generates a
$Z$ boson plus two jets, including the $q\bar{q} \to Z q \bar{q}$, $gg
\to Z q\bar{q}$, $q\bar{q} \to Zgg$, $qq \to Zqq$, and $gq \to Z gq$
processes. The $Z$ is defined to be jet 1; it is treated in the narrow
resonance approximation and is decayed isotropically. The quarks,
antiquarks, and gluons are defined to be jets 2 and 3 and are
symmetrized in the usual way.

\subsection{QCD Radiative Corrections\label{qcdshower}}

      After the primary hard scattering is generated, QCD radiative
corrections are added to allow the possibility of many jets. This is
essential to get the correct event structure, especially at high
energy.

      Consider the emission of one extra gluon from an initial or a
final quark line,
$$
q(p) \to q(p_1) + g(p_2)
$$
From QCD perturbation theory, for small $p^2$ the cross section is
given by the lowest order cross section multiplied by a factor
$$
\sigma = \sigma_0  \alpha_s(p^2)/(2\pi p^2) P(z)
$$
where $z=p_1/p$ and $P(z)$ is an Altarelli-Parisi function. The same
form holds for the other allowed branchings,
\begin{eqnarray*}
g(p) &\to& g(p_1) + g(p_2) \\
g(p) &\to& q(p_1) + \bar q(p_2)
\end{eqnarray*}
These factors represent the collinear singularities of perturbation
theory, and they produce the leading log QCD scaling violations for the
structure functions and the jet fragmentation functions. They also
determine the shape of a QCD jet, since the jet $M^2$ is of order
$\alpha_s p_t^2$ and hence small.

      The branching approximation consists of keeping just these
factors which dominate in the collinear limit but using exact,
non-collinear kinematics. Thus higher order QCD is reduced to a
classical cascade process, which is easy to implement in a Monte Carlo
program. To avoid infrared and collinear singularities, each parton in
the cascade is required to have a mass (spacelike or timelike) greater
than some cutoff $t_c$. The assumption is that all physics at lower
scales is incorporated in the nonperturbative model for hadronization.
In ISAJET the cutoff is taken to be a rather large value,
$(6\,\GeV)^2$, because independent fragmentation is used for the jet 
fragmentation; a low cutoff would give too many hadrons from
overlapping partons. It turns out that the branching approximation not
only incorporates the correct scaling violations and jet structure but
also reproduces the exact three-jet cross section within factors of
order 2 over all of phase space.

      This approximation was introduced for final state radiation by
Fox and Wolfram. The QCD cascade is determined by the probability for
going from mass $t_0$ to mass $t_1$ emitting no resolvable radiation.
For a resolution cutoff $z_c < z < 1-z_c$, this is given by a simple
expression,
$$      
P(t_0,t_1)=\left(\alpha_s(t_0)/\alpha_s(t_1)\right)^{2\gamma(z_c)/b_0}
$$
where
$$
\gamma(z_c)=\int_{z_c}^{1-z_c} dz\,P(z),\qquad
b_0=(33-2n_f)/(12\pi)
$$
Clearly if $P(t_0,t_1)$ is the integral probability, then $dP/dt_1$ is
the probability for the first radiation to occur at $t_1$. It is
straightforward to generate this distribution and then iteratively to
correct it to get a cutoff at fixed $t_c$ rather than at fixed $z_c$.

      For the initial state it is necessary to take account of the
spacelike kinematics and of the structure functions. Sjostrand has
shown how to do this by starting at the hard scattering and evolving
backwards, forcing the ordering of the spacelike masses $t$. The
probability that a given step does not radiate can be derived from the
Altarelli-Parisi equations for the structure functions. It has a form
somewhat similar to $P(t_0,t_1)$ but involving a ratio of the structure
functions for the new and old partons. It is possible to find a bound
for this ratio in each case and so to generate a new $t$ and $z$ as for
the final state. Then branchings for which the ratio is small are
rejected in the usual Monte Carlo fashion. This ratio suppresses the
radiation of very energetic partons. It also forces the branching $g
\to t + \bar t$ for a $t$ quark if the $t$ structure function vanishes
at small momentum transfer.

      At low energies, the branching of an initial heavy quark into a
gluon sometimes fails; these events are discarded and a warning is
printed.

      Since $t_c$ is quite large, the radiation of soft gluons is cut
off. To compensate for this, equal and opposite transverse boosts are
made to the jet system and to the beam jets after fragmentation with a
mean value
$$
\langle p_t^2\rangle = (.1\,\GeV) \sqrt{Q^2}
$$
The dependence on $Q^2$ is the same as the cutoff used for DRELLYAN and
the coefficient is adjusted to fit the $p_t$ distribution for the $W$.

      Radiation of gluons from gluinos and scalar quarks is also
included in the same approximation, but the production of gluino or
scalar quark pairs from gluons is ignored. Very little radiation is
expected for heavy particles produced near threshold.

      Radiation of photons, $W$'s, and $Z$'s from final state quarks is
treated in the same approximation as QCD radiation except that the
coupling constant is fixed. Initial state electroweak radiation is not
included; it seems rather unimportant. The $W^+$'s, $W^-$'s and $Z$'s
are decayed into the modes allowed by the \verb|WPMODE|, \verb|WMMODE|,
and \verb|Z0MODE| commands respectively. {\it Warning:} The branching
ratios implied by these commands are not included in the cross section
because an arbitrary number of $W$'s and $Z$'s can in principle be
radiated.

\subsection{Jet Fragmentation:}

      Quarks and gluons are fragmented into hadrons using the
independent fragmentation ansatz of Field and Feynman. For a quark
$q$, a new quark-antiquark pair $q_1 \bar q_1$ is generated with
$$
u : d : s = .43 : .43 : .14
$$
A meson $q \bar q_1$ is formed carrying a fraction $z$ of the momentum,
$$
E' + p_z' = z (E + p_z)
$$
and having a transverse momentum $p_t$ with $\langle p_t \rangle =
0.35\,\GeV$. Baryons are included by generating a diquark with
probability 0.10 instead of a quark; adjacent diquarks are not
allowed, so no exotic mesons are formed. For light quarks $z$ is
generated with the splitting function
$$
f(z) = 1-a + a(b+1)(1-z)^b, \qquad
a = 0.96, b = 3
$$
while for heavy quarks the Peterson form
$$
f(z) = x (1-x)^2 / ( (1-x)^2 + \epsilon x )^2
$$
is used with $\epsilon = .80 / m_c^2$ for $c$ and $\epsilon = .50 /
m_q^2$ for $q = b, t, y, x$. These values of $\epsilon$ have been
determined by fitting PEP, PETRA, and LEP data with ISAJET and should
not be compared with values from other fits. Hadrons with longitudinal
momentum less than zero are discarded. The procedure is then iterated
for the new quark $q_1$ until all the momentum is used. A gluon is
fragmented like a randomly selected $u$, $d$, or $s$ quark or
antiquark. 

      In the fragmentation of gluinos and scalar quarks, supersymmetric
hadrons are not distinguished from partons. This should not matter
except possibly for very light masses. The Peterson form for $f(x)$ is
used with the same value of epsilon as for heavy quarks, $\epsilon =
0.5 / m^2$.

      Independent fragmentation correctly describes the fast hadrons in
a jet, but it fails to conserve energy or flavor exactly. Energy
conservation is imposed after the event is generated by boosting the
hadrons to the appropriate rest frame, rescaling all of the
three-momenta, and recalculating the energies.

\subsection{Beam Jets}

      There is now experimental evidence that beam jets are different in
minimum bias events and in hard scattering events. ISAJET therefore uses
similar a algorithm but different parameters in the two cases.

      The standard models of particle production are based on pulling
pairs of particles out of the vacuum by the QCD confining field,
leading naturally to only short-range rapidity correlations and to
essentially Poisson multiplicity fluctuations. The minimum bias data
exhibit KNO scaling and long-range correlations. A natural explanation
of this was given by the model of Abramovskii, Kanchelli and Gribov.
In their model the basic amplitude is a single cut Pomeron with
Poisson fluctuations around an average multiplicity $\langle n
\rangle$, but unitarity then produces graphs giving $K$ cut Pomerons
with multiplicity $K\langle n \rangle$.

      A simplified version of the AKG model is used in ISAJET. The
number of cut Pomerons is chosen with a distribution adjusted to fit the
data. For a minimum bias event this distribution is
$$
P(K) = ( 1 + 4 K^2 ) \exp{-1.8 K}
$$
while for hard scattering
$$
P(1) \to 0.1 P(1),\quad  P(2) \to 0.2 P(2),\quad  P(3) \to 0.5 P(3)
$$
For each side of each event an $x_0$ for the leading baryon is selected
with a distribution varying from flat for $K = 1$ to like that for
mesons for large K:
$$
f(x) = N(K) (1- x_0)^c(K),\qquad c(K) = 1/K + ( 1 - 1/K ) b(s)
$$
The $x_i$ for the cut Pomerons are generated uniformly and then
rescaled to $1-x_0$. Each cut Pomeron is then hadronized in its own
center of mass using a modified independent fragmentation model with
an energy dependent splitting function to reproduce the rise in
$dN/dy$:
$$
f(x) = 1 - a  +  a(b(s) + 1)^ b(s),\qquad 
b(s) = b_0 + b_1  \log(s)
$$
The energy dependence is put into $f(x)$ rather than $P(K)$ because in
the AKG scheme the single particle distribution comes only from the
single chain. The probabilities for different flavors are taken to be
$$
u : d : s = .46 : .46 : .08
$$
to reproduce the experimental $K/\pi$ ratio.
\newpage
\section{Sample Jobs\label{SAMPLE}}

      The simplest ISAJET job reads a user-supplied parameter file and
writes a data file and a listing file. The following is an example of
a parameter file which generates each type of event:
\begin{verbatim}
SAMPLE TWOJET JOB
800.,100,2,50/
TWOJET
PT
50,100,50,100/
JETTYPE1
'GL'/
JETTYPE2
'UP','UB','DN','DB','ST','SB'/
END
SAMPLE DRELLYAN JOB
800.,100,2,50/
DRELLYAN
QMW
80,100/
WTYPE
'W+','W-'/
END
SAMPLE MINBIAS JOB
800.,100,2,50/
MINBIAS
END
SAMPLE WPAIR JOB
800.,100,2,50/
WPAIR
PT
50,100,50,100/
JETTYPE1
'W+','W-','Z0'/
JETTYPE2
'W+','W-','Z0'/
WMODE1
'E+','E-','NUS'/
WMODE2
'QUARKS'/
END
SAMPLE HIGGS JOB FOR SSC
40000,100,1,1/
HIGGS
QMH
400,1600/
HMASS
800/
JETTYPE1
'Z0'/
JETTYPE2
'Z0'/
WMODE1
'MU+','MU-'/
WMODE2
'E+','E-'/
PT
50,20000,50,20000/
END
SAMPLE SUSY JOB
1800,100,1,10/
SUPERSYM
PT
50,100,50,100/
JETTYPE1
'GLSS','SQUARKS'/
JETTYPE2
'GLSS','SQUARKS'/
GAUGINO
60,1,40,40/
SQUARK
80.3,80.3,80.5,81.6,85,110/
FORCE
29,30,1,-1/
FORCE
21,29,1/
FORCE
22,29,2/
FORCE
23,29,3/
FORCE
24,29,4/
FORCE
25,29,5/
FORCE
26,29,6/
END
SAMPLE MSSM JOB FOR TEVATRON
1800.,100,1,1/
SUPERSYM
BEAMS
'P','AP'/
MSSMA
200,-200,500,2/
MSSMB
200,200,200,200,200/
MSSMC
200,200,200,200,200,0,0,0/
JETTYPE1
'GLSS'/
JETTYPE2
'SQUARKS'/
PT
100,300,100,300/
END
SAMPLE MSSM SUGRA JOB FOR LHC
14000,100,1,10/
SUPERSYM
PT
50,500,50,500/
SUGRA
247,302,-617.5,10,-1/
TMASS
175/
END
SAMPLE SUGRA HIGGS JOB USING DEFAULT QMH RANGE
14000,100,20,50/
HIGGS
SUGRA
200,200,0,2,+1/
HTYPE
'HA0'/
JETTYPE1
'GAUGINOS','SLEPTONS'/
JETTYPE2
'GAUGINOS','SLEPTONS'/
END
SAMPLE E+E- TO SUGRA JOB WITH POLARIZED BEAMS AND BREM/BEAMSTRAHLUNG
500.,100,1,1/
E+E-
SUGRA
125,125,0,3,1/
TMASS
175,-1,1/
EPOL
-.9,0./
EEBEAM
200.,500.,.1072,.12/
JETTYPE1
'ALL'/
JETTYPE2
'ALL'/
NTRIES
10000/
END
SAMPLE WH JOB
2000,100,0,0/
WHIGGS
BEAMS
'P','AP'/
HMASS
100./
JETTYPE1
'W+','W-','HIGGS'/
JETTYPE2
'W+','W-','HIGGS'/
WMODE1
'ALL'/
WMODE2
'ALL'/
PT
10,300,10,300/
END 
SAMPLE EXTRA DIMENSIONS JOB
14000,100,1,100/
EXTRADIM
QMW
5,1000/
QTW
500,1000/
EXTRAD
2,1000,.FALSE./
END
SAMPLE ZJJ JOB AT LHC
14000,100,1,100/
ZJJ
PT
20,7000,20,7000,20,7000/
MIJLIM
0,0,20,7000/
MTOT
100,500/
NSIGMA
200/
NTRIES
10000/
END
STOP
\end{verbatim}
\noindent See Section~\ref{INPUT} of this manual for a complete list
of the possible commands in a parameter file. Note that all input to
ISAJET must be in {\it UPPER} case only.

      Subroutine RDTAPE is supplied to read events from an ISAJET data
file, which is a machine-dependent binary file. It restores the event
data to the FORTRAN common blocks described in Section~\ref{OUTPUT}.
The skeleton of an analysis job using HBOOK and PAW from the CERN
Program Library is provided in patch ISAPLT but is not otherwise
supported. A Zebra output format based on code from the D0
Collaboration is also provided in patch ISAZEB; see the separate
documentation in patch ISZTEXT.

\subsection{DEC VMS}

      On a VAX or ALPHA running VMS, ISAJET can be compiled by
executing the .COM file contained in P=ISAUTIL,D=MAKEVAX. Extract this
deck as ISAMAKE.COM and type
\begin{verbatim}
@ISAMAKE
\end{verbatim}
This will run YPATCHY with the pilot patches described in
Section~\ref{PATCHY} and the VAX flag to extract the source code,
decay table, and documentation. The source code is compiled and made
into a library, which is linked with the following main program,
\begin{verbatim}
      PROGRAM ISARUN
C          MAIN PROGRAM FOR ISAJET
      OPEN(UNIT=1,STATUS='OLD',FORM='FORMATTED',READONLY)
      OPEN(UNIT=2,STATUS='NEW',FORM='UNFORMATTED')
      OPEN(UNIT=3,STATUS='OLD',FORM='FORMATTED')
      OPEN(UNIT=4,STATUS='NEW',FORM='FORMATTED')
      CALL ISAJET(-1,2,3,4)
      STOP
      END
\end{verbatim}
to produce ISAJET.EXE. Two other executables, ISASUSY.EXE and
ISASUGRA.EXE, will also be produced to calculate SUSY masses and decay
modes without generating events. Temporary files can be removed by
typing
\begin{verbatim}
@ISAMAKE CLEAN
\end{verbatim}

      Create an input file \verb|JOBNAME.PAR| following the examples
above or the instructions in Section~\ref{INPUT} and run ISAJET with
the command
\begin{verbatim}
@ISAJET JOBNAME
\end{verbatim}
using the ISAJET.COM file contained P=ISAUTIL,D=RUNVAX. This will
create a binary output file \verb|JOBNAME.DAT| and a listing file
\verb|JOBNAME.LIS|. Analyze the output data using the commands
described in Section~\ref{TAPE}.

      There is also an simple interactive interface to ISAJET which
will prompt the user for commands, write a parameter file, and
optionally execute it.

\subsection{IBM VM/CMS}

      On an IBM mainframe running VM/CMS, run YPATCHY with the pilot
patches described in Section~\ref{PATCHY} and the IBM flag to extract
the source code, decay table, and documentation. Compile the source
code and link it with the main program
\begin{verbatim}
      PROGRAM ISARUN
C          MAIN PROGRAM FOR ISAJET
      OPEN(UNIT=1,STATUS='OLD',FORM='FORMATTED')
      OPEN(UNIT=2,STATUS='NEW',FORM='UNFORMATTED')
      OPEN(UNIT=3,STATUS='OLD',FORM='FORMATTED')
      OPEN(UNIT=4,STATUS='NEW',FORM='FORMATTED')
      CALL ISAJET(-1,2,3,4)
      STOP
      END
\end{verbatim}
to make ISAJET MODULE.

      Create a file called \verb|JOBNAME INPUT| containing ISAJET
input commands following the examples above or the instructions in
Section~\ref{INPUT}. Then run ISAJET using ISAJET EXEC, which is
contained in P=ISAUTIL,D=RUNIBM.  The events will be produced on
\verb|JOBNAME DATA A| and the listing on \verb|JOBNAME OUTPUT A|.

\subsection{Unix}

      The Makefile contained in P=ISAUTIL,D=MAKEUNIX has been tested
on DEC Ultrix, Hewlett Packard HP-UX, IBM RS/6000 AIX, Linux, Silicon
Graphics IRIX, Sun SunOS, and Sun Solaris. It should work with minor
modifications on almost any Unix system with /bin/csh, \verb|ypatchy|
or \verb|nypatchy|, and a reasonable Fortran 77 compiler. Extract the
Makefile and edit it, changing the installation parameters to reflect
your system. Note in particular that CERNlib is usually compiled with
underscores postpended to all external names; you must choose the
appropriate compiler option if you intend to link with it. Then type
\begin{verbatim}
make
\end{verbatim}
This should produce an executable \verb|isajet.x| for the event
generator, which links the code with the following main program:
\begin{verbatim}
      PROGRAM RUNJET
      CHARACTER*60 FNAME
      READ 1000, FNAME
1000  FORMAT(A)
      PRINT 1020, FNAME
1020  FORMAT(1X,'Data file      = ',A)
      OPEN(2,FILE=FNAME,STATUS='NEW',FORM='UNFORMATTED')
      READ 1000, FNAME
      PRINT 1030, FNAME
1030  FORMAT(1X,'Parameter file = ',A)
      OPEN(3,FILE=FNAME,STATUS='OLD',FORM='FORMATTED')
      READ 1000, FNAME
      PRINT 1040, FNAME
1040  FORMAT(1X,'Listing file   = ',A)
      OPEN(4,FILE=FNAME,STATUS='NEW',FORM='FORMATTED')
      READ 1000, FNAME
      OPEN(1,FILE=FNAME,STATUS='OLD',FORM='FORMATTED')
      CALL ISAJET(-1,2,3,4)
      STOP
      END
\end{verbatim}
Two other executables, \verb|isasusy.x| and \verb|isasugra.x|, will
also be produced to calculate SUSY masses and decay modes without
generating events. Type
\begin{verbatim}
make clean
\end{verbatim}
to delete the temporary files.

      Most Unix systems do not allow two jobs to read the same decay
table file at the same time. The shell script in P=ISAUTIL,D=RUNUNIX
copies the decay table to a temporary file to avoid this problem.
Extract this file as \verb|isajet|. Create an input file
\verb|jobname.par| following the examples above or the instructions in
Section~\ref{INPUT} and run ISAJET with the command
\begin{verbatim}
isajet jobname
\end{verbatim}
This will create a binary output file \verb|jobname.dat| and a listing
file \verb|jobname.lis|. Analyze the output data using the commands
described in Section~\ref{TAPE}.

      This section only describes running ISAJET as a standalone
program and generating output in machine-dependent binary form. The
user may elect to analyze events as they are generated; this is
discussed in Section~\ref{MAIN} of this manual.
\newpage
\section{Patchy and PAM Organization\label{PATCHY}}

      Patchy is a code management system developed at CERN and used to
maintain the CERN Library. It is used to provide versions of ISAJET for
a wide variety of computers. Instructions for using PATCHY are
available from \verb|http://wwwinfo.cern.ch/asdoc/Welcome.html|.

      A master source file in Patchy is called a ``PAM.'' The ISAJET
PAM contains all the source code and documentation plus Patchy
commands to include common blocks and to select the desired version. It
is divided into the following patches: 

      \verb|ISACDE|: contains all common blocks, etc. These are divided
into decks based on their usage.

      \verb|ISADATA|: contains block data ALDATA. This must always be
loaded when using ISAJET.

      \verb|ISAJET|: contains the code for generating events. Each
subroutine is in a separate deck with the same name.

      \verb|ISASSRUN|: contains the main program for ISASUSY, which
prompts for input parameters and prints out all the decay modes. It is
selected by \verb|*ISASUSY|.

      \verb|ISASUSY|: contains code to calculate all the decay widths
and branching fractions in the minimal supersymmetric model.

      \verb|ISATAPE|: contains the code for reading and writing tapes,
again with each subroutine on a separate deck.

      \verb|ISARUN|: contains a main program and a simple interactive
interface.  It is selected by \verb|IF=INTERACT|.

      \verb|ISAZEB|: contains Zebra format output routines, an
alternative to the ISATAPE routines.

      \verb|ISZRUN|: contains the analog of ISAPLT for the Zebra
format. 

      \verb|ISAPLT|: contains a simple calorimeter simulation and the
skeleton of a histogramming job using HBOOK.

      \verb|ISATEXT|: contains the instructions for using ISAJET, i.e.
the text of this document. It also includes the documentation for
ISASUSY.

      \verb|ISZTEXT|: contains the instructions for the Zebra output
routines and a description of the Zebra banks.

      \verb|ISADECAY|: contains the input decay table.

      The code is actually maintained using RCS on a Silicon Graphics
computer at BNL. Patchy is used primarily to handle common blocks and
machine dependent code.

      The input to YPATCHY must contain both \verb|+USE| cards, which
define the wanted program version, and \verb|+EXE| cards, which
determine which patches or decks are written to the ASM file. To
facilitate this selection, the ISAJET PAM contains the following pilot
patches:

      \verb|*ISADECAY|: USE selects ISADECAY and all corrections to it.

      \verb|*ISAJET|: USE selects ISACDE, ISADATA, ISAJET, ISATAPE,
ISARUN and all corrections to them. Note that ISARUN is not actually
selected without \verb|+USE,INTERACT|.

      \verb|*ISAPLT|: USE selects ISACDE, ISAPLT, and all corrections
to them.

      \verb|*ISASUSY|: USE selects CDESUSY, ISASUSY, and ISASSRUN to
create a program to calculate all the MSSM decay modes.

      \verb|*ISATEXT|: USE selects ISACDE, ISATEXT, and all corrections
to them. 

      \verb|*ISAZEB|: USE selects ISAJET with a Zebra output format.

      \verb|*ISZRUN|: USE selects the Zebra analysis package.

      Patches are provided to select the machine dependent features for
specific computers or operating systems:

      \verb|ANSI|: ANSI standard Fortran (no time or date functions)

      \verb|APOLLO|: APOLLO -- only tested by CERN

      \verb|CDC|: CDC 7600 and 60-bit CYBER (obsolete)

      \verb|CRAY|: CRAY with UNICOS

      \verb|DECS|: DEC Station with Ultrix 

      \verb|ETA|: ETA 10 running Unix System V (obsolete)

      \verb|HPUX|: HP/9000 7xx running Unix System V

      \verb|IBM|: IBM 370 and 30xx running VM/CMS 

      \verb|IBMRT|: IBM RS/6000 running AIX 3.x or 4.x

      \verb|LINUX|: PC running Linux with f2c/gcc or g77 compiler

      \verb|SGI|: Silicon Graphics running IRIX

      \verb|SUN|: Sun Sparcstation running SUNOS or Solaris

      \verb|VAX|: DEC VAX or Alpha running VMS

\noindent These patches in turn select a variety of patches and IF
flags, allowing one to select more specific features, as indicated
below. (Replace \verb|&| by \verb|+| everywhere.)
\begin{verbatim}
&PATCH,ANSI.                      GENERIC ANSI FORTRAN.
&USE,DOUBLE.                      DOUBLE PRECISION.
&USE,STDIO.                       STANDARD FORTRAN 77 TAPE INPUT/OUTPUT.
&USE,MOVEFTN.                     FORTRAN REPLACEMENT FOR MOVLEV.
&USE,RANFFTN,IF=-CERN.            FORTRAN RANF.
&USE,RANFCALL.                    STANDARD RANSET AND RANGET CALLS.
&USE,NORANLUX,IF=-RANLUX.         NO RANLUX RANDOM NUMBERS.
&USE,NOCERN,IF=-CERN.             NO CERN LIBRARY.
&EOD

&PATCH,APOLLO.
&DECK,BLANKDEK.
&USE,DOUBLE.                      DOUBLE PRECISION.
&USE,STDIO.                       STANDARD FORTRAN 77 TAPE INPUT/OUTPUT.
&USE,MOVEFTN.                     FORTRAN REPLACEMENT FOR MOVLEV.
&USE,RANFFTN,IF=-CERN.            FORTRAN RANF.
&USE,RANFCALL.                    STANDARD RANSET AND RANGET CALLS.
&USE,NORANLUX,IF=-RANLUX.         NO RANLUX RANDOM NUMBERS.
&USE,NOCERN,IF=-CERN.             NO CERN LIBRARY.
&USE,IMPNONE.                     IMPLICIT NONE
&EOD.

&PATCH,CDC.                       CDC 7600 OR CYBER 175.
&USE,SINGLE.                      SINGLE PRECISION.
&USE,LEVEL2.                      LEVEL 2 STORAGE.
&USE,CDCPACK.                     PACK 2 WORDS PER WORD FOR INPUT/OUTPUT.
&USE,RANFCALL.                    STANDARD RANSET AND RANGET CALLS.
&USE,NOCERN,IF=-CERN.             NO CERN LIBRARY.
&EOD

&PATCH,CRAY.                      CRAY XMP OR 2.
&USE,SINGLE.                      SINGLE PRECISION.
&USE,STDIO.                       STANDARD FORTRAN 77 TAPE INPUT/OUTPUT.
&USE,MOVEFTN.                     FORTRAN REPLACEMENT FOR MOVLEV.
&USE,NOCERN,IF=-CERN.             NO CERN LIBRARY.
&EOD

&PATCH,DECS.                      DEC STATION (ULTRIX)
&USE,SUN.
&EOD

&PATCH,ETA.                       ETA-10.
&USE,SINGLE.                      SINGLE PRECISION.
&USE,STDIO.                       STANDARD FORTRAN 77 TAPE INPUT/OUTPUT.
&USE,MOVEFTN.                     FORTRAN REPLACEMENT FOR MOVLEV.
&USE,RANFCALL.                    STANDARD RANSET AND RANGET CALLS.
&USE,NOCERN,IF=-CERN.             NO CERN LIBRARY.
&EOD

&PATCH,HPUX.                      HP/9000 7XX RUNNING UNIX.
&USE,DOUBLE.                      DOUBLE PRECISION.
&USE,STDIO.                       STANDARD FORTRAN 77 TAPE INPUT/OUTPUT.
&USE,MOVEFTN.                     FORTRAN REPLACEMENT FOR MOVLEV.
&USE,RANFFTN,IF=-CERN.            FORTRAN RANF.
&USE,RANFCALL.                    STANDARD RANSET AND RANGET CALLS.
&USE,NORANLUX,IF=-RANLUX.         NO RANLUX RANDOM NUMBERS.
&USE,NOCERN,IF=-CERN.             NO CERN LIBRARY.
&USE,IMPNONE.                     IMPLICIT NONE
&EOD

&PATCH,IBM.                       IBM 370 OR 30XX.
&USE,DOUBLE.                      DOUBLE PRECISION.
&USE,STDIO.                       STANDARD FORTRAN 77 TAPE INPUT/OUTPUT.
&USE,MOVEFTN.                     FORTRAN REPLACEMENT FOR MOVLEV.
&USE,RANFFTN,IF=-CERN.            FORTRAN RANF.
&USE,RANFCALL.                    STANDARD RANSET AND RANGET CALLS.
&USE,NORANLUX,IF=-RANLUX.         NO RANLUX RANDOM NUMBERS.
&USE,NOCERN,IF=-CERN.             NO CERN LIBRARY.
&EOD

&PATCH,IBMRT.                     IBM RS/6000 WITH AIX 3.1
&USE,DOUBLE.                      DOUBLE PRECISION.
&USE,STDIO.                       STANDARD FORTRAN 77 TAPE INPUT/OUTPUT.
&USE,MOVEFTN.                     FORTRAN REPLACEMENT FOR MOVLEV.
&USE,RANFFTN,IF=-CERN.            FORTRAN RANF.
&USE,RANFCALL.                    STANDARD RANSET AND RANGET CALLS.
&USE,NORANLUX,IF=-RANLUX.         NO RANLUX RANDOM NUMBERS.
&USE,NOCERN,IF=-CERN.             NO CERN LIBRARY.
&USE,IMPNONE.                     IMPLICIT NONE
&EOD

&PATCH,LINUX.                     IBM PC WITH LINUX 1.X
&USE,DOUBLE.                      DOUBLE PRECISION.
&USE,STDIO.  STANDARD FORTRAN 77 TAPE INPUT/OUTPUT.
&USE,MOVEFTN.                     FORTRAN REPLACEMENT FOR MOVLEV.
&USE,RANFFTN,IF=-CERN.            FORTRAN RANF.
&USE,RANFCALL.                    STANDARD RANSET AND RANGET CALLS.
&USE,NORANLUX,IF=-RANLUX.         NO RANLUX RANDOM NUMBERS.
&USE,NOCERN,IF=-CERN.             NO CERN LIBRARY.
&USE,IMPNONE.                     IMPLICIT NONE
&EOD

&PATCH,SGI.
SILICON GRAPHICS 4D/XX.
&USE,DOUBLE.                      DOUBLE PRECISION.
&USE,STDIO.                       STANDARD FORTRAN 77 TAPE INPUT/OUTPUT.
&USE,MOVEFTN.                     FORTRAN REPLACEMENT FOR MOVLEV.
&USE,RANFFTN,IF=-CERN.            FORTRAN RANF.
&USE,RANFCALL.                    STANDARD RANSET AND RANGET CALLS.
&USE,NORANLUX,IF=-RANLUX.         NO RANLUX RANDOM NUMBERS.
&USE,NOCERN,IF=-CERN.             NO CERN LIBRARY.
&EOD

&PATCH,SUN.                       SUN (SPARC)
&USE,DOUBLE.                      DOUBLE PRECISION.
&USE,STDIO.                       STANDARD FORTRAN 77 TAPE INPUT/OUTPUT.
&USE,MOVEFTN.                     FORTRAN REPLACEMENT FOR MOVLEV.
&USE,RANFFTN,IF=-CERN.            FORTRAN RANF.
&USE,RANFCALL.                    STANDARD RANSET AND RANGET CALLS.
&USE,NORANLUX,IF=-RANLUX.         NO RANLUX RANDOM NUMBERS.
&USE,NOCERN,IF=-CERN.             NO CERN LIBRARY.
&EOD

&PATCH,VAX.                       DEC VAX 11/780 OR 8600.
&USE,DOUBLE.                      DOUBLE PRECISION.
&USE,STDIO.                       STANDARD FORTRAN 77 TAPE INPUT/OUTPUT.
&USE,MOVEFTN.                     FORTRAN REPLACEMENT FOR MOVLEV.
&USE,RANFFTN,IF=-CERN.            FORTRAN RANF.
&USE,RANFCALL.                    STANDARD RANSET AND RANGET CALLS.
&USE,NORANLUX,IF=-RANLUX.         NO RANLUX RANDOM NUMBERS.
&USE,NOCERN,IF=-CERN.             NO CERN LIBRARY.
&USE,IMPNONE.                     IMPLICIT NONE
&EOD
\end{verbatim}

      An empty patch INTERACT selects a main program and an interactive
interface which will prompt the user for parameters and do some error
checking. A patch CERN allows ISAJET to take the random number generator
RANF and several other routines from the CERN Library; to use this
include the Patchy command
\begin{verbatim}
&USE,CERN.
\end{verbatim}
Similarly, a patch PDFLIB enables the interface to the PDFLIB parton
distribution compilation by H. Plothow-Besch:
\begin{verbatim}
&USE,PDFLIB
\end{verbatim}
The only internal links with PDFLIB are calls to the routines PDFSET,
PFTOPDG, and DXPDF, and the common blocks W50510 and W50517,
\begin{verbatim}
C          Copy of PDFLIB common block
      COMMON/W50510/IFLPRT
      INTEGER IFLPRT
      SAVE /W50510/
C          Copy of PDFLIB common block
      COMMON/W50517/N6
      INTEGER N6
      SAVE /W50517/
\end{verbatim}
which are used to specify the level of output messages and the logical
unit number for them. A patch RANLUX selects the RANLUX random number
generator instead of the 48-bit congruential generator RANF:
\begin{verbatim}
&USE,RANLUX
\end{verbatim}
RANLUX has a very long period, and each 32-bit integer seed gives an
independent sequence.

      In general it should be sufficient to run YPATCHY with the
following cradle (replace \verb|&| with \verb|+| everywhere):
\begin{verbatim}
&USE,(*ISAJET,*ISATEXT,*ISADECAY,*ISAPLT).     CHOOSE ONE.
&USE,ANSI,DECS,HPUX,IBM,IBMRT,SGI,SUN,....     CHOOSE ONE.
&[USE,INTERACT].                               FOR INTERACTIVE MODE.
&[USE,CERN.]                                   FOR CERN LIBRARY.
&[USE,HBOOK3.]                                 HBOOK 3 FOR ISAPLT.
&EXE.
&PAM.
&QUIT.
\end{verbatim}

      The input to YPATCHY can also contain changes by the user. It is
suggested that these not be made permanent parts of the PAM to avoid
possible conflicts with later corrections.
\newpage
\section{Main Program\label{MAIN}}

      A main program for ISAJET is not supplied in the \verb|isajet.car|
file. To generate events and write them to disk, the user should provide
a main program which opens the files and then calls subroutine ISAJET.
In the following sample, i,j,m,n are arbitrary unit numbers.

      Main program for VMS:
\begin{verbatim}
      PROGRAM RUNJET
C
C          MAIN PROGRAM FOR ISAJET ON BNL VAX CLUSTER.
C
      OPEN(UNIT=i,FILE='$2$DUA14:[ISAJET.ISALIBRARY]DECAY.DAT',
     $STATUS='OLD',FORM='FORMATTED',READONLY)
      OPEN(UNIT=j,FILE='myjob.dat',STATUS='NEW',FORM='UNFORMATTED')
      OPEN(UNIT=m,FILE='myjob.par',STATUS='OLD',FORM='FORMATTED')
      OPEN(UNIT=n,FILE='myjob.lis',STATUS='NEW',FORM='FORMATTED')
C
      CALL ISAJET(+-i,+-j,m,n)
C
      STOP
      END
\end{verbatim}

      Main program for IBM (VM/CMS)
\begin{verbatim}
      PROGRAM RUNJET
C
C          MAIN PROGRAM FOR ISAJET ON IBM ASSUMING FILES HAVE BEEN
C          OPENED WITH FILEDEF.
C
      CALL ISAJET(+-i,+-j,m,n)
C
      STOP
      END
\end{verbatim}

      Main program for Unix; this is created by the standard 
\verb|Makefile|:
\begin{verbatim}
      PROGRAM RUNJET
C
C          Main program for ISAJET on Unix
C
      CHARACTER*60 FNAME
C
C          Open user files
      READ 1000, FNAME
1000  FORMAT(A)
      PRINT 1020, FNAME
1020  FORMAT(1X,'Data file      = ',A)
      OPEN(2,FILE=FNAME,STATUS='NEW',FORM='UNFORMATTED')
      READ 1000, FNAME
      PRINT 1030, FNAME
1030  FORMAT(1X,'Parameter file = ',A)
      OPEN(3,FILE=FNAME,STATUS='OLD',FORM='FORMATTED')
      READ 1000, FNAME
      PRINT 1040, FNAME
1040  FORMAT(1X,'Listing file   = ',A)
      OPEN(4,FILE=FNAME,STATUS='NEW',FORM='FORMATTED')
C          Open decay table
      READ 1000, FNAME
      OPEN(1,FILE=FNAME,STATUS='OLD',FORM='FORMATTED')
C
C          Run ISAJET
      CALL ISAJET(-1,2,3,4)
C
      STOP
      END
\end{verbatim}

      The arguments of ISAJET are tape numbers for files, all of which
should be opened by the main program.

      \verb|TAPEi|: Decay table (formatted). A positive sign prints
the decay table on the output listing. A negative sign suppress
printing of the decay table.

      \verb|TAPEj|: Output file for events (unformatted). A positive
sign writes out both resonances and stable particles. A negative sign
writes out only stable particles.

      \verb|TAPEm|: Commands as defined in Section 6 (formatted).

      \verb|TAPEn|: Output listing (formatted).

\noindent In the sample jobs in Section 3, TAPEm is the default
Fortran input, and TAPEn is the default Fortran output.

\subsection{Interactive Interface}

      To use the interactive interface, replace the call to ISAJET in
the above main program by
\begin{verbatim}
      CALL ISASET(+-i,+-j,m,n)
      CALL ISAJET(+-i,+-j,m,n)
\end{verbatim}
ISASET calls DIALOG, which prompts the user for possible commands,
does a limited amount of error checking, and writes a command file on
TAPEm. This command file is rewound for execution by ISAJET. A main
program is included in patch ISARUN to open the necessary files and to
call ISASET and ISAJET.

\subsection{User Control of Event Loop}

      If the user wishes to integrate ISAJET with another program and
have control over the event generation, he can call the driving
subroutines himself. The driving subroutines are:

      \verb|ISAINI(+-i,+-j,m,n)|: initialize ISAJET. The arguments are
the same as for subroutine ISAJET.

      \verb|ISABEG(IFL)|: begin a run. IFL is a return flag: IFL=0
for a good set of commands; IFL=1001 for a STOP; any other value means
an error.

      \verb|ISAEVT(I,OK,DONE)| generate event I. Logical flag OK
signifies a good event (almost always .TRUE.); logical flag DONE
signifies the end of a run.

      \verb|ISAEND|: end a run.

\noindent There are also subroutines provided to write standard ISAJET
records, or Zebra records if the Zebra option is selected:

      \verb|ISAWBG| to write a begin-of-run record, should be called
immediately after ISABEG

      \verb|ISAWEV| to write an event record, should be called
immediately after ISAEVT

      \verb|ISAWND| to write an end-of-run record, should be called
immediately after ISAEND

      The control of the event loop is somewhat complicated to
accomodate multiple evolution and fragmentation as described in
Section 11. Note in particular that after calling ISAEVT one should
process or write out the event only if OK=.TRUE. The check on the DONE
flag is essential if one is doing multiple evolution and
fragmentation. The following example indicates how events might be
generated, analyzed, and discarded (replace \verb|&| by \verb|+|
everywhere):
\begin{verbatim}
      PROGRAM SAMPLE
C
&SELF,IF=IMPNONE
      IMPLICIT NONE
&SELF
&CDE,ITAPES
&CDE,IDRUN
&CDE,PRIMAR
&CDE,ISLOOP
C
      INTEGER JTDKY,JTEVT,JTCOM,JTLIS,IFL,ILOOP
      LOGICAL OK,DONE
      SAVE ILOOP
C--------------------------------------------------------------------- 
C>         Open files as above
C>         Call user initialization
C--------------------------------------------------------------------- 
C
C          Initialize ISAJET
C
      CALL ISAINI(-i,0,m,n)
    1 IFL=0
      CALL ISABEG(IFL)
      IF(IFL.NE.0) STOP
C
C          Event loop
C
      ILOOP=0
  101 CONTINUE
        ILOOP=ILOOP+1
C          Generate one event - discard if .NOT.OK
        CALL ISAEVT(ILOOP,OK,DONE)
        IF(OK) THEN
C--------------------------------------------------------------------- 
C>         Call user analysis for event
C--------------------------------------------------------------------- 
        ENDIF
      IF(.NOT.DONE) GO TO 101
C
C          Calculate cross section and luminosity
C
      CALL ISAEND
C--------------------------------------------------------------------- 
C>         Call user summary
C--------------------------------------------------------------------- 
      GO TO 1
      END
\end{verbatim}

\subsection{Multiple Event Streams}

      It may be desirable to generate several different kinds of events
simultaneously to study pileup effects. While normally one would want
to superimpose minimum bias or low-pt jet events on a signal of
interest, other combinations might also be interesting. It would be
very inefficient to reinitialize ISAJET for each event. Therefore, a
pair of subroutines is provided to save and restore the context, i.e.
all of the initialization information, in an array. The syntax is
\begin{verbatim}
      CALL CTXOUT(NC,VC,MC)
      CALL CTXIN(NC,VC,MC)
\end{verbatim}
where VC is a real array of dimension MC and NC is the number of words
used, about 20000 in the standard case. If NC exceeds MC, a warning is
printed and the job is terminated. The use of these routines is
illustrated in the following example, which opens the files with names
read from the standard input and then superimposes on each event of
the signal sample three events of a pileup sample. It is assumed that
a large number of events is specified in the parameter file for the
pileup sample so that it does not terminate.
\begin{verbatim}
      PROGRAM SAMPLE
C
C          Example of generating two kinds of events.
C
      CHARACTER*60 FNAME
      REAL VC1(20000),VC2(20000)
      LOGICAL OK1,DONE1,OK2,DONE2
      INTEGER NC1,NC2,IFL,ILOOP,I2,ILOOP2
C
C          Open decay table
      READ 1000, FNAME
1000  FORMAT(A)
      OPEN(1,FILE=FNAME,STATUS='OLD',FORM='FORMATTED')
C          Open user files
      READ 1000, FNAME
      OPEN(3,FILE=FNAME,STATUS='OLD',FORM='FORMATTED')
      READ 1000, FNAME
      OPEN(4,FILE=FNAME,STATUS='NEW',FORM='FORMATTED')
      READ 1000,FNAME
      OPEN(13,FILE=FNAME,STATUS='OLD',FORM='FORMATTED')
      READ 1000,FNAME
      OPEN(14,FILE=FNAME,STATUS='NEW',FORM='FORMATTED')
C
C          Initialize ISAJET
      CALL ISAINI(-1,0,3,4)
      CALL CTXOUT(NC1,VC1,20000)
      CALL ISAINI(-1,0,13,14)
      IFL=0
      CALL ISABEG(IFL)
      IF(IFL.NE.0) STOP1
      CALL CTXOUT(NC2,VC2,20000)
      ILOOP2=0
      CALL user_initialization_routine
C
1     IFL=0
      CALL CTXIN(NC1,VC1,20000)
      CALL ISABEG(IFL)
      CALL CTXOUT(NC1,VC1,20000)
      IF(IFL.NE.0) GO TO 999
      ILOOP=0
C
C          Main event
C
101   CONTINUE
        ILOOP=ILOOP+1
        CALL CTXIN(NC1,VC1,20000)
        CALL ISAEVT(ILOOP,OK1,DONE1)
        CALL CTXOUT(NC1,VC1,20000)
        IF(.NOT.OK1) GO TO 101
        CALL user_analysis_routine
C
C          Pileup
C
        CALL CTXIN(NC2,VC2,20000)
        I2=0
201     CONTINUE
          ILOOP2=ILOOP2+1
          CALL ISAEVT(ILOOP2,OK2,DONE2)
          IF(OK2) I2=I2+1
          IF(DONE2) STOP2
          CALL user_analysis_routine
        IF(I2.LT.3) GO TO 201
        CALL CTXOUT(NC2,VC2,20000)
C
      IF(.NOT.DONE1) GO TO 101
C
C          Calculate cross section and luminosity
C
      CALL CTXIN(NC1,VC1,20000)
      CALL ISAEND
      GO TO 1
C
999   CALL CTXIN(NC2,VC2,20000)
      CALL ISAEND
      CALL user_termination_routine
      STOP
      END
\end{verbatim}
It is possible to superimpose arbitrary combinations of events,
including events of the same reaction type with different parameters.
In general the number of events would be selected randomly based on the
cross sections and the luminosity.

      At this time CTXOUT and CTXIN cannot be used with the Zebra
output routines.

\subsection{Main Programs for ISASUSY and ISASUGRA\label{sugrun}}

     In addition to the event generator, \verb|isajet.car| contains two
programs to calculate SUSY masses and decay modes: ISASUSY, which
accepts weak scale parameters, and ISASUGRA, which calculates the weak
scale parameters from those at some high scale. The main programs, SSRUN
and SUGRUN respectively, are included. They both prompt for interactive
input and then call the appropriate ISAJET subroutines. The output is
formatted and printed by SSPRT and SUGPRT respectively. Executables for
ISASUSY and ISASUGRA are built by the Unix \verb|Makefile| and VMS
\verb|isamake.com|. 

     It is fairly straightforward to modify these routines to scan SUSY
parameters, but given the variety of possible scans, no attempt has been
made to provide code for this. If only masses are needed, SSMSSM can be
modified to remove the calls to the routines that calculate branching
ratios.
\newpage
\section{Input\label{INPUT}}

\subsection{Input Format}

      ISAJET is controlled by commands read from the specified input
file by subroutine READIN. (In the interactive version, this file is
first created by subroutine DIALOG.) Syntax errors will generate a
message and stop execution. Based on these commands, subroutine LOGIC
will setup limits for all variables and check for inconsistencies.
Several runs with different parameters can be combined into one job.
The required input format is:
\begin{verbatim}
Title
Ecm,Nevent,Nprint,Njump/
Reaction
(Optional parameters)
END
(Optional additional runs)
STOP
\end{verbatim}
with all lines starting in column 1 and typed in {\it upper} case. These
lines are explained below.

      Title line: Up to 80 characters long. If the first four letters
are STOP, control is returned to main program. If the first four letters
are SAME, the parameters from previous run are used excepting those
which are explicitly changed.

      Ecm line: This line must always be given even if the title is
SAME. It must give the center of mass energy (Ecm) and the number of
events (Nevent) to be generated. One may also specify the number of
events to be printed (Nprint) and the increment (Njump) for printing.
The first event is always printed if Nprint $>$ 0. For example:
\begin{verbatim}
800.,1000,10,100/
\end{verbatim}
generates 1000 events at $E_{\rm cm} = 800\,\GeV$ and prints 10
events. The events printed are: 1,100,200,\dots. Note that an event
typically takes several pages of output. This line is read with a list
directed format (READ*).

     After Nprint events have been printed, a single line containing the
run number, the event number, and the random number seed is printed
every Njump events (if Njump is nonzero). This seed can be used to start
a new job with the given event if in the new run NSIGMA is set equal to
zero:
\begin{verbatim}
SEED
value/
NSIGMA
0/
\end{verbatim}
In general the same events will only be generated on the same type of
computer.

      Reaction line: This line must be given unless title is SAME, when
it must be omitted. It selects the type of events to be generated. The
present version can generate TWOJET, E+E-, DRELLYAN, MINBIAS, WPAIR,
SUPERSYM, HIGGS, PHOTON, TCOLOR, or WHIGGS events. This line is read
with an A8 format.

\subsection{Optional Parameters}

      Each optional parameter requires two lines.
The first line is a keyword specifying the parameter and the second
line gives the values for the parameter. The parameters can be given in
any order. Numerical values are read with a list directed format
(READ*), jet and particle types are read with a character format and
must be enclosed in quotes, and logical flags with an L1 format. All
momenta are in GeV and all angles are in radians.

      The parameters can be classified in several groups:
\begin{center}
\begin{tabular}{lllll}
\hline\hline
Jet Limits: & W/H Limits: & Decays:     & Constants:  & Other: \\
\hline
JETTYPE1    & HTYPE       & FORCE       & AMSB        & BEAMS \\
JETTYPE2    & PHIW        & FORCE1      & CUTJET      & EPOL \\
JETTYPE3    & QMH         & NOB         & CUTOFF      & EEBEAM \\
MIJLIM      & QMW         & NODECAY     & EXTRAD      & EEBREM \\
MTOT        & QTW         & NOETA       & FRAGMENT    & NPOMERON \\
P           & THW         & NOEVOLVE    & GAUGINO     & NSIGMA \\
PHI         & WTYPE       & NOFRGMNT    & GMSB        & NTRIES \\
PT          & XW          & NOGRAV      & GMSB2       & PDFLIB \\
TH          & YW          & NOPI0       & HMASS       & SEED \\
X           &             & NOTAU       & HMASSES     & STRUC \\
Y           &             &             & LAMBDA      & WFUDGE \\
WMODE1      &             &             & MGVTNO      & WMMODE \\
WMODE2      &             &             & MSSMA       & WPMODE \\
            &             &             & MSSMB       & Z0MODE \\
            &             &             & MSSMC       & \\
            &             &             & MSSMD       & \\
            &             &             & MSSME       & \\
            &             &             & NUSUG1      & \\
            &             &             & NUSUG2      & \\
            &             &             & NUSUG3      & \\
            &             &             & NUSUG4      & \\
            &             &             & NUSUG5      & \\
            &             &             & SIGQT       & \\
            &             &             & SIN2W       & \\
            &             &             & SLEPTON     & \\
            &             &             & SQUARK      & \\
            &             &             & SSBCSC      & \\
            &             &             & SUGRA       & \\
            &             &             & SUGRHN      & \\
            &             &             & TCMASS      & \\
            &             &             & TMASS       & \\
            &             &             & WMASS       & \\
\hline\hline
\end{tabular}
\end{center}

      It may be helpful to know that the TWOJET, WPAIR, PHOTON,
SUPERSYM, and WHIGGS processes use the same controlling routines and
so share many of the same variables.  In particular, PT limits should
normally be set for these processes, and JETTYPE1 and JETTYPE2 are
used to select the reactions. Similarly, the DRELLYAN, HIGGS, and
TCOLOR processes use the same controlling routines since they all
generate s-channel resonances. The mass limits for these processes are
set by QMW.  Normally the QMW limits will surround the $W^\pm$, $Z^0$,
or Higgs mass, but this is not required.  (QMH acts like QMW for the
Higgs process.) For historical reasons, JETTYPE1 and JETTYPE2 are used
to select the W decay modes in DRELLYAN, while WMODE1 and WMODE2 select
the W decay modes for WPAIR, HIGGS, and WHIGGS. Also, QTW can be used
to generate DRELLYAN events with non-zero transverse momentum, whereas
HIGGS automatically fixes QTW to be zero. (Of course, non-zero
transverse momentum will be generated by gluon radiation.)

      For example the lines
\begin{verbatim}
P
40.,50.,10.,100./
\end{verbatim}
would set limits for the momentum of jet 1 between 40 and 50 GeV, and
for jet 2 between 10 and 100 GeV. As another example the lines
\begin{verbatim}
WTYPE
'W+'/
\end{verbatim}
would specify that for DRELLYAN events only W+ events will be generated.
If for a kinematic variable only the lower limit is specified then that
parameter is fixed to the given value. Thus the lines
\begin{verbatim}
P
40.,,10./
\end{verbatim}
will fix the momentum for jet 1 to be 40 GeV and for jet 2 to be 10
GeV. If only the upper limit is specified then the default value is used
for the lower limit. Jet 1 or jet 2 parameters for DRELLYAN events refer
to the W decay products and cannot be fixed. If QTW is fixed to 0, then
standard Drell-Yan events are generated.

      A complete list of keywords and their default values follows.

\newpage
\begin{center}
\begin{tabular}{lll}
\hline\hline
Keyword                &                   & Explanation                    \\
Values                 & Default values    &                                \\
\hline
AMSB                   &                   & Anomaly-mediated SUSY breaking \\
$m_0$,$m_{3/2}$,$\tan\beta$,$\sgn\mu$ & none & scalar mass, gravitino mass, \\
                       &                   & VEV ratio, sign                \\
                       &                   &                                \\
BEAMS                  &                   & Initial beams. Allowed are     \\
type$_1$,type$_2$      & 'P','P'           & 'P','AP','N','AN'.             \\
                       &                   &                                \\
CUTJET                 &                   & Cutoff mass for QCD jet        \\
$\mu_c$                & 6.                & evolution.                     \\
                       &                   &                                \\
CUTOFF                 &                   & Cutoff $qt^2=\mu^2Q^\nu$ for   \\
$\mu^2$, $\nu$         & .200,1.0          & DRELLYAN events.               \\
                       &                   &                                \\
EEBEAM                 &                   & impose brem/beamstrahlung      \\
$\sqrt{\hat{s}}_{min}$, $\sqrt{\hat{s}}_{max}$, $\Upsilon$, $\sigma_z$ &
none & min and max subprocess energy, \\
                       & & beamstrahlung parameter $\Upsilon$ \\
                       & & longitudinal beam size $\sigma_z$ in mm \\
                       &                   &                                \\
EEBREM                 &                 & impose bremsstrahlung for $e^+e^-$ \\
$\sqrt{\hat{s}}_{min}$, $\sqrt{\hat{s}}_{max}$ & none & min and max subprocess 
energy \\
                       &                   &                                \\
EPOL                   &              & Polarization of $e^-$ ($e^+$) beam, \\
$P_L(e^-),P_L(e^+)$    & 0,0               & $P_L(e)=(n_L-n_R)/(n_L-n_R)$,  \\
                       &                   & so that $-1 \le P_L \le 1$     \\
                       &                   &                                \\
EXTRAD                 &                   & Parameters for EXTRADIM process\\
$\delta$,$M_D$,UVCUT   & None              & UVCUT is logical flag          \\
                       &                   &                                \\
FORCE                  &                   & Force decay of particles,      \\
$i,i_1,...,i_5$/       & None              & $\pm i \to \pm(i1+...+i5)$.    \\
                       &                   & Can call 20 times.             \\
                       &                   & See note for $i$ = quark.      \\
                       &                   &                                \\
FORCE1                 &                   & Force decay $i \to i1+...+i5$. \\
$i,i_1,...,i_5$/       & None              & Can call 40 times.             \\
                       &                   & See note for $i$ = quark.      \\
                       &                   &                                \\
FRAGMENT               &                   & Fragmentation parameters.      \\
$P_{ud}$,\dots         & .4,\dots          & See also SIGQT, etc.           \\
                       &                   &                                \\
GAUGINO                &                   & Masses for $\tilde g$, 
$\tilde\gamma$,                                                             \\
$m_1$,$m_2$,$m_3,m_4$  & 50,0,100,100      & $\tilde W^+$, and $\tilde Z^0$ \\
\hline\hline
\end{tabular}
\end{center}

\newpage
\begin{center}
\begin{tabular}{lll}
\hline\hline
GMSB                   &                   & GMSB messenger SUSY breaking,  \\
$\Lambda_m$,$M_m$,$N_5$ & none             & mass, number of $5+\bar5$, VEV \\
$\tan\beta$,$\sgn\mu$,$C_{\rm gr}$ &       & ratio, sign, gravitino scale   \\
                       &                   &                                \\
GMSB2                  &                   & non-minimal GMSB parameters    \\
$\slashchar{R}$,$\delta M_{H_d}^2$,$\delta M_{H_u}^2$,$D_Y(M)$ & 1,0,0,0 & 
gaugino mass multiplier \\
$N_{5_1}$,$N_{5_2}$,$N_{5_3}$ & $N_5$     & Higgs mass shifts, D-term mass$^2$\\
                       &                   & indep. gauge group messengers  \\
                       &                   &                                \\
HMASS                  & 0                 & Mass for standard Higgs.       \\
$m$                    &                   &                                \\
                       &                   &                                \\
HMASSES                &                   & Higgs meson masses for         \\
$m_1$,\dots,$m_9$      & 0,...,0           & charges 0,0,0,0,0,1,1,2,2.     \\
                       &                   &                                \\
HTYPE                  &                   & One MSSM Higgs type ('HL0',    \\
'HL0'/ or...           & none              & 'HH0', or 'HA0')               \\
                       &                   &                                \\
JETTYPE1               &                   & )Select types for jets:        \\
'GL','UP',...          & 'ALL'             & )'ALL'; 'GL'; 'QUARKS'='UP',   \\
                       &                   & )'UB','DN','DB','ST','SB',     \\
JETTYPE2               &                   & )'CH','CB','BT','BB','TP',     \\
'GL','UP',...          & 'ALL'             & )'TB','X','XB','Y','YB';       \\
                       &                   & )'LEPTONS'='E-','E+','MU-',    \\
JETTYPE3               &                   & )'MU+','TAU-','TAU+'; 'NUS';   \\
'GL','UP',...          & 'ALL'             & )'GM','W+','W-','Z0'           \\
                       &                   & ) See note for SUSY types.     \\
                       &                   &                                \\
LAMBDA                 &                   & QCD scale                      \\
$\Lambda$              & .2                &                                \\
                       &                   &                                \\
MGVTNO                 &                   & Gravitino mass -- ignored for  \\
$M_{\rm gravitino}$    & $10^{20}$~GeV     & GMSB model                     \\
                       &                   &                                \\
MIJLIM                 &                   & Multimet mass limits           \\
$i$,$j$,$M_{\rm min}$,$M_{\rm max}$ & 0,0,$1\,\GeV$,$1\,\GeV$ &             \\
                       &                   &                                \\
MSSMA                  &                   & MSSM parameters --             \\
$m(\tilde g)$,$\mu$,   & Required          & Gluino mass, $\mu$, $A$ mass,  \\
$m(A)$,$\tan\beta$     &                   & $\tan\beta$                    \\
                       &                   &                                \\
MSSMB                  &                   & MSSM 1st generation --         \\
$m(q_1)$,$m(d_r)$,$m(u_r)$, & Required     & Left and right soft squark and \\
$m(l_1)$,$m(e_r)$      &                   & slepton masses                 \\
\hline\hline
\end{tabular}
\end{center}

\newpage
\begin{center}
\begin{tabular}{lll}
\hline\hline
MSSMC                  &                   & MSSM 3rd generation --         \\
$m(q_3)$,$m(b_r)$,$m(t_r)$,  & Required    & Soft squark masses, slepton    \\
$m(l_3)$,$m(\tau_r)$,  &                   & masses, and squark and slepton \\
$A_t$,$A_b$,$A_\tau$   &                   & mixings                        \\
                       &                   &                                \\
MSSMD                  &                   & MSSM 2nd generation --         \\
$m(q_2)$,$m(s_r)$,$m(c_r)$,  & from MSSMB  & Left and right soft squark and \\
$m(l_2)$,$m(mu_r)$     &                   & slepton masses                 \\
                       &                   &                                \\
MSSME                  &                   & MSSM gaugino masses --         \\
$M_1$,$M_2$            & MSSMA + GUT       & Default is to scale from gluino\\
                       &                   &                                \\
MTOT                   &                   & Mass range for multiparton     \\
$M_{\rm min}$,$M_{\rm max}$ & None         & processes                      \\
                       &                   &                                \\
NOB                    &                   & Suppress B decays to use       \\
TRUE or FALSE          & FALSE             & external package.              \\
                       &                   &                                \\
NODECAY                &                   & Suppress all decays.           \\
TRUE or FALSE          & FALSE             &                                \\
                       &                   &                                \\
NOETA                  &                   & Suppress eta decays.           \\
TRUE or FALSE          & FALSE             &                                \\
                       &                   &                                \\
NOEVOLVE               &                   & Suppress QCD evolution and     \\
TRUE or FALSE          & FALSE             & hadronization.                 \\
                       &                   &                                \\
NOGRAV                 &                   & Suppress gravitino decays in   \\
TRUE or FALSE          & FALSE             & GMSB model                     \\
                       &                   &                                \\
NOHADRON               &                   & Suppress hadronization of      \\
TRUE or FALSE          & FALSE             & jets and beam jets.            \\
                       &                   &                                \\
NONUNU                 &                   & Suppress $Z^0$ neutrino decays.\\
TRUE or FALSE          & FALSE             &                                \\
                       &                   &                                \\
NOPI0                  &                   &Suppress $\pi^0$ decays.        \\
TRUE or FALSE          & FALSE             &                                \\
                       &                   &                                \\
NOTAU                  &                   & Suppress tau decays to use     \\
TRUE or FALSE          & FALSE             & external package.              \\
                       &                   &                                \\
NPOMERON               &                   & Allow $n_1<n<n_2$ cut pomerons.\\
$n_1$,$n_2$            & 1,20              & Controls beam jet mult.        \\
\hline\hline
\end{tabular}
\end{center}

\newpage
\begin{center}
\begin{tabular}{lll}
\hline\hline
NSIGMA                 &                   & Generate n unevolved events    \\
$n$                    & 20                & for SIGF calculation.          \\
                       &                   &                                \\
NTRIES                 &                   & Stop if after n tries          \\
$n$                    & 1000              & cannot find a good event.      \\
                       &                   &                                \\
NUSUG1                 &                   & Optional non-universal SUGRA   \\
$M_1$,$M_2$,$M_3$      & none              & gaugino masses                 \\
                       &                   &                                \\
NUSUG2                 &                   & Optional non-universal SUGRA   \\
$A_t$,$A_b$,$A_\tau$   & none              & $A$ terms                      \\
                       &                   &                                \\
NUSUG3                 &                   & Optional non-universal SUGRA   \\
$M_{H_d}$,$M_{H_u}$    & none              & Higgs masses                   \\
                       &                   &                                \\
NUSUG4                 &                   & Optional non-universal SUGRA   \\
$M_{u_L}$,$M_{d_R}$,$M_{u_R}$, & none      & 1st/2nd generation masses      \\
$M_{e_L}$,$M_{e_R}$    &                   &                                \\
                       &                   &                                \\
NUSUG5                 &                   & Optional non-universal SUGRA   \\
$M_{t_L}$,$M_{b_R}$,$M_{t_R}$, & none      & 3rd generation masses          \\
$M_{\tau_L}$,$M_{\tau_R}$ &                &                                \\
                       &                   &                                \\
P                      &                   & Momentum limits for jets.      \\
$p_{\rm min}(1)$,\dots,$p_{\rm max}(3)$ & 
1.,$0.5E_{\rm cm}$                         &                                \\
                       &                   &                                \\
PDFLIB                 &                   & CERN PDFLIB parton distribution\\
'name$_1$',val$_1$,\dots & None            & parameters. See PDFLIB manual. \\
                       &                   &                                \\
PHI                    &                   & Phi limits for jets.           \\
$\phi_{\rm min}(1)$,\dots,$\phi_{\rm max}(3)$ & 0,$2\pi$ &                  \\
                       &                   &                                \\
PHIW                   &                   & Phi limits for W.              \\
$\phi_{\rm min}$,$\phi_{\rm max}$ & 
0,$2\pi$                                   &                                \\
                       &                   &                                \\
PT or PPERP            &                   & $p_t$ limits for jets.         \\
$p_{t,{\rm min}}(1)$,\dots,$p_{t,{\rm max}}(3)$  & 
$.05E_{\rm cm}$,$.2E_{\rm cm}$             & Default for TWOJET only.       \\
                       &                   &                                \\
QMH                    &                   & Mass limits for Higgs.         \\
$q_{\rm min}$,$q_{\rm max}$ & 
$.05E_{\rm cm}$,$.2E_{\rm cm}$             & Equivalent to QMW.             \\
                       &                   &                                \\
QMW                    &                   & Mass limits for $W$.           \\
$q_{\rm min}$,$q_{\rm max}$ & 
$.05E_{\rm cm}$,$.2E_{\rm cm}$             &                                \\
\hline\hline
\end{tabular}
\end{center}

\newpage
\begin{center}
\begin{tabular}{lll}
\hline\hline
QTW                    &                   & $q_t$ limits for $W$. Fix 
$q_t=0$                                                                     \\
$q_{t,{\rm min}}$,$q_{t,{\rm max}}$ & 
.1,$.025E_{\rm cm}$                        & for standard Drell-Yan.        \\
                       &                   &                                \\
SEED                   &                   & Seed $<$281474976710656.D0 for \\
real/integer           & 0                 & RANF or $<$$2^{31}$ for RANLUX.\\
                       &                   &                                \\
SIGQT                  &                   & Internal $k_t$ parameter for   \\
$\sigma$               & .35               & jet fragmentation.             \\
                       &                   &                                \\
SIN2W                  &                   & Weinberg angle. See WMASS.     \\
$\sin^2(\theta_W)$     & .232              &                                \\
                       &                   &                                \\
SLEPTON                &                   & Masses for $\tilde \nu_e$,
$\tilde e$, $\tilde\nu_\mu$, $\tilde\mu$, $\tilde\nu_\tau$, $\tilde\tau$    \\
$m_1$,\dots,$m_6$      & 100,\dots,101.8   &                                \\
                       &                   &                                \\
SQUARK                 &                   & Masses for $\tilde u$,
$\tilde d$, $\tilde s$, $\tilde c$, $\tilde b$, $\tilde t$                  \\
$m_1$,\dots,$m_6$      & 100.3,...,240.    &                                \\
                       &                   &                                \\
SSBCSC                 &                   & Alternate mass scale for RGE   \\
$M$                    & $M_{GUT}$         & boundary conditions.           \\
                       &                   &                                \\
STRUC                  &                   & Structure functions. CTEQ5L,   \\
name                   & 'CTEQ5L'          & CTEQ3L, CTEQ2L, EHLQ, OR DO    \\
                       &                   &                                \\
SUGRA                  &                   & Minimal supergravity parameters\\
$m_0$,$m_{1/2}$,$A_0$, & none              & scalar M, gaugino M, trilinear \\
$\tan\beta$,$\sgn\mu$  &                   & breaking term, vev ratio, +-1  \\
TH or THETA            &                   & Theta limits for jets. Do not  \\
$\theta_{\rm min}(1)$,\dots,$\theta_{\rm max}(3)$ & 0,$\pi$ & also set Y.   \\
                       &                   &                                \\
SUGRHN                 &                   & SUGRA see-saw $\nu$-effect     \\
$m_{\nu_\tau}$,$M_N$,$A_n$,$m_{\tilde\nu_R}$ & $0,1E20,0,0$ & nu-mass, 
int. scale, \\
                       &                   & GUT scale nu SSB terms         \\
                       &                   &                                \\
THW                    &                   & Theta limits for W. Do not     \\
$\theta_{\rm min}$,$\theta_{\rm max}$ & 0,$\pi$ & also set YW.              \\
                       &                   &                                \\
TCMASS                 &                   & Technicolor mass and width.    \\
$m$,$\Gamma$           & 1000,100          &                                \\
                       &                   &                                \\
TMASS                  &                   & t, y, and x quark masses.      \\
$m_t$,$m_y$,$m_x$      & 180.,-1.,-1.      &                                \\
\hline\hline
\end{tabular}
\end{center}

\newpage
\begin{center}
\begin{tabular}{lll}
\hline\hline
WFUDGE                 &                   & Fudge factor for DRELLYAN      \\
factor                 & 1.85              & evolution scale.               \\
                       &                   &                                \\
WMASS                  &                   & W and Z masses. See SIN2W.     \\
$M_W$,$M_Z$            & 80.2, 91.19       &                                \\
                       &                   &                                \\
WMMODE                 &                   & Decay modes for $W^-$ in parton\\
'UP',\dots,'TAU+'      & 'ALL'             & cascade. See JETTYPE.          \\
                       &                   &                                \\
WMODE1                 &                   & )                              \\
'UP','UB',\dots        & 'ALL'             & )Decay modes for WPAIR.        \\
                       &                   & )Same code for quarks and      \\
WMODE2                 &                   & )leptons as JETTYPE.           \\
'UP','UB',\dots        & 'ALL'             & )                              \\
                       &                   &                                \\
WPMODE                 &                   & Decay modes for $W^+$ in parton\\
'UP',\dots,'TAU+'      & 'ALL'             & cascade. See JETTYPE.          \\
                       &                   &                                \\
WTYPE                  &                   & Select W type: W+,W-,GM,Z0.    \\
type$_1$,type$_2$      & 'GM','Z0'         & Do not mix W+,W- and GM,Z0.    \\
                       &                   &                                \\
X                      &                   & Feynman x limits for jets.     \\
$x_{\rm min}(1)$,\dots,$x_{\rm max}(3)$ & 
$-1$,1                                     &                                \\
                       &                   &                                \\
XGEN                   &                   & Jet fragmentation, Peterson    \\
a(1),\dots,a(8)        & .96,3,0,.8,.5,... & with $\epsilon=a(n)/m^2$, 
$n=4$-8.                                                                    \\
                       &                   &                                \\
XGENSS                 &                   & Fragmentation of GLSS, UPSS,   \\
a(1),\dots,a(7)        & .5,.5,...         & etc. with $\epsilon=a(n)/m**2$ \\
                       &                   &                                \\
XW                     &                   & Feynman x limits for W.        \\
$x_{\rm min}$,$x_{\rm max}$ & 
$-1$,1                                     &                                \\
                       &                   &                                \\
Y                      &                   & Y limits for each jet.         \\
$y_{\rm min}(1)$,\dots,$y_{\rm max}(3)$ & from PT & Do not also set TH.     \\
                       &                   &                                \\
YW                     &                   & Y limits for W.                \\
$y_{\rm min}$,$y_{\rm max}$ & from QTW,QMW & Do not set both YW and THW.    \\
                       &                   &                                \\
Z0MODE                 &                   & Decay modes for $Z^0$ in parton\\
'UP',\dots,'TAU+'      & 'ALL'             & cascade. See JETTYPE.          \\
\hline\hline
\end{tabular}
\end{center}

\newpage
\subsection{Kinematic and Parton-type Parameters}

      While the TWOJET PT limits and the DRELLYAN QMW limits are
formally optional parameters, they are set by default to be fractions of
$\sqrt{s}$. Thus, for example, the parameter file
\begin{verbatim}
DEFAULT TWOJET JOB
14000,100,1,100/
TWOJET
END
STOP
\end{verbatim}
will execute, but it will generate jets between 5\% and 20\% of
$\sqrt{s}$, which is probably not what is wanted. Similarly, the
parameter file
\begin{verbatim}
DEFAULT DRELLYAN JOB
14000,100,1,100/
DRELLYAN
END
STOP
\end{verbatim}
will generate $\gamma + Z$ events with masses between 5\% and 20\% of
$\sqrt{s}$, not masses around the $Z$ mass, and transverse momenta
between $1\,{\rm GeV}$ and 2.5\% of $\sqrt{s}$.

      Normally the user should set PT limits for TWOJET, PHOTON, WPAIR,
SUPERSYM, and WHIGGS events and QMW and QTW limits for DRELLYAN,
HIGGS, and TCOLOR events. If these limits are not set, they will be
selected as fractions of $E_{\rm cm}$. This can give nonsense. For
TWOJET the $p_t$ range should usually be less than about a factor of
two except for $b$ and $t$ jets at low $p_t$ to produce uniform
statistics. For $W^+$, $W^-$, or $Z^0$ events or for Higgs events the
QMW (QMH) range should usually include the mass. But one can select
different limits to study, e.g., virtual $W$ production or the effect
of a lighter or heavier Higgs on WW scattering. If only $t$ decays are
selected, then the lower QMW limit must be above the $t$ threshold.
For standard Drell-Yan events QTW should be fixed to zero,
\begin{verbatim}
QTW
0/
\end{verbatim}
Transverse momenta will then be generated by initial state gluon
radiation. A range of QTW can also be given. For SUPERSYM either the
masses and decay modes should be specified, or the MSSM, SUGRA, GMSB, or
AMSB parameters should be given. For fourth generation quarks it is
necessary to specify the quark masses.

      Note that if the limits given cover too large a kinematic range,
the program can become very inefficient, since it makes a fit to the
cross section over the specified range. NTRIES has to be increased if
narrow limits are set for X, XW or for jet 1 and jet 2 parameters in
DRELLYAN events. For larger ranges several runs can be combined together
using the integrated cross section per event SIGF/NEVENT as the weight.
This cross section is calculated for each run by Monte Carlo integration
over the specified kinematic limits and is printed at the end of the
run. It is corrected for JETTYPEi, WTYPE, and WMODEi selections; it
cannot be corrected for branching ratios of forced decays or for WPMODE,
WMMODE, or Z0MODE selections, since these can affect an arbitrary number
of particles.

      To generate events over a large range, it is much more efficient
to combine several runs. This is facilitated by using the special job
title SAME as described above. Note that SAME cannot be used to combine
standard DRELLYAN events (QTW fixed equal to 0) and DRELLYAN events with
nonzero QTW.

      The cross sections for multiparton final states in general have
infrared and collinear singularities. To obtain sensible results, it
is in general essential to set limits both on the $p_T$ of each final
parton using PT and on the mass of each pair of partons using MIJLIM.
The default lower limits are all $1\,{\rm GeV}$. Using these default
limits without thought will likely give absurd results.

      For TWOJET, DRELLYAN, and most other processes, the JETTYPEi and
WTYPEi keywords should be used to select the subprocesses to be
included. For $e^+ e^- \to W^+ W^-$, $Z^0 Z^0$, use FORCE and FORCE1
instead of WMODEi to select the $W$ decay modes. Note that these {\it
do not} change the calculated cross section. (In the E+E- process, the
$W$ and $Z$ decays are currently treated as particle decays, whereas in
the WPAIR and HIGGS processes they are treated as $2 \to 4$ parton
processes.)

      For HIGGS with $W^+W^-$ or $Z^0Z^0$ decays allowed it is
generally necessary to set PT limits for the W's, e.g.
\begin{verbatim}
PT
50,20000,50,20000/
\end{verbatim} 
If this is not done, then the default lower limit of 1 GeV is used,
and the $t$-channel exchanges will dominate, as they should in the
effective $W$ approximation. Depending on the other parameters, the
program may fail to generate an event in NTRIES tries.

\subsection{SUSY Parameters}

      SUPERSYM (SUSY) by default generates just gluinos and squarks in
pairs. There are no default masses or decay modes. Masses can be set
using GAUGINO, SQUARK, SLEPTON, and HMASSES. Decay modes can be
specified with FORCE or by modifying the decay table. Left and right
squarks are distinguished but assumed to be degenerate, except for
stops. Since version 7.11, types must be selected with JETTYPEi using
the supersymmetric names, e.g.
\begin{verbatim}
JETTYPE1
'GLSS','UPSSL','UPSSR'/
\end{verbatim}
Use of the corresponding standard model names, e.g.
\begin{verbatim}
JETTYPE1
'GL','UP'/
\end{verbatim}
and generation of pure photinos, winos, and zinos are no longer
supported.

      If MSSMA, MSSMB and MSSMC are given, then the specified parameters
are used to calculate all the masses and decay modes with the ISASUSY
package assuming the minimal supersymmetric extension of the standard
model (MSSM). There are no default values, so you must specify values
for each MSSMi, i=A-C. MSSMD can optionally be used to set the second
generation squark and slepton parameters; if it is omitted, then the
first generation ones are used. MSSME can optionally be used to set the
U(1) and SU(2) gaugino masses; if it is omitted, then the grand
unification values are used. The parameters and the use of the MSSM is
preserved if the title is SAME. FORCE can be used to override the
calculated branching ratios. 

      The MSSM option also generates charginos and neutralinos with
cross sections based on the MSSM mixing angles in addition to squarks
and sleptons. These can be selected with JETTYPEi; the complete list of
supersymmetric options is:
\begin{verbatim}
'GLSS',
'UPSSL','UBSSL','DNSSL','DBSSL','STSSL','SBSSL','CHSSL','CBSSL',
'BTSS1','BBSS1','TPSS1','TBSS1',
'UPSSR','UBSSR','DNSSR','DBSSR','STSSR','SBSSR','CHSSR','CBSSR',
'BTSS2','BBSS2','TPSS2','TBSS2',
'W1SS+','W1SS-','W2SS+','W2SS-','Z1SS','Z2SS','Z3SS','Z4SS',
'NUEL','ANUEL','EL-','EL+','NUML','ANUML',MUL-','MUL+','NUTL',
'ANUTL','TAU1-','TAU1+','ER-','ER+','MUR-','MUR+','TAU2-','TAU2+',
'Z0','HL0','HH0','HA0','H+','H-',
'SQUARKS','GAUGINOS','SLEPTONS','ALL'.
\end{verbatim}
Note that mixing between $L$ and $R$ stop states results in 1 (light)
and 2 (heavy) stop, sbottom and stau eigenstates, which depend on the
input parameters of left- and right- scalar masses, plus $A$ terms,
$\mu$ and $\tan\beta$. The last four JETTYPE's generate respectively
all allowed combinations of squarks and antisquarks, all combinations
of charginos and neutralinos, all combinations of sleptons and
sneutrinos, and all SUSY particles. 

      For SUSY Higgs pair production or associated production in E+E-,
select the appropriate JETTYPE's, e.g.
\begin{verbatim}
JETTYPE1
'Z0'/
JETTYPE2
'HL0'/
\end{verbatim}

As usual, this gives only half the cross section. For single production
of neutral SUSY Higgs in $pp$ and $\bar pp$ reactions, use the HIGGS
process together with the MSSMi, SUGRA, GMSB, or AMSB keywords. You must
specify one and only one Higgs type using
\begin{verbatim}
HTYPE
'HL0' or 'HH0' or 'HA0'/     <<<<< One only!
\end{verbatim}
If no QMH range is given, one is calculated using $M \pm 5 \Gamma$ for
the selected Higgs. Decays into quarks, leptons, gauge bosons, lighter
Higgs bosons, and SUSY particles are generated using the on-shell
branching ratios from ISASUSY. You can use JETTYPEi to select the
allowed Higgs modes and WMODEi to select the allowed decays of W and Z
bosons. Since heavy SUSY Higgs bosons couple weakly to W pairs, WW
fusion and WW scattering are not included. 

      SUGRA can be used instead of MSSMi to generate MSSM decays with
parameters determined from $m_0$, $m_{1/2}$, $A_0$, $\tan\beta$, and
$\sgn\mu=\pm1$ in the minimal supergravity framework. The NUSUGi
keywords can optionally be used to specify additional parameters for
non-universal SUGRA models, while SUGRHN is used to specify the
parameterf of an optional right-handed neutrino. Similarly, the GMSB
keyword is used to specify the $\Lambda$, $M_m$, $N_5$, $\tan\beta$,
$\sgn\mu=\pm1$, and $C_{\rm grav}$ parameters of the minimal Gauge
Mediated SUSY Breaking model. GMSB2 can optionally be used to specify
additional parameters of non-minimal GMSB models. The AMSB keyword is
used to specify $m_0$, $m_{3/2}$, $\tan\beta$, and $\sgn\mu$ for the
minimal Anomaly Mediated SUSY Breaking model. Note that $m_{3/2}$ is
much larger than the weak scale, typically 50~TeV.

      WHIGGS is used to generate $W$ plus neutral Higgs events. For the
Standard Model the JETTYPE is \verb|HIGGS|. If any of the SUSY models
is specified, then the appropriate SUSY Higgs type should be used,
most likely \verb|HL0|. In either case WMODEi is used to specify the
$W$ decay modes. The Higgs is treated as a particle; its decay modes
can be set using FORCE.

\subsection{Forced Decay Modes}

      The FORCE keyword requires special care. Its list must contain the
numerical particle IDENT codes, e.g.
\begin{verbatim}
FORCE
140,130,-120/
\end{verbatim}
The charge-conjugate mode is also forced for its antiparticle. Thus the
above example forces both $\bar D^0 \to K^+ \pi^-$ and $D^0 \to K^-
\pi^+$. If only a specific decay is wanted one should use the FORCE1
command; e.g.
\begin{verbatim}
FORCE1
140,130,-120/
\end{verbatim}
only forces $\bar D^0 \to K^+ \pi^-$.

      To force a heavy quark decay one must generally separately force
each hadron containing it. If the decay is into three leptons or quarks,
then the real or virtual W propagator is inserted automatically. Since
Version 7.30, top and fourth generation quarks are treated as
particles and decayed directly rather than first being made into
hadrons. Thus for example
\begin{verbatim}
FORCE1
6,-12,11,5/
\end{verbatim}
forces all top quarks to decay into an positron, neutrino and a
b-quark (which will be hadronized). For the physical top mass, the
positron and neutrino will come from a real W. Note that forcing $t
\to W^+ b$ and $W^+ \to e^+ \nu_e$ does {\it not} give the same
result; the first uses the correct $V-A$ matrix element, while the
second decays the $W$ according to phase space.

      Forced modes included in the decay table or generated by ISASUSY
will automatically be put into the correct order and will use the
correct matrix element. Modes not listed in the decay table are
allowed, but caution is advised because a wrong decay mode can cause
an infinite loop or other unexpected effects.

      FORCE (FORCE1) can be called at most 20 (40) times in any run plus
all subsequent 'SAME' runs. If it is called more than once for a given
parent, all calls are listed, and the last call is used. Note that FORCE
applies to particles only, but that for gamma, W+, W-, Z0 and
supersymmetric particles the same IDENT codes are used both as jet types
and as particles.

\subsection{Parton Distributions}

      The default parton distributions are fit CTEQ5L from the CTEQ
Collaboration using lowest order QCD. The CTEQ3L, CTEQ2L, and the old
EHLQ and Duke-Owens distributions can be selected using the STRUC
keyword.

      If PDFLIB support is enabled (see Section 4), then any of the
distributions in the PDFLIB compilation by H. Plothow-Besch can be
selected using the PDFLIB keyword and giving the proper parameters,
which are identical to those described in the PDFLIB manual and are
simply passed to the routine PDFSET. For example, to select fit 29
(CTEQ3L) by the CTEQ group, leaving all other parameters with their
default values, use
\begin{verbatim}
PDFLIB
'CTEQ',29D0/
\end{verbatim}
Note that the fit-number and the other parameters are of type DOUBLE
PRECISION (REAL on 64-bit machines). There is no internal passing of
parameters except for those which control the printing of messages.

\subsection{Multiparton Processes}

      For multiparton final states one should in general set limits
on the total mass \verb|MTOT| of the final state, on the minimum
\verb|PT| of each light parton, and on the minimum mass \verb|MIMLIM|
of each pair of light partons. Limits for \verb|PT| are set in the
ususal way. Limits for the mass $M_{ij}$ of partons $i,j$ are set using
\begin{verbatim}
MIJLIM
i,j,Mmin,Mmax
\end{verbatim}
If $i=j=0$, the limit is applied to all jet pairs. For example the
following parameter file generates \verb|ZJJ| events at the LHC with a
mimimum $p_T$ of $20\,\GeV$ and a minimum mass of $20\,\GeV$ for all
jet pairs:
\begin{verbatim}
GENERATE ZJJ with PTMIN = 20 GEV AND MMIN = 20 GEV
14000,100,1,100/
ZJJ
PT
20,7000,20,7000,20,7000/
MIJLIM
0,0,20,7000/
MTOT
100,500/
NSIGMA
200/
NTRIES
10000/
END
STOP
\end{verbatim}
The default lower limits for \verb|PT| and \verb|MIJLIM| are
$1\,\GeV$. While these limits are sufficient to make the cross
sections finite, they will in general not give physically sensible
results. Thus, {\it the user must think carefully about what limits
should be set.}
\newpage
\section{Output\label{OUTPUT}}

      The output tape or file contains three types of records. A
beginning record is written by a call to ISAWBG before generating a set
of events; an event record is written by a call to ISAWEV for each
event; and an end record is written for each run by a call to ISAWND.
These subroutines load the common blocks described below into a single
\begin{verbatim}
COMMON/ZEVEL/ZEVEL(1024) 
\end{verbatim}
and write it out when it is full. A subroutine RDTAPE, described in
the next section, inverts this process so that the user can analyze
the event.

      ZEVEL is written out to TAPEj by a call to BUFOUT. For the CDC
version IF = PAIRPAK is selected; BUFOUT first packs two words from
ZEVEL into one word in 
\begin{verbatim}
COMMON/ZVOUT/ZVOUT(512) 
\end{verbatim}
using subroutine PAIRPAK and then does a buffer out of ZVOUT to TAPEj.
Typically at least two records are written per event. For all other
computers IF=STDIO is selected, and ZEVEL is written out with a
standard FORTRAN unformatted write.

\subsection{Beginning Record}

      At the start of each run ISAWBG is called. It writes out the
following common blocks:
\begin{verbatim}
      COMMON/DYLIM/QMIN,QMAX,QTMIN,QTMAX,YWMIN,YWMAX,XWMIN,XWMAX,THWMIN,
     2  THWMAX,PHWMIN,PHWMAX
     3  ,SETLMQ(12)
      SAVE /DYLIM/
      LOGICAL SETLMQ
      EQUIVALENCE(BLIM1(1),QMIN)
      REAL      QMIN,QMAX,QTMIN,QTMAX,YWMIN,YWMAX,XWMIN,XWMAX,THWMIN,
     +          THWMAX,PHWMIN,PHWMAX,BLIM1(12)
\end{verbatim}
\begin{tabular}{lcl}
QMIN,QMAX          &=& $W$ mass limits\\
QTMIN,QTMAX        &=& $W$ $q_t$ limits\\
YWMIN,YWMAX        &=& $W$ $\eta$ rapidity limits\\
XWMIN,XWMAX        &=& $W$ $x_F$ limits\\
THWMIN,THWMAX      &=& $W$ $\theta$ limits\\
PHWMIN,PHWMAX      &=& $W$ $\phi$ limits\\
\end{tabular}

\begin{verbatim}
      COMMON/IDRUN/IDVER,IDG(2),IEVT,IEVGEN
      SAVE /IDRUN/
      INTEGER   IDVER,IDG,IEVT,IEVGEN
\end{verbatim}
\begin{tabular}{lcl}
IDVER              &=& program version\\
IDG(1)             &=& run date (10000$\times$month+100$\times$day+year)\\
IDG(2)             &=& run time (10000$\times$hour+100$\times$minute+second)\\
IEVT               &=& event number\\
\end{tabular}

\begin{verbatim}
C          Jet limits
      INTEGER MXLIM
      PARAMETER (MXLIM=8)
      INTEGER MXLX12
      PARAMETER (MXLX12=12*MXLIM)
      COMMON/JETLIM/PMIN(MXLIM),PMAX(MXLIM),PTMIN(MXLIM),PTMAX(MXLIM),
     $YJMIN(MXLIM),YJMAX(MXLIM),PHIMIN(MXLIM),PHIMAX(MXLIM),
     $XJMIN(MXLIM),XJMAX(MXLIM),THMIN(MXLIM),THMAX(MXLIM),
     $SETLMJ(12*MXLIM)
      SAVE /JETLIM/
      COMMON/FIXPAR/FIXP(MXLIM),FIXPT(MXLIM),FIXYJ(MXLIM),
     $FIXPHI(MXLIM),FIXXJ(MXLIM),FIXQM,FIXQT,FIXYW,FIXXW,FIXPHW
      SAVE /FIXPAR/
      COMMON/SGNPAR/CTHS(2,MXLIM),THS(2,MXLIM),YJS(2,MXLIM),XJS(2,MXLIM)  
      SAVE /SGNPAR/
      REAL      PMIN,PMAX,PTMIN,PTMAX,YJMIN,YJMAX,PHIMIN,PHIMAX,XJMIN,
     +          XJMAX,THMIN,THMAX,BLIMS(12*MXLIM),CTHS,THS,YJS,XJS
      LOGICAL SETLMJ
      LOGICAL FIXQM,FIXQT,FIXYW,FIXXW,FIXPHW
      LOGICAL FIXP,FIXPT,FIXYJ,FIXPHI,FIXXJ
      EQUIVALENCE(BLIMS(1),PMIN(1))
\end{verbatim}
\begin{tabular}{lcl}
PMIN,PMAX          &=& jet momentum limits\\
PTMIN,PTMAX        &=& jet $p_t$ limits\\
YJMIN,YJMAX        &=& jet $\eta$ rapidity limits\\
PHIMIN,PHIMAX      &=& jet $\phi$ limits\\
THMIN,THMAX        &=& jet $\theta$ limits\\
\end{tabular}

\begin{verbatim}
      INTEGER MXKEYS
      PARAMETER (MXKEYS=20)
      COMMON/KEYS/IKEYS,KEYON,KEYS(MXKEYS)
      COMMON/XKEYS/REAC
      SAVE /KEYS/,/XKEYS/
      LOGICAL KEYS
      LOGICAL KEYON
      CHARACTER*8 REAC
      INTEGER   IKEYS
\end{verbatim}
\begin{tabular}{lcl}
KEYON              &=& normally TRUE, FALSE if no good reaction\\
KEYS               &=& TRUE if reaction I is chosen\\
                   && 1 for TWOJET\\
                   && 2 for E+E-\\
                   && 3 for DRELLYAN\\
                   && 4 for MINBIAS\\
                   && 5 for SUPERSYM\\
                   && 6 for WPAIR\\
REAC               &=& character reaction code\\
\end{tabular}

\begin{verbatim}
      COMMON/PRIMAR/NJET,SCM,HALFE,ECM,IDIN(2),NEVENT,NTRIES,NSIGMA
      SAVE /PRIMAR/
      INTEGER   NJET,IDIN,NEVENT,NTRIES,NSIGMA
      REAL      SCM,HALFE,ECM
\end{verbatim}
\begin{tabular}{lcl}
NJET               &=& number of jets per event\\
SCM                &=& square of com energy\\
HALFE              &=& beam energy\\
ECM                &=& com energy\\
IDIN               &=& ident code for initial beams\\
NEVENT             &=& number of events to be generated\\
NTRIES             &=& maximum number of tries for good jet parameters\\
NSIGMA             &=& number of extra events to determine SIGF\\
\end{tabular}

\begin{verbatim}
      INTEGER MXGOQ,MXGOJ
      PARAMETER (MXGOQ=85,MXGOJ=8)
      COMMON/Q1Q2/GOQ(MXGOQ,MXGOJ),GOALL(MXGOJ),GODY(4),STDDY,
     $GOWW(25,2),ALLWW(2),GOWMOD(25,MXGOJ)
      SAVE /Q1Q2/
      LOGICAL GOQ,GOALL,GODY,STDDY,GOWW,ALLWW,GOWMOD
\end{verbatim}
\begin{tabular}{lcl}
GOQ(I,K)           &=& TRUE if quark type I allowed for jet k\\
                   && I = 1  2  3  4  5  6  7  8  9 10 11 12 13\\
                   && \ \ $\Rightarrow$ $g$ $u$ $\bar u$ $d$ $\bar d$ $s$ 
                      $\bar s$ $c$ $\bar c$ $b$ $\bar b$ $t$ $\bar t$\\
                   && I = 14   15 16 17  18   19  20  21  22   23   24   25\\
                   && \ \ $\Rightarrow$ $\nu_e$ $\bar\nu_e$ $e^-$ $e^+$ 
                      $\nu_\mu$ $\bar\nu_\mu$ $\mu^-$ $\mu^+$ $\nu_\tau$ 
                      $\bar\nu_\tau$ $\tau^-$ $\tau^+$\\
GOALL(K)           &=& TRUE if all jet types allowed\\
GODY(I)            &=& TRUE if $W$ type I is allowed.\\
                    I= 1  2  3  4\\
                      GM W+ W- Z0\\
STDDY              &=& TRUE if standard DRELLYAN\\
GOWW(I,K)          &=& TRUE if I is allowed in the decay of K for WPAIR.\\
ALLWW(K)           &=& TRUE if all allowed in the decay of K for WPAIR.\\
\end{tabular}

\begin{verbatim}
      COMMON/QCDPAR/ALAM,ALAM2,CUTJET,ISTRUC
      SAVE /QCDPAR/
      INTEGER   ISTRUC
      REAL      ALAM,ALAM2,CUTJET
\end{verbatim}
\begin{tabular}{lcl}
ALAM               &=& QCD scale $\Lambda$\\
ALAM2              &=& QCD scale $\Lambda^2$\\
CUTJET             &=& cutoff for generating secondary partons\\
ISTRUC             &=& 3 for Eichten (EHLQ), \\
                   &=& 4 for Duke (DO) \\
                   &=& 5 for CTEQ 2L\\
                   &=& 6 for CTEQ 3L\\
                   &=& $-999$ for PDFLIB\\
\end{tabular}

\begin{verbatim}
      COMMON/QLMASS/AMLEP(100),NQLEP,NMES,NBARY
      SAVE /QLMASS/
      INTEGER   NQLEP,NMES,NBARY
      REAL      AMLEP
\end{verbatim}
\begin{tabular}{lcl}
AMLEP(6:8)         &=& $t$,$y$,$x$ masses, only elements written\\
\end{tabular}

\subsection{Event Record}

      For each event ISAWEV is called. It writes out the following
common blocks:
\begin{verbatim}
      COMMON/FINAL/NKINF,SIGF,ALUM,ACCEPT,NRECS
      SAVE /FINAL/
      INTEGER   NKINF,NRECS
      REAL      SIGF,ALUM,ACCEPT
\end{verbatim}
\begin{tabular}{lcl}
SIGF              &=& integrated cross section, only element written\\
\end{tabular}

\begin{verbatim}
      COMMON/IDRUN/IDVER,IDG(2),IEVT,IEVGEN
      SAVE /IDRUN/
      INTEGER   IDVER,IDG,IEVT,IEVGEN
\end{verbatim}
\begin{tabular}{lcl}
IDVER              &=& program version\\
IDG                &=& run identification\\
IEVT               &=& event number\\
\end{tabular}

\begin{verbatim}
      COMMON/JETPAR/P(3),PT(3),YJ(3),PHI(3),XJ(3),TH(3),CTH(3),STH(3)
     1 ,JETTYP(3),SHAT,THAT,UHAT,QSQ,X1,X2,PBEAM(2)
     2 ,QMW,QW,QTW,YW,XW,THW,QTMW,PHIW,SHAT1,THAT1,UHAT1,JWTYP
     3 ,ALFQSQ,CTHW,STHW,Q0W
     4 ,INITYP(2),ISIGS,PBEAMS(5)
      SAVE /JETPAR/
      INTEGER   JETTYP,JWTYP,INITYP,ISIGS
      REAL      P,PT,YJ,PHI,XJ,TH,CTH,STH,SHAT,THAT,UHAT,QSQ,X1,X2,
     +          PBEAM,QMW,QW,QTW,YW,XW,THW,QTMW,PHIW,SHAT1,THAT1,UHAT1,
     +          ALFQSQ,CTHW,STHW,Q0W,PBEAMS
\end{verbatim}
\begin{tabular}{lcl}
P                  &=& jet momentum $\vert\vec p\vert$\\
PT                 &=& jet $p_t$\\
YJ                 &=& jet $\eta$ rapidity\\
PHI                &=& jet $\phi$\\
XJ                 &=& jet $x_F$\\
TH                 &=& jet $\theta$\\
CTH                &=& jet $\cos(\theta)$\\
STH                &=& jet $\sin(\theta)$\\
JETTYP             &=& jet type. The code is listed under /Q1Q2/ above\\
                   &&  {\it continued\dots}\\
\end{tabular}

\begin{tabular}{lcl}
SHAT,THAT,UHAT     &=& hard scattering $\hat s$, $\hat t$, $\hat u$\\
QSQ                &=& effective $Q^2$\\
X1,X2              &=& initial parton $x_F$\\
PBEAM              &=& remaining beam momentum\\
QMW                &=& $W$ mass\\
QW                 &=& $W$ momentum\\
QTW                &=& $W$ transverse momentum\\
YW                 &=& $W$ rapidity\\
XW                 &=& $W$ $x_F$\\
THW                &=& $W$ $\theta$\\
QTMW               &=& $\sqrt{q_{t,W}^2+Q^2}$\\
PHIW               &=& $W$ $\phi$\\
SHAT1,THAT1,UHAT1  &=& invariants for $W$ decay\\
JWTYP              &=& $W$ type. The code is listed under /Q1Q2/ above.\\
ALFQSQ             &=& QCD coupling $\alpha_s(Q^2)$\\
CTHW               &=& $W$ $\cos(\theta)$\\
STHW               &=& $W$ $\sin(\theta)$\\
Q0W                &=& $W$ energy\\
\end{tabular}

\begin{verbatim}
      INTEGER   MXJSET,JPACK
      PARAMETER (MXJSET=400,JPACK=1000)
      COMMON/JETSET/NJSET,PJSET(5,MXJSET),JORIG(MXJSET),JTYPE(MXJSET),
     $JDCAY(MXJSET)
      SAVE /JETSET/
      INTEGER   NJSET,JORIG,JTYPE,JDCAY
      REAL      PJSET
\end{verbatim}
\begin{tabular}{lcl}
NJSET              &=& number of partons\\
PJSET(1,I)         &=& $p_x$ of parton I\\
PJSET(2,I)         &=& $p_y$ of parton I\\
PJSET(3,I)         &=& $p_z$ of parton I\\
PJSET(4,I)         &=& $p_0$ of parton I\\
PJSET(5,I)         &=& mass of parton I\\
JORIG(I)           &=& JPACK*JET+K if I is a decay product of K.\\
                   && IF K=0 then I is a primary parton.\\
                   && (JET = 1,2,3 for final jets.)\\
                   && (JET = 11,12 for initial jets.)\\
JTYPE(I)           &=& IDENT code for parton I\\
JDCAY(I)           &=& JPACK*K1+K2 if K1 and K2 are decay products of I.\\
                   &&  If JDCAY(I)=0 then I is a final parton\\
MXJSET             &=& dimension for /JETSET/ arrays.\\
JPACK              &=& packing integer for /JETSET/ arrays.\\
\end{tabular}

\begin{verbatim}
      INTEGER   MXSIGS,IOPAK
      PARAMETER (MXSIGS=3000,IOPAK=100)
      COMMON/JETSIG/SIGMA,SIGS(MXSIGS),NSIGS,INOUT(MXSIGS),SIGEVT
      SAVE /JETSIG/
      INTEGER   NSIGS,INOUT
      REAL      SIGMA,SIGS,SIGEVT
\end{verbatim}
\begin{tabular}{lcl}
SIGMA              &=& cross section summed over types\\
SIGS(I)            &=& cross section for reaction I (not written)\\
NSIGS              &=& number of nonzero cross sections (not written)\\
INOUT(I)           &=& packed partons for process I (not written)\\
MXSIGS             &=& dimension for JETSIG arrays (not written)\\
SIGEVT             &=& partial cross section for selected channel\\
\end{tabular}

\begin{verbatim}
      INTEGER   MXPTCL,IPACK
      PARAMETER (MXPTCL=4000,IPACK=10000)
      COMMON/PARTCL/NPTCL,PPTCL(5,MXPTCL),IORIG(MXPTCL),IDENT(MXPTCL)
     1,IDCAY(MXPTCL)
      SAVE /PARTCL/
      INTEGER   NPTCL,IORIG,IDENT,IDCAY
      REAL      PPTCL
\end{verbatim}
\begin{tabular}{lcl}
NPTCL              &=& number of particles\\
PPTCL(1,I)         &=& $p_x$ for particle I\\
PPTCL(2,I)         &=& $p_y$ for particle I\\
PPTCL(3,I)         &=& $p_z$ for particle I\\
PPTCL(4,I)         &=& $p_0$ for particle I\\
PPTCL(5,I)         &=& mass for particle I\\
IORIG(I)           &=& IPACK*JET+K if I is a decay product of K.\\
                   &=& -(IPACK*JET+K) if I is a primary particle from\\
                   &&  parton K in /JETSET/.\\
                   &=& 0 if I is a primary beam particle.\\
                   && (JET = 1,2,3 for final jets.)\\
                   && (JET = 11,12 for initial jets.)\\
IDENT(I)           &=& IDENT code for particle I\\
IDCAY(I)           &=& IPACK*K1+K2 if decay products are K1-K2 inclusive.\\
                   && If IDCAY(I)=0 then particle I is stable.\\
MXPTCL             &=& dimension for /PARTCL/ arrays.\\
IPACK              &=& packing integer for /PARTCL/ arrays.\\
\end{tabular}

\begin{verbatim}
      COMMON/PINITS/PINITS(5,2),IDINIT(2)
      SAVE /PINITS/
      INTEGER   IDINIT
      REAL      PINITS
\end{verbatim}
\begin{tabular}{lcl}
PINITS(1,I)        &=& $p_x$ for initial parton I\\
PINITS(2,I)        &=& $p_y$ for initial parton I\\
PINITS(3,I)        &=& $p_z$ for initial parton I\\
PINITS(4,I)        &=& $p_0$ for initial parton I\\
PINITS(5,I)        &=& mass for initial parton I\\
IDINIT(I)          &=& IDENT for initial parton I\\
\end{tabular}

\begin{verbatim}
      INTEGER MXJETS
      PARAMETER (MXJETS=10)
      COMMON/PJETS/PJETS(5,MXJETS),IDJETS(MXJETS),QWJET(5),IDENTW 
     $,PPAIR(5,4),IDPAIR(4),JPAIR(4),NPAIR,IFRAME(MXJETS)
      SAVE /PJETS/
      INTEGER   IDJETS,IDENTW,IDPAIR,JPAIR,NPAIR,IFRAME
      REAL      PJETS,QWJET,PPAIR
\end{verbatim}
\begin{tabular}{lcl}
PJETS(1,I)         &=& $p_x$ for jet I\\
PJETS(2,I)         &=& $p_y$ for jet I\\
PJETS(3,I)         &=& $p_z$ for jet I\\
PJETS(4,I)         &=& $p_0$ for jet I\\
PJETS(5,I)         &=& mass for jet I\\
IDJETS(I)          &=& IDENT code for jet I\\
QWJET(1)           &=& $p_x$ for $W$\\
QWJET(2)           &=& $p_y$ for $W$\\
QWJET(3)           &=& $p_z$ for $W$\\
QWJET(4)           &=& $p_0$ for $W$\\
QWJET(5)           &=& mass for $W$\\
IDENTW             &=& IDENT CODE for $W$\\
PPAIR(1,I)         &=& $p_x$ for WPAIR decay product I\\
PPAIR(2,I)         &=& $p_y$ for WPAIR decay product I\\
PPAIR(3,I)         &=& $p_z$ for WPAIR decay product I\\
PPAIR(4,I)         &=& $p_0$ for WPAIR decay product I\\
PPAIR(5,I)         &=& mass for WPAIR decay product I\\
IDPAIR(I)          &=& IDENT code for WPAIR product I\\
JPAIR(I)           &=& JETTYPE code for WPAIR product I\\
NPAIR              &=& 2 for $W^\pm\gamma$ events, 4 for $WW$ events\\
\end{tabular}

\begin{verbatim}
      COMMON/TOTALS/NKINPT,NWGEN,NKEEP,SUMWT,WT
      SAVE /TOTALS/
      INTEGER   NKINPT,NWGEN,NKEEP
      REAL      SUMWT,WT
\end{verbatim}
\begin{tabular}{lcl}
NKINPT             &=& number of kinematic points generated.\\
NWGEN              &=& number of W+jet events accepted.\\
NKEEP              &=& number of events kept.\\
SUMWT              &=& sum of weighted cross sections.\\
WT                 &=& current weight. (SIGMA$\times$WT = event weight.)\\
\end{tabular}

\begin{verbatim}
      COMMON/WSIG/SIGLLQ
      SAVE /WSIG/
      REAL      SIGLLQ
\end{verbatim}
\begin{tabular}{lcl}
SIGLLQ             &=& cross section for $W$ decay.\\
\end{tabular}

      Of course irrelevant common blocks such as /WSIG/ for TWOJET
events are not written out.

\subsection{End Record} 

      At the end of a set ISAWND is called. It writes out the
following common block:
\begin{verbatim}
      COMMON/FINAL/NKINF,SIGF,ALUM,ACCEPT,NRECS
      SAVE /FINAL/
      INTEGER   NKINF,NRECS
      REAL      SIGF,ALUM,ACCEPT
\end{verbatim}
\begin{tabular}{lcl}
NKINF             &=& number of points generated to calculate SIGF\\
SIGF              &=& integrated cross section for this run\\
ALUM              &=& equivalent luminosity for this run\\
ACCEPT            &=& ratio of events kept over events generated\\
NRECS             &=& number of physical records for this run\\
\end{tabular}

      Events within a given run have uniform weight. Separate runs can
be combined together using SIGF/NEVENT as the weight per event. This
gives a true cross section in mb units.

      The user can replace subroutines ISAWBG, ISAWEV, and ISAWND to
write out the events in a different format or to update histograms
using HBOOK or any similar package.
\newpage
\section{File Reading\label{TAPE}}

      The FORTRAN instruction
\begin{verbatim}
      CALL RDTAPE(IDEV,IFL)
\end{verbatim}
will read a beginning record, an end record or an event (which can be
more than one record). IDEV is the tape number and
\begin{verbatim}
      IFL=0  for a good read,
      IFL=-1 for an end of file.
\end{verbatim}
The information is restored to the common blocks described above. The
type of record is contained in
\begin{verbatim}
      COMMON/RECTP/IRECTP,IREC
      SAVE /RECTP/
      INTEGER   IRECTP,IREC
\end{verbatim}
\begin{tabular}{lcl}
IRECTP            &=& 100 for an event record\\
IRECTP            &=& 200 for a beginning record\\
IRECTP            &=& 300 for an end record\\
IREC              &=& no. of physical records in event record, 0 
                      otherwise\\
\end{tabular}

      The parton momenta from the primary hard scattering are
contained in /PJETS/. The parton momenta generated by the QCD cascade
are contained in /JETSET/. The hadron momenta both from the QCD jets
and from the beam jets are contained in /PARTCL/. The final hadron
momenta and the associated pointers should be used to calculate the
jet momenta, since they are changed both by the QCD cascade and by
hadronization. Particles with IDCAY=0 are stable, while the others are
resonances.

      The weight per event needed to produce a weighted histogram in
millibarn units is SIGF/NEVENT. The integrated cross section SIGF is
calculated by Monte Carlo integration during the run for the given
kinematic limits and JETTYPE, WTYPE, and WMODE selections. Any of three
methods can be used to find the value of SIGF:

      (1) The current value, which is written out with each event, can
be used. To prevent enormous fluctuations at the beginning of a run,
NSIGMA extra primary parton events are generated first. The default
value, NSIGMA = 20, gives negligible overhead but may not be large
enough for good accuracy.

      (2) The value SIGF calculated with the full statistics of the run
can be obtained by reading through the tape until an end record
(IRECTP=300) is found. After SIGF is saved with a different name, the
first event record for the run can be found by backspacing the tape
NRECS times.

      (3) Unweighted histograms can be made for the run and the weight
added after the end record is found. An implementation of this using
special features of HBOOK is contained in ISAPLT.

      The functions AMASS(IDENT), CHARGE(IDENT), and LABEL(IDENT) are
available to determine the mass, charge, and character label in A8
format. Subroutine FLAVOR returns the quark content of any hadron and
may be useful to convert IDENT codes to other schemes. CALL PRTEVT(0)
prints an event.
\newpage
\section{Decay Table\label{DECAY}}

      ISAJET uses an external table of decay modes. Particles can be
put into the table in arbitrary order, but all modes for each particle
must be grouped together. The table is rewound and read in before each
run with a READ* format. Beginning with Version 7.41, the decay table
must begin with a comment of the form
\begin{verbatim}
' ISAJET     V7.41   11-JAN-1999 20:41:57'
\end{verbatim}
If this does not match the internal version number, a warning is
printed. After this initial line, each entry must have the form
\begin{verbatim}
IDENT,MELEM,CBR,ID1,ID2,ID3,ID4,ID5/
\end{verbatim}
where IDENT is the code for the parent particle, MELEM specifies the
decay matrix element, CBR is the cumulative branching ratio, and
ID1,\dots,ID5 are the IDENT codes for the decay products. The
currently defined values of MELEM are:
\begin{center}
\begin{tabular}{cl}
\hline
MELEM   &\quad Matrix Element \\
\hline
0       &\quad Phase Space \\
1       &\quad Dalitz decay \\
2       &\quad $\omega/\phi$ decay \\
3       &\quad $V-A$ decay \\
4       &\quad top decay: $V-A$ plus $W$ propagator \\
5       &\quad $\tau \to \ell \nu \bar \nu$ \\
6       &\quad $\tau \to \nu \pi$, $\nu K$ \\
7       &\quad $\tau \to \nu \rho$, $\nu a_1$ \\
\hline
\end{tabular}
\end{center}
The parent IDENT must be positive; the charge conjugate mode is used
for the antiparticle. The values of CBR must of course be positive and
monotonically increasing for each mode, with the last value being 1.00
for each parent IDENT. The last parent IDENT code must be zero. Care
should be taken in adding new modes, since there is no checking for
validity. In some cases order is important; note in particular that
quarks and gluons must always appear last so that they can be removed
and fragmented into hadrons.

      The format of the decay table for Versions 7.41 and later is
incompatible with that for Versions 7.40 and earlier. Using an
obsolete decay table will produce incorrect results.

      The decay table is contained in patch ISADECAY.
\newpage
\section{IDENT Codes\label{IDENT}}

      ISAJET uses a numerical ident code for particle types. Quarks
and leptons are numbered in order of mass:
\begin{verbatim}
         UP     = 1             NUE    = 11
         DN     = 2             E-     = 12
         ST     = 3             NUM    = 13
         CH     = 4             MU-    = 14
         BT     = 5             NUT    = 15
         TP     = 6             TAU-   = 16
\end{verbatim}
with a negative sign for antiparticles. Arbitrary conventions are:
\begin{verbatim}
         GL     = 9
         GM     = 10
         KS     = 20
         KL     =-20
         W+     = 80
         Z0     = 90
\end{verbatim}
The supersymmetric particle IDENT codes distinguish between the
partners of left and right handed fermions and include the Higgs
sector of the minimal supersymmetric model:
\begin{verbatim}
         UPSSL ... TPSS1 = 21 ... 26
         NUEL ... TAU1-  = 31 ... 36
         UPSSR ... TPSS2 = 41 ... 46
         NUER ... TAU2-  = 51 ... 56
         GLSS  = 29
         Z1SS  = 30            Z2SS  = 40
         Z3SS  = 50            Z4SS  = 60
         W1SS+ = 39            W2SS+ = 49

         HL0   = 82            HH0   = 83
         HA0   = 84            H+    = 86
\end{verbatim}
Finally, the gravitino and graviton are
\begin{verbatim}
         GVSS  = 91            GRAV  = 92
\end{verbatim}
The same symbol is used for the graviton and its (possible) Kaluza-Klein
excitations.

      The code for a meson is a compound integer +-JKL, where J.LE.K are
the quarks and L is the spin. The sign is for the J quark. Glueball
IDENT codes have not been selected, but the choice GL=9 clearly allows
990, 9990, etc. Flavor singlet mesons are ordered by mass,
\begin{verbatim}
         PI0    = 110
         ETA    = 220
         ETAP   = 330
         ETAC   = 440
\end{verbatim}
which is natural for the heavy quarks. Similarly, the code for a
baryon is a compound integer +-IJKL formed from the three quarks I,J,K
and a spin label L=0,1. The code for a diquark is +-IJ00. Additional
states are distinguished by a fifth integer, e.g., 
\begin{verbatim}
         A1+    = 10121
\end{verbatim}
These and a few J=2 mesons are used in some of the B decays.

      A routine PRTLST is provided to print out a complete list of valid
IDENT codes and associated information. The usage is
      CALL PRTLST(LUN, AMY, AMX)
where LUN is the unit number and AMY and AMX are the masses of the Y and
X quarks respectively. This routine should be linked with the ISAJET
library and with ALDATA.

      The complete list of ident codes follows. (Hadrons containing $t$
quarks are defined but are no longer listed since the $t$ quark is
treated as a particle.)
\begin{verbatim}
      IDENT     LABEL           MASS    CHARGE
          1     UP        .30000E+00       .67
         -1     UB        .30000E+00      -.67
          2     DN        .30000E+00      -.33
         -2     DB        .30000E+00       .33
          3     ST        .50000E+00      -.33
         -3     SB        .50000E+00       .33
          4     CH        .16000E+01       .67
         -4     CB        .16000E+01      -.67
          5     BT        .49000E+01      -.33
         -5     BB        .49000E+01       .33
          6     TP        .17500E+03       .67
         -6     TB        .17500E+03      -.67

          9     GL       0.               0.00

         10     GM       0.               0.00

         11     NUE      0.               0.00
        -11     ANUE     0.               0.00
         12     E-        .51100E-03     -1.00
        -12     E+        .51100E-03      1.00
         13     NUM      0.               0.00
        -13     ANUM     0.               0.00
         14     MU-       .10566E+00     -1.00
        -14     MU+       .10566E+00      1.00
         15     NUT      0.               0.00
        -15     ANUT     0.               0.00
         16     TAU-      .18070E+01     -1.00
        -16     TAU+      .18070E+01      1.00

         20     KS        .49767E+00      0.00
        -20     KL        .49767E+00      0.00

         21     UPSSL     none            0.67
        -21     UBSSL     none           -0.67
         22     DNSSL     none           -0.33
        -22     DBSSL     none            0.33
         23     STSSL     none           -0.33
         23     SBSSL     none            0.33
         24     CHSSL     none            0.67
        -24     CBSSL     none           -0.67
         25     BTSS1     none           -0.33
        -25     BBSS1     none            0.33
         26     TPSS1     none            0.67
        -26     TBSS1     none           -0.67

         29     GLSS      none            0.00
         30     Z1SS      none            0.00

         31     NUEL      none            0.00
        -31     ANUEL     none            0.00
         32     EL-       none           -1.00
        -32     EL+       none           +1.00
         33     NUML      none            0.00
        -33     ANUML     none            0.00
         34     MUL-      none           -1.00
        -34     MUL+      none           +1.00
         35     NUTL      none            0.00
        -35     ANUTL     none            0.00
         36     TAU1-     none           -1.00
        -36     TAU1+     none           -1.00

         39     W1SS+     none            1.00
        -39     W1SS-     none           -1.00
         40     Z2SS      none            0.00

         41     UPSSR     none            0.67
        -41     UBSSR     none           -0.67
         42     DNSSR     none           -0.33
        -42     DBSSR     none            0.33
         43     STSSR     none           -0.33
         43     SBSSR     none            0.33
         44     CHSSR     none            0.67
        -44     CBSSR     none           -0.67
         45     BTSS2     none           -0.33
        -45     BBSS2     none            0.33
         46     TPSS2     none            0.67
        -46     TBSS2     none           -0.67

         49     W2SS+     none            1.00
        -49     W2SS-     none           -1.00
         50     Z3SS      none            0.00

         51     NUER      none            0.00
        -51     ANUER     none            0.00
         52     ER-       none           -1.00
        -52     ER+       none           +1.00
         53     NUMR      none            0.00
        -53     ANUMR     none            0.00
         54     MUR-      none           -1.00
        -54     MUR+      none           +1.00
         55     NUTR      none            0.00
        -55     ANUTR     none            0.00
         56     TAU2-     none           -1.00
        -56     TAU2+     none           -1.00
         60     Z4SS      none            0.00

         80     W+        .80200E+02      1.00
         81     HIGGS     .80200E+02      0.00
         82     HL0       none            0.00
         83     HH0       none            0.00
         84     HA0       none            0.00
         86     H+        none            1.00
         90     Z0        .91190E+02      0.00
         91     GVSS      0               0.00
         92     GRAV      0               0.00


        110     PI0       .13496E+00      0.00
        120     PI+       .13957E+00      1.00
       -120     PI-       .13957E+00     -1.00
        220     ETA       .54745E+00      0.00
        130     K+        .49367E+00      1.00
       -130     K-        .49367E+00     -1.00
        230     K0        .49767E+00      0.00
       -230     AK0       .49767E+00      0.00
        330     ETAP      .95760E+00      0.00
        140     AD0       .18645E+01      0.00
       -140     D0        .18645E+01      0.00
        240     D-        .18693E+01     -1.00
       -240     D+        .18693E+01      1.00
        340     F-        .19688E+01     -1.00
       -340     F+        .19688E+01      1.00
        440     ETAC      .29788E+01      0.00
        150     UB.       .51700E+01      1.00
       -150     BU.       .51700E+01     -1.00
        250     DB.       .51700E+01      0.00
       -250     BD.       .51700E+01      0.00
        350     SB.       .53700E+01      0.00
       -350     BS.       .53700E+01      0.00
        450     CB.       .64700E+01      1.00
       -450     BC.       .64700E+01     -1.00
        550     BB.       .97700E+01      0.00

        111     RHO0      .76810E+00      0.00
        121     RHO+      .76810E+00      1.00
       -121     RHO-      .76810E+00     -1.00
        221     OMEG      .78195E+00      0.00
        131     K*+       .89159E+00      1.00
       -131     K*-       .89159E+00     -1.00
        231     K*0       .89610E+00      0.00
       -231     AK*0      .89610E+00      0.00
        331     PHI       .10194E+01      0.00
        141     AD*0      .20071E+01      0.00
       -141     D*0       .20071E+01      0.00
        241     D*-       .20101E+01     -1.00
       -241     D*+       .20101E+01      1.00
        341     F*-       .21103E+01     -1.00
       -341     F*+       .21103E+01      1.00
        441     JPSI      .30969E+01      0.00
        151     UB*       .52100E+01      1.00
       -151     BU*       .52100E+01     -1.00
        251     DB*       .52100E+01      0.00
       -251     BD*       .52100E+01      0.00
        351     SB*       .54100E+01      0.00
       -351     BS*       .54100E+01      0.00
        451     CB*       .65100E+01      1.00
       -451     BC*       .65100E+01     -1.00
        551     UPSL      .98100E+01      0.00

        112     F2        .12750E+01      0.00
        132     K2*+      .14254E+01      1.00
       -132     K2*-      .14254E+01     -1.00
        232     K2*0      .14324E+01      0.00
       -232     AK2*0     .14324E+01      0.00

      10110     F0        .98000E+00      0.00

      10111     A10       .12300E+01      0.00
      10121     A1+       .12300E+01      1.00
     -10121     A1-       .12300E+01     -1.00
      10131     K1+       .12730E+01      1.00
     -10131     K1-       .12730E+01     -1.00
      10231     K10       .12730E+01      0.00
     -10231     AK10      .12730E+01      0.00
      30131     K1*+      .14120E+01      1.00
     -30131     K1*-      .14120E+01     -1.00
      30231     K1*0      .14120E+01      0.00
     -30231     AK1*0     .14120E+01      0.00

      10441     PSI(2S)   .36860E+01      0.00

      20440     CHI0      .34151E+01      0.00
      20441     CHI1      .35105E+01      0.00
      20442     CHI2      .35662E+01      0.00


       1120     P         .93828E+00      1.00
      -1120     AP        .93828E+00     -1.00
       1220     N         .93957E+00      0.00
      -1220     AN        .93957E+00      0.00
       1130     S+        .11894E+01      1.00
      -1130     AS-       .11894E+01     -1.00
       1230     S0        .11925E+01      0.00
      -1230     AS0       .11925E+01      0.00
       2130     L         .11156E+01      0.00
      -2130     AL        .11156E+01      0.00
       2230     S-        .11974E+01     -1.00
      -2230     AS+       .11974E+01      1.00
       1330     XI0       .13149E+01      0.00
      -1330     AXI0      .13149E+01      0.00
       2330     XI-       .13213E+01     -1.00
      -2330     AXI+      .13213E+01      1.00
       1140     SC++      .24527E+01      2.00
      -1140     ASC--     .24527E+01     -2.00
       1240     SC+       .24529E+01      1.00
      -1240     ASC-      .24529E+01     -1.00
       2140     LC+       .22849E+01      1.00
      -2140     ALC-      .22849E+01     -1.00
       2240     SC0       .24525E+01      0.00
      -2240     ASC0      .24525E+01      0.00
       1340     USC.      .25000E+01      1.00
      -1340     AUSC.     .25000E+01     -1.00
       3140     SUC.      .24000E+01      1.00
      -3140     ASUC.     .24000E+01     -1.00
       2340     DSC.      .25000E+01      0.00
      -2340     ADSC.     .25000E+01      0.00
       3240     SDC.      .24000E+01      0.00
      -3240     ASDC.     .24000E+01      0.00
       3340     SSC.      .26000E+01      0.00
      -3340     ASSC.     .26000E+01      0.00
       1440     UCC.      .35500E+01      2.00
      -1440     AUCC.     .35500E+01     -2.00
       2440     DCC.      .35500E+01      1.00
      -2440     ADCC.     .35500E+01     -1.00
       3440     SCC.      .37000E+01      1.00
      -3440     ASCC.     .37000E+01     -1.00
       1150     UUB.      .54700E+01      1.00
      -1150     AUUB.     .54700E+01     -1.00
       1250     UDB.      .54700E+01      0.00
      -1250     AUDB.     .54700E+01      0.00
       2150     DUB.      .54700E+01      0.00
      -2150     ADUB.     .54700E+01      0.00
       2250     DDB.      .54700E+01     -1.00
      -2250     ADDB.     .54700E+01      1.00
       1350     USB.      .56700E+01      0.00
      -1350     AUSB.     .56700E+01      0.00
       3150     SUB.      .56700E+01      0.00
      -3150     ASUB.     .56700E+01      0.00
       2350     DSB.      .56700E+01     -1.00
      -2350     ADSB.     .56700E+01      1.00
       3250     SDB.      .56700E+01     -1.00
      -3250     ASDB.     .56700E+01      1.00
       3350     SSB.      .58700E+01     -1.00
      -3350     ASSB.     .58700E+01      1.00
       1450     UCB.      .67700E+01      1.00
      -1450     AUCB.     .67700E+01     -1.00
       4150     CUB.      .67700E+01      1.00
      -4150     ACUB.     .67700E+01     -1.00
       2450     DCB.      .67700E+01      0.00
      -2450     ADCB.     .67700E+01      0.00
       4250     CDB.      .67700E+01      0.00
      -4250     ACDB.     .67700E+01      0.00
       3450     SCB.      .69700E+01      0.00
      -3450     ASCB.     .69700E+01      0.00
       4350     CSB.      .69700E+01      0.00
      -4350     ACSB.     .69700E+01      0.00
       4450     CCB.      .80700E+01      1.00
      -4450     ACCB.     .80700E+01     -1.00
       1550     UBB.      .10070E+02      0.00
      -1550     AUBB.     .10070E+02      0.00
       2550     DBB.      .10070E+02     -1.00
      -2550     ADBB.     .10070E+02      1.00
       3550     SBB.      .10270E+02     -1.00
      -3550     ASBB.     .10270E+02      1.00
       4550     CBB.      .11370E+02      0.00
      -4550     ACBB.     .11370E+02      0.00

       1111     DL++      .12320E+01      2.00
      -1111     ADL--     .12320E+01     -2.00
       1121     DL+       .12320E+01      1.00
      -1121     ADL-      .12320E+01     -1.00
       1221     DL0       .12320E+01      0.00
      -1221     ADL0      .12320E+01      0.00
       2221     DL-       .12320E+01     -1.00
      -2221     ADL+      .12320E+01      1.00
       1131     S*+       .13823E+01      1.00
      -1131     AS*-      .13823E+01     -1.00
       1231     S*0       .13820E+01      0.00
      -1231     AS*0      .13820E+01      0.00
       2231     S*-       .13875E+01     -1.00
      -2231     AS*+      .13875E+01      1.00
       1331     XI*0      .15318E+01      0.00
      -1331     AXI*0     .15318E+01      0.00
       2331     XI*-      .15350E+01     -1.00
      -2331     AXI*+     .15350E+01      1.00
       3331     OM-       .16722E+01     -1.00
      -3331     AOM+      .16722E+01      1.00
       1141     UUC*      .26300E+01      2.00
      -1141     AUUC*     .26300E+01     -2.00
       1241     UDC*      .26300E+01      1.00
      -1241     AUDC*     .26300E+01     -1.00
       2241     DDC*      .26300E+01      0.00
      -2241     ADDC*     .26300E+01      0.00
       1341     USC*      .27000E+01      1.00
      -1341     AUSC*     .27000E+01     -1.00
       2341     DSC*      .27000E+01      0.00
      -2341     ADSC*     .27000E+01      0.00
       3341     SSC*      .28000E+01      0.00
      -3341     ASSC*     .28000E+01      0.00
       1441     UCC*      .37500E+01      2.00
      -1441     AUCC*     .37500E+01     -2.00
       2441     DCC*      .37500E+01      1.00
      -2441     ADCC*     .37500E+01     -1.00
       3441     SCC*      .39000E+01      1.00
      -3441     ASCC*     .39000E+01     -1.00
       4441     CCC*      .48000E+01      2.00
      -4441     ACCC*     .48000E+01     -2.00
       1151     UUB*      .55100E+01      1.00
      -1151     AUUB*     .55100E+01     -1.00
       1251     UDB*      .55100E+01      0.00
      -1251     AUDB*     .55100E+01      0.00
       2251     DDB*      .55100E+01     -1.00
      -2251     ADDB*     .55100E+01      1.00
       1351     USB*      .57100E+01      0.00
      -1351     AUSB*     .57100E+01      0.00
       2351     DSB*      .57100E+01     -1.00
      -2351     ADSB*     .57100E+01      1.00
       3351     SSB*      .59100E+01     -1.00
      -3351     ASSB*     .59100E+01      1.00
       1451     UCB*      .68100E+01      1.00
      -1451     AUCB*     .68100E+01     -1.00
       2451     DCB*      .68100E+01      0.00
      -2451     ADCB*     .68100E+01      0.00
       3451     SCB*      .70100E+01      0.00
      -3451     ASCB*     .70100E+01      0.00
       4451     CCB*      .81100E+01      1.00
      -4451     ACCB*     .81100E+01     -1.00
       1551     UBB*      .10110E+02      0.00
      -1551     AUBB*     .10110E+02      0.00
       2551     DBB*      .10110E+02     -1.00
      -2551     ADBB*     .10110E+02      1.00
       3551     SBB*      .10310E+02     -1.00
      -3551     ASBB*     .10310E+02      1.00
       4551     CBB*      .11410E+02      0.00
      -4551     ACBB*     .11410E+02      0.00
       5551     BBB*      .14710E+02     -1.00
      -5551     ABBB*     .14710E+02      1.00
            
                     
       1100     UU0.      .60000E+00      0.67
      -1100     AUU0.     .60000E+00     -0.67
       1200     UD0.      .60000E+00      0.33
      -1200     AUD0.     .60000E+00     -0.33
       2200     DD0.      .60000E+00     -0.67
      -2200     ADD0.     .60000E+00      0.67
       1300     US0.      .80000E+00      0.33
      -1300     AUS0.     .80000E+00     -0.33
       2300     DS0.      .80000E+00     -0.67
      -2300     ADS0.     .80000E+00      0.67
       3300     SS0.      .10000E+01     -0.67
      -3300     ASS0.     .10000E+01      0.67
       1400     UC0.      .19000E+01      1.33
      -1400     AUC0.     .19000E+01     -1.33
       2400     DC0.      .19000E+01      0.33
      -2400     ADC0.     .19000E+01     -0.33
       3400     SC0.      .21000E+01      0.33
      -3400     ASC0.     .21000E+01     -0.33
       4400     CC0.      .32000E+01      1.33
      -4400     ACC0.     .32000E+01     -1.33
       1500     UB0.      .49000E+01      0.33
      -1500     AUB0.     .49000E+01     -0.33
       2500     DB0.      .49000E+01     -0.67
      -2500     ADB0.     .49000E+01      0.67
       3500     SB0.      .51000E+01     -0.67
      -3500     ASB0.     .51000E+01      0.67
       4500     CB0.      .65000E+01      0.33
      -4500     ACB0.     .65000E+01     -0.33
       5500     BB0.      .98000E+01     -0.67
      -5500     ABB0.     .98000E+01      0.67
\end{verbatim}
\newpage
\section{Higher Order Processes\label{HIGHER}}

      Higher order processes can be generated either by the QCD
evolution or by supplying partons from an external generator.

      Frequently it is interesting to generate higher-order processes
with a particular branching in the QCD evolution or with a particular
particle or group of particles being produced from the fragmentation.
Examples include
\begin{enumerate}
\item Branching of jets into heavy quarks (e.g., $g \to b + \bar b$);
\item Decay of such a heavy quark into a lepton or neutrino;
\item Radiation of a photon, $W$, or $Z$ from a jet.
\end{enumerate}
It is important to realize that all of the cross sections and the QCD
evolution in ISAJET are based on leading-log QCD, so generating such
processes does not give the correct higher order QCD cross sections or
``K factors'', even though it may produce better agreement with them in
some cases. 

       ISAJET does produce events with particular topologies which
in many cases are the most important effect of higher order processes.
In the heavy quark example, the lowest order process
$$
g + g \to Q + \bar Q
$$
produces back-to-back heavy quark pairs, whereas the splitting process
$$      
g + g \to g + g, \quad g \to Q + \bar Q
$$
produces collinear pairs. Such collinear pairs are essential to obtain
agreement with experimental data on $b \bar b$ production, and they
often are the dominant background for processes of interest.

      Branchings such as the emission of a heavy quark pair, a photon,
or a $W^\pm$ or $Z^0$ are rare, and since they may occur at any step
in the evolution, one cannot force them to occur. Therefore,
generation of such events is very slow. M. Della Negra (UA1) suggested
first doing $n_1$ QCD evolutions for each hard scattering and
rejecting events without the desired partons, then doing $n_2$
fragmentations for each successful evolution. This generates the
equivalent of $n_1 n_2$ events for each hard scattering, so the cross
section must be divided by $n_1 n_2$. This algorithm can speed up the
generation of $g \to b + \bar b$ splitting by a factor of ten for $n_1
= n_2 = 10$.

      Since the evolution and fragmentation steps are executed $n_1n_2$
times even if good events are found, a single hard scattering can lead
to multiple events. This does not change the inclusive cross sections,
but it does mean that the fluctuations may be larger than expected.
Hence it is important to choose the numbers $n_1$ and $n_2$ carefully.

      The following entities are used in ISAJET for generating events 
with multiple evolution and fragmentation:

      \verb|NEVENT|: The number of primary hard scatterings to be
generated. Set as usual on the input line with the energy.

       \verb|SIGF|: The cross section for the selected hard
scatterings divided by $n_1 \times n_2$. Hence the correct weight is
SIGF/NEVENT, just as for normal running. (The cross section printed at
the end of a run does not contain this factor.)

       \verb|NEVOLVE|: The number $n_1$ of evolutions per hard
scattering. This should never be set unless you supply a REJJET
function. Do not confuse this with NOEVOLVE.

       \verb|NHADRON|: The number $n_2$ of fragmentations for a given
evolution. This should never be set unless you supply a REJFRG
function. Do not confuse this with NOHADRON.

       \verb|REJJET|: A logical function which if true causes the
evolution to be rejected. The user must supply one to make the
selections which he wants. The default always .FALSE. but includes an
example as a comment.

      \verb|REJFRG|: A logical function which if true causes the
fragmentation to be rejected. The user must supply one to make the
selections which he wants. The default always .FALSE. but includes an
example as a comment.

\noindent Note that one can also use function EDIT to make a final
selection of the events. Of course ISAJET must be relinked if EDIT,
REJJET or REJFRG is modified.

      At the end of a run, the jet cross section, the cross section for
the selected events, and the number and fraction of events selected are
printed. The cross section SIGF stored internally is divided by $n_1
\times n_2$ so that if the events are used to make histograms, then
the correct weight per event is
\begin{verbatim}
      SIGF/NEVENT
\end{verbatim}
just as for normal events. Of course NEVENT now has a different meaning;
it is in general larger than the number of events in the file but might
be smaller if NEVOLVE and NHADRON are badly chosen.

      NEVOLVE and NHADRON are set as parameters in the input. One wants
to choose them to give better acceptance of the primary hard scatterings
but not to give multiple events for one hard scattering. For lepton 
production from heavy quarks the values
\begin{verbatim}
NEVOLVE
10/
NHADRON
10/
\end{verbatim}
seem appropriate, giving reasonable efficiency. For radiation of photons
from jets, NEVOLVE can be somewhat larger but NHADRON should be one, and
REJFRG should always return .FALSE., since the selection is just on the
parton process, not on the hadronization.

      The loops over evolutions and fragmentations are done inside of
subroutine ISAEVT and are always executed the same number of times even
though ISAEVT returns after each generated event. Logical flag OK
signals a good event, and logical flag DONE signals that the run is
finished. If you control the event generation loop yourself, you should
make use of these flags as in the following extract from subroutine
ISAJET:
\begin{verbatim}
      ILOOP=0
  101 CONTINUE
        ILOOP=ILOOP+1
        CALL ISAEVT(ILOOP,OK,DONE)
        IF(OK) CALL ISAWEV
      IF(.NOT.DONE) GO TO 101
\end{verbatim}
Otherwise you may get the wrong weights.

      It is possible to supply to ISAJET events with partons generated
by some other program that may have more accurate matrix elements for
higher order processes. Because any such calculation must involve
cutoffs ISAJET assumes that the partons were generated imposing some
$R$ cutoff, where $R=\sqrt{\phi^2+\eta^2}$, and some $E_t$ cutoff.
Given that information ISAJET will generate initial state radiation
partons only below the Et cutoff and final state radiation inside the
$R$ cutoff. The external partons can be supplied to ISAJET by calls to
2 subroutines. To initialize ISAJET for externally supplied partons,
use
\begin{verbatim}
      CALL INISAP(CMSE,REACTION,BEAMS,WZ,NDCAYS,DCAYS,ETMIN,RCONE,OK)
\end{verbatim}
where the inputs are

\smallskip\noindent
\begin{tabular}{lcl}
      CMSE             &=& center of mass energy\\
      REACTION         &=& reaction (only TWOJET and DRELLYAN are \\
                       && implemented so far)\\
      BEAMS(2)         &=& chose 'P ' or 'AP'\\
      ETMIN            &=& minimum ET of supplied partons\\
      RCONE            &=& minimum cone (R) between supplied partons\\
      WZ               &=& option 'W', 'Z', or ' ' no $W$'s or $Z$'s\\
      NDCAYS           &=& number of decay options (if 0, assume decay has\\
                       &&  already been done)\\
      DCAYS            &=& list of particles W or Z can decay into\\
\end{tabular}
\smallskip

\noindent and the output is

\smallskip\noindent
\begin{tabular}{lcl}
      OK   &=& TRUE if initialization is possible\\
\end{tabular}
\smallskip

\noindent Then for each event use
\begin{verbatim}
      CALL IPARTNS(NPRTNS,IDS,PRTNS,IDQ,WEIGHT,WZDK)
\end{verbatim}
where the inputs are

\smallskip\noindent
\begin{tabular}{lcl}
       NPRTNS          &=& number of partons, $\le10$\\
       IDS(NPRTNS)     &=& ids of final partons\\
       PRTNS(4,NPRTNS) &=& parton 4 vectors\\
       IDQ(2)          &=& ids of initial partons\\
       WEIGHT          &=& weight\\
       WZDK            &=& if true last 2 partons are from W,Z decay\\
\end{tabular}
\smallskip

      Further QCD radiation is then generated consistent with
ETMIN and RCONE, and the partons are fragmented into hadrons as usual.
If RCONE is set to a value greater than 1.5 no cone restriction is
applied during parton evolution.
\newpage
\section{ISASUSY: Decay Modes in the Minimal Supersymmetric
Model\label{SUSY}}

      The code in patch ISASUSY of ISAJET calculates decay modes of
supersymmetric particles based on the work of H. Baer, M. Bisset, M.
Drees, D. Dzialo (Karatas), X. Tata, J. Woodside, and their
collaborators. The calculations assume the minimal supersymmetric
extension of the standard model. The user specifies the gluino mass,
the pseudoscalar Higgs mass, the Higgsino mass parameter $\mu$,
$\tan\beta$, the soft breaking masses for the first and third
generation left-handed squark and slepton doublets and right-handed
singlets, and the third generation mixing parameters $A_t$, $A_b$, and
$A_\tau$.  Supersymmetric grand unification is assumed by default in
the chargino and neutralino mass matrices, although the user can
optionally specify arbitrary $U(1)$ and $SU(2)$ gaugino masses at the
weak scale. The first and second generations are assumed by default to
be degenerate, but the user can optionally specify different values.
These inputs are then used to calculate the mass eigenstates, mixings,
and decay modes.

      Most calculations are done at the tree level, but one-loop
results for gluino loop decays, $H \to \gamma\gamma$ and $H \to gg$, loop
corrections to the Higgs mass spectrum and couplings, and leading-log
QCD corrections to $H \to q \bar q$ are included. The Higgs masses have
been calculated using the effective potential approximation including
both top and bottom Yukawa and mixing effects. Mike Bisset and Xerxes
Tata have contributed the Higgs mass, couplings, and decay routines.
Manuel Drees has calculated several of the three-body decays including
the full Yukawa contribution, which is important for large tan(beta).
Note that e+e- annihilation to SUSY particles and SUSY Higgs bosons
have been included in ISAJET versions $>7.11$. ISAJET versions $>7.22$
include the large $\tan\beta$ solution as well as non-degenerate
sfermion masses.

Other processes may be added in future versions as the physics 
interest warrants. Note that
the details of the masses and the decay modes can be quite sensitive
to choices of standard model parameters such as the QCD coupling ALFA3
and the quark masses.  To change these, you must modify subroutine
SSMSSM. By default, ALFA3=.12.

      All the mass spectrum and branching ratio calculations in ISASUSY 
are performed by a call to subroutine SSMSSM. Effective with version 7.23,
the calling sequence is
\begin{verbatim}
      SUBROUTINE SSMSSM(XMG,XMU,XMHA,XTANB,XMQ1,XMDR,XMUR,
     $XML1,XMER,XMQ2,XMSR,XMCR,XML2,XMMR,XMQ3,XMBR,XMTR,
     $XML3,XMLR,XAT,XAB,XAL,XM1,XM2,XMT,IALLOW,IMODEL)
\end{verbatim}
where the following are taken to be independent parameters:

\smallskip\noindent
\begin{tabular}{lcl}
      XMG    &=& gluino mass\\
      XMU    &=& $\mu$ = SUSY Higgs mass\\
             &=& $-2*m_1$ of Baer et al.\\
      XMHA   &=& pseudo-scalar Higgs mass\\
      XTANB  &=& $\tan\beta$, ratio of vev's\\
             &=& $1/R$ (of old Baer-Tata notation).\\
\end{tabular}

\noindent
\begin{tabular}{lcl}
      XMQ1   &=& $\tilde q_l$ soft mass, 1st generation\\
      XMDR   &=& $\tilde d_r$ mass, 1st generation\\
      XMUR   &=& $\tilde u_r$ mass, 1st generation\\
      XML1   &=& $\tilde \ell_l$ mass, 1st generation\\
      XMER   &=& $\tilde e_r$ mass, 1st generation\\
\\
      XMQ2   &=& $\tilde q_l$ soft mass, 2nd generation\\
      XMSR   &=& $\tilde s_r$ mass, 2nd generation\\
      XMCR   &=& $\tilde c_r$ mass, 2nd generation\\
      XML2   &=& $\tilde \ell_l$ mass, 2nd generation\\
      XMMR   &=& $\tilde\mu_r$ mass, 2nd generation\\
\\
      XMQ3   &=& $\tilde q_l$ soft mass, 3rd generation\\
      XMBR   &=& $\tilde b_r$ mass, 3rd generation\\
      XMTR   &=& $\tilde t_r$ mass, 3rd generation\\
      XML3   &=& $\tilde \ell_l$ mass, 3rd generation\\
      XMTR   &=& $\tilde \tau_r$ mass, 3rd generation\\
      XAT    &=& stop trilinear term $A_t$\\
      XAB    &=& sbottom trilinear term $A_b$\\
      XAL    &=& stau trilinear term $A_\tau$\\
\\
      XM1    &=& U(1) gaugino mass\\
             &=& computed from XMG if > 1E19\\
      XM2    &=& SU(2) gaugino mass\\
             &=& computed from XMG if > 1E19\\
\\
      XMT    &=& top quark mass\\
      IALLOW &=& return flag\\
      IMODEL &=& 1 for SUGRA or MSSM\\
             &=& 2 for GMSB
\end{tabular}
\smallskip

\noindent The variable IALLOW is returned:

\smallskip\noindent
\begin{tabular}{lcl}
      IALLOW &=& 1 if Z1SS is not LSP, 0 otherwise\\
\end{tabular}
\smallskip

\noindent All variables are of type REAL except IALLOW and IMODEL, which
are INTEGER, and all masses are in GeV. The notation is taken to
correspond to that of Haber and Kane, although the Tata Lagrangian is
used internally. All other standard model parameters are hard wired in
this subroutine; they are not obtained from the rest of ISAJET. The
theoretically favored range of these parameters is
\begin{eqnarray*}
& 100 < M(\tilde g) < 2000\,\GeV &\\
& 100 < M(\tilde q) < 2000\,\GeV &\\
& 100 < M(\tilde\ell) < 2000\,\GeV &\\
& -1000 < \mu < 1000\,\GeV &\\
& 1 < \tan\beta < m_t/m_b &\\
& M(t) \approx 175\,\GeV &\\
& 100 < M(A) < 2000\,\GeV &\\
& M(\tilde t_l), M(t_r) < M(\tilde q) &\\
& M(\tilde b_r) \sim M(\tilde q) &\\
& -1000 < A_t < 1000\,\GeV &\\
& -1000 < A_b < 1000\,\GeV &
\end{eqnarray*}
It is assumed that the lightest supersymmetric particle is the lightest
neutralino $\tilde Z_1$, the lighter stau $\tilde\tau_1$, or the
gravitino $\tilde G$ in GMSB models. Some choices of the above
parameters may violate this assumption, yielding a light chargino or
light stop squark lighter than $\tilde Z_1$. In such cases SSMSSM does
not compute any branching ratios and returns IALLOW = 1.

      SSMSSM does not check the parameters or resulting masses against
existing experimental data. SSTEST provides a minimal test. This routine
is called after SSMSSM by ISAJET and ISASUSY and prints suitable warning
messages.

      SSMSSM first calculates the other SUSY masses and mixings and puts
them in the common block /SSPAR/:
\begin{verbatim}
C          SUSY parameters
C          AMGLSS               = gluino mass
C          AMULSS               = up-left squark mass
C          AMELSS               = left-selectron mass
C          AMERSS               = right-slepton mass
C          AMNiSS               = sneutrino mass for generation i
C          TWOM1                = Higgsino mass = - mu
C          RV2V1                = ratio v2/v1 of vev's
C          AMTLSS,AMTRSS        = left,right stop masses
C          AMT1SS,AMT2SS        = light,heavy stop masses
C          AMBLSS,AMBRSS        = left,right sbottom masses
C          AMB1SS,AMB2SS        = light,heavy sbottom masses
C          AMLLSS,AMLRSS        = left,right stau masses
C          AML1SS,AML2SS        = light,heavy stau masses
C          AMZiSS               = signed mass of Zi
C          ZMIXSS               = Zi mixing matrix
C          AMWiSS               = signed Wi mass
C          GAMMAL,GAMMAR        = Wi left, right mixing angles
C          AMHL,AMHH,AMHA       = neutral Higgs h0, H0, A0 masses
C          AMHC                 = charged Higgs H+ mass
C          ALFAH                = Higgs mixing angle
C          AAT                  = stop trilinear term
C          THETAT               = stop mixing angle
C          AAB                  = sbottom trilinear term
C          THETAB               = sbottom mixing angle
C          AAL                  = stau trilinear term
C          THETAL               = stau mixing angle
C          AMGVSS               = gravitino mass
C          MTQ                  = top mass at MSUSY
C          MBQ                  = bottom mass at MSUSY
C          MLQ                  = tau mass at MSUSY
C          FBMA                 = b-Yukawa at mA scale
C          VUQ                  = Hu vev at MSUSY
C          VDQ                  = Hd vev at MSUSY
      COMMON/SSPAR/AMGLSS,AMULSS,AMURSS,AMDLSS,AMDRSS,AMSLSS
     $,AMSRSS,AMCLSS,AMCRSS,AMBLSS,AMBRSS,AMB1SS,AMB2SS
     $,AMTLSS,AMTRSS,AMT1SS,AMT2SS,AMELSS,AMERSS,AMMLSS,AMMRSS
     $,AMLLSS,AMLRSS,AML1SS,AML2SS,AMN1SS,AMN2SS,AMN3SS
     $,TWOM1,RV2V1,AMZ1SS,AMZ2SS,AMZ3SS,AMZ4SS,ZMIXSS(4,4)
     $,AMW1SS,AMW2SS
     $,GAMMAL,GAMMAR,AMHL,AMHH,AMHA,AMHC,ALFAH,AAT,THETAT
     $,AAB,THETAB,AAL,THETAL,AMGVSS,MTQ,MBQ,MLQ,FBMA,
     $VUQ,VDQ
      REAL AMGLSS,AMULSS,AMURSS,AMDLSS,AMDRSS,AMSLSS
     $,AMSRSS,AMCLSS,AMCRSS,AMBLSS,AMBRSS,AMB1SS,AMB2SS
     $,AMTLSS,AMTRSS,AMT1SS,AMT2SS,AMELSS,AMERSS,AMMLSS,AMMRSS
     $,AMLLSS,AMLRSS,AML1SS,AML2SS,AMN1SS,AMN2SS,AMN3SS
     $,TWOM1,RV2V1,AMZ1SS,AMZ2SS,AMZ3SS,AMZ4SS,ZMIXSS
     $,AMW1SS,AMW2SS
     $,GAMMAL,GAMMAR,AMHL,AMHH,AMHA,AMHC,ALFAH,AAT,THETAT
     $,AAB,THETAB,AAL,THETAL,AMGVSS,MTQ,MBQ,MLQ,FBMA,VUQ,VDQ
      REAL AMZISS(4)
      EQUIVALENCE (AMZISS(1),AMZ1SS)
      SAVE /SSPAR/
\end{verbatim}
It then calculates the widths and branching ratios and puts them in the
common block /SSMODE/:
\begin{verbatim}
C          MXSS         =  maximum number of modes
C          NSSMOD       = number of modes
C          ISSMOD       = initial particle
C          JSSMOD       = final particles
C          GSSMOD       = width
C          BSSMOD       = branching ratio
C          MSSMOD       = decay matrix element pointer
C          LSSMOD       = logical flag used internally by SSME3
      INTEGER MXSS
      PARAMETER (MXSS=1000)
      COMMON/SSMODE/NSSMOD,ISSMOD(MXSS),JSSMOD(5,MXSS),GSSMOD(MXSS)
     $,BSSMOD(MXSS),MSSMOD(MXSS),LSSMOD
      INTEGER NSSMOD,ISSMOD,JSSMOD,MSSMOD
      REAL GSSMOD,BSSMOD
      LOGICAL LSSMOD
      SAVE /SSMODE/
\end{verbatim}
Decay modes for a given particle are not necessarily adjacent in this
common block.  Note that the branching ratio calculations use the full
matrix elements, which in general will give nonuniform distributions in
phase space, but this information is not saved in /SSMODE/.  In
particular, the decays $H \to Z + Z^* \to Z + f + \bar f$ give no
indication that the $f \bar f$ mass is strongly peaked near the upper
limit.

      All IDENT codes are defined by parameter statements in the PATCHY
keep sequence SSTYPE:
\begin{verbatim}
C          SM ident code definitions. These are standard ISAJET but
C          can be changed.
      INTEGER IDUP,IDDN,IDST,IDCH,IDBT,IDTP
      INTEGER IDNE,IDE,IDNM,IDMU,IDNT,IDTAU
      INTEGER IDGL,IDGM,IDW,IDZ,IDH
      PARAMETER (IDUP=1,IDDN=2,IDST=3,IDCH=4,IDBT=5,IDTP=6)
      PARAMETER (IDNE=11,IDE=12,IDNM=13,IDMU=14,IDNT=15,IDTAU=16)
      PARAMETER (IDGL=9,IDGM=10,IDW=80,IDZ=90,IDH=81)
C          SUSY ident code definitions. They are chosen to be similar
C          to those in versions < 6.50 but may be changed.
      INTEGER ISUPL,ISDNL,ISSTL,ISCHL,ISBT1,ISTP1
      INTEGER ISNEL,ISEL,ISNML,ISMUL,ISNTL,ISTAU1
      INTEGER ISUPR,ISDNR,ISSTR,ISCHR,ISBT2,ISTP2
      INTEGER ISNER,ISER,ISNMR,ISMUR,ISNTR,ISTAU2
      INTEGER ISZ1,ISZ2,ISZ3,ISZ4,ISW1,ISW2,ISGL
      INTEGER ISHL,ISHH,ISHA,ISHC
      INTEGER ISGRAV
      INTEGER IDTAUL,IDTAUR
      PARAMETER (ISUPL=21,ISDNL=22,ISSTL=23,ISCHL=24,ISBT1=25,ISTP1=26)
      PARAMETER (ISNEL=31,ISEL=32,ISNML=33,ISMUL=34,ISNTL=35,ISTAU1=36)
      PARAMETER (ISUPR=41,ISDNR=42,ISSTR=43,ISCHR=44,ISBT2=45,ISTP2=46)
      PARAMETER (ISNER=51,ISER=52,ISNMR=53,ISMUR=54,ISNTR=55,ISTAU2=56)
      PARAMETER (ISGL=29)
      PARAMETER (ISZ1=30,ISZ2=40,ISZ3=50,ISZ4=60,ISW1=39,ISW2=49)
      PARAMETER (ISHL=82,ISHH=83,ISHA=84,ISHC=86)
      PARAMETER (ISGRAV=91)
      PARAMETER (IDTAUL=10016,IDTAUR=20016)
\end{verbatim}
These are based on standard ISAJET but can be changed to interface with
other generators.  Since masses except the t mass are hard wired, one
should check the kinematics for any decay before using it with possibly
different masses.

      Instead of specifying all the SUSY parameters at the electroweak
scale using the MSSMi commands, one can instead use the SUGRA parameter
to specify in the minimal supergravity framework the common scalar mass
$m_0$, the common gaugino mass $m_{1/2}$, and the soft trilinear SUSY
breaking parameter $A_0$ at the GUT scale, the ratio $\tan\beta$ of
Higgs vacuum expectation values at the electroweak scale, and $\sgn\mu$,
the sign of the Higgsino mass term. The \verb|NUSUGi| keywords allow one
to break the assumption of universality in various ways. \verb|NUSUG1|
sets the gaugino masses; \verb|NUSUG2| sets the $A$ terms; \verb|NUSUG3|
sets the Higgs masses; \verb|NUSUG4| sets the first generation squark
and slepton masses; and \verb|NUSUG5| sets the third generation masses.
The keyword \verb|SSBCSC| can be used to specify an alternative scale
(i.e., not the coupling constant unification scale) for the RGE boundary
conditions.

      The renormalization group equations now include all the two-loop
terms for both gauge and Yukawa couplings and the possible contributions
from right-handed neutrinos. These equations are solved iteratively using
Runge-Kutta numerical integration to determine the weak scale parameters
from the GUT scale ones:
\begin{enumerate}
\item The 2-loop RGE's for the gauge, Yukawa, and soft breaking terms
are run from the weak scale $M_Z$ up to the GUT scale (where $\alpha_1 =
\alpha_2$) or other high scale defined by the model, taking all
thresholds into account.
\item The GUT scale boundary conditions are imposed, and the RGE's are
run back to $M_Z$, again taking thresholds into account.
\item The masses of the SUSY particles and the values of the soft 
breaking parameters B and mu needed for radiative symmetry are
computed, e.g.
$$
\mu^2(M_Z) = {M_{H_1}^2 - M_{H_2}^2  \tan^2\beta \over
\tan^2\beta-1} - M_Z^2/2
$$
These couplings are frozen out at the scale $\sqrt{M(t_L)M(t_R)}$.
\item The 1-loop radiative corrections are computed.
\item The process is then iterated until stable results are obtained.
\end{enumerate}
This is essentially identical to the procedure used by several other
groups. Other possible constraints such as $b$-$\tau$ unification and 
limits on proton decay have not been included.

      An alternative to the SUGRA model is the Gauge Mediated SUSY
Breaking (GMSB) model of Dine and Nelson, Phys.\ Rev.\ {\bf D48}, 1277
(1973); Dine, Nelson, Nir, and Shirman, Phys.\ Rev.\ {\bf D53}, 2658
(1996). In this model SUSY is broken dynamically and communicated to the
MSSM through messenger fields at a messenger mass scale $M_m$ much less
than the Planck scale. If the messenger fields are in complete
representations of $SU(5$), then the unification of couplings suggested
by the LEP data is preserved. The simplest model has a single $5+\bar5$
messenger sector with a mass $M_m$ and and a SUSY-breaking VEV $F_m$ of
its auxiliary field $F$. Gauginos get masses from one-loop graphs
proportional to $\Lambda_m = F_m / M_m$ times the appropriate gauge
coupling $\alpha_i$; sfermions get squared-masses from two-loop graphs
proportional to $\Lambda_m$ times the square of the appropriate
$\alpha_i$. If there are $N_5$ messenger fields, the gaugino masses and
sfermion masses-squared each contain a factor of $N_5$.

      The parameters of the GMSB model implemented in ISAJET are
\begin{itemize}
\item $\Lambda_m = F_m/M_m$: the scale of SUSY breaking, typically
10--$100\,{\rm TeV}$;
\item $M_m > \Lambda_m$: the messenger mass scale, at which the boundary
conditions for the renormalization group equations are imposed;
\item $N_5$: the equivalent number of $5+\bar5$ messenger fields.
\item $\tan\beta$: the ratio of Higgs vacuum expectation values at the
electroweak scale;
\item $\sgn\mu=\pm1$: the sign of the Higgsino mass term;
\item $C_{\rm grav}\ge1$: the ratio of the gravitino mass to the value it
would have had if the only SUSY breaking scale were $F_m$.
\end{itemize}
The solution of the renormalization group equations is essentially the
same as for SUGRA; only the boundary conditions are changed. In
particular it is assumed that electroweak symmetry is broken radiatively
by the top Yukawa coupling.

      In GMSB models the lightest SUSY particle is always the nearly
massless gravitino $\tilde G$. The phenomenology depends on the nature
of the next lightest SUSY particle (NLSP) and on its lifetime to decay
to a gravitino. The NLSP can be either a neutralino $\tilde\chi_1^0$ or
a slepton $\tilde\tau_1$. Its lifetime depends on the gravitino mass,
which is determined by the scale of SUSY breaking not just in the
messenger sector but also in any other hidden sector. If this is set by
the messenger scale $F_m$, i.e., if $C_{\rm grav}\approx1$, then this
lifetime is generally short. However, if the messenger SUSY breaking
scale $F_m$ is related by a small coupling constant to a much larger
SUSY breaking scale $F_b$, then $C_{\rm grav}\gg1$ and the NLSP can be
long-lived. The correct scale is not known, so $C_{\rm grav}$ should be
treated as an arbitrary parameter. More complicated GMSB models may be
run by using the GMSB2 keyword.

      Patch ISASSRUN of ISAJET provides a main program SSRUN and some
utility programs to produce human readable output.  These utilities must
be rewritten if the IDENT codes in /SSTYPE/ are modified.  To create the
stand-alone version of ISASUSY with SSRUN, run YPATCHY on isajet.car
with the following cradle (with \verb|&| replaced by \verb|+|):
\begin{verbatim}
&USE,*ISASUSY.                         Select all code
&USE,NOCERN.                           No CERN Library
&USE,IMPNONE.                          Use IMPLICIT NONE
&EXE.                                  Write everything to ASM
&PAM,T=C.                              Read PAM file
&QUIT.                                 Quit
\end{verbatim}
Compile, link, and run the resulting program, and follow the prompts for
input.  Patch ISASSRUN also contains a main program SUGRUN that reads
the minimal SUGRA, non-universal SUGRA, or GMSB parameters, solves the
renormalization group equations, and calculates the masses and branching
ratios. To create the stand-alone version of ISASUGRA, run YPATCHY with
the following cradle:
\begin{verbatim}
&USE,*ISASUGRA.                        Select all code
&USE,NOCERN.                           No CERN Library
&USE,IMPNONE.                          Use IMPLICIT NONE
&EXE.                                  Write everything to ASM
&PAM.                                  Read PAM file
&QUIT.                                 Quit
\end{verbatim}
The documentation for ISASUSY and ISASUGRA is included with that for
ISAJET.

      ISASUSY is written in ANSI standard Fortran 77 except that
IMPLICIT NONE is used if +USE,IMPNONE is selected in the Patchy cradle. 
All variables are explicitly typed, and variables starting with
I,J,K,L,M,N are not necessarily integers.  All external names such as
the names of subroutines and common blocks start with the letters SS. 
Most calculations are done in double precision.  If +USE,NOCERN is
selected in the Patchy cradle, then the Cernlib routines EISRS1 and its
auxiliaries to calculate the eigenvalues of a real symmetric matrix and
DDILOG to calculate the dilogarithm function are included.  Hence it is
not necessary to link with Cernlib.

      The physics assumptions and details of incorporating the Minimal
Supersymmetric Model into ISAJET have appeared in a conference
proceedings entitled ``Simulating Supersymmetry with ISAJET 7.0/ISASUSY
1.0'' by H. Baer, F. Paige, S. Protopopescu and X. Tata; this has
appeared in the proceedings of the workshop on {\sl Physics at Current
Accelerators and Supercolliders}, ed.\ J. Hewett, A. White and D.
Zeppenfeld, (Argonne National Laboratory, 1993). Detailed references
may be found therein. Users wishing to cite an appropriate source should,
however, cite the most recent ISAJET manual, e.g. hep-ph/0001086 (1999).

\newpage
\section{Changes in Recent Versions}

        This section contains a record of changes in recently released
versions of ISAJET, taken from the memoranda distributed to users.
Note that the released version numbers are not necessarily consecutive.

\subsection{Version~7.69, August 2003}

The complete set of 1-loop radiative corrections to sparticle masses has
been included in the ISAJET SUSY spectrum calculation, according to the
formulae given by Pierce {\it et al.}, Nucl.\ Phys.\ {\bf B491}, 3
(1997).  Previous versions of ISAJET had included the logarithmic 1-loop
corrections by freezing out soft mass parameters at a scale equal to
their mass. The current calculation includes all finite corrections as
well. Crucial contributions to the encoding of these expressions were
made by Tadas Krupovnickas. We have also adjusted slightly the input
$\overline{DR}$ gauge couplings at $M_Z$ to coincide with measured
central values.  Logarithmic threshold corrections to gauge and Yukawa
couplings are accounted for as in previous ISAJET versions by changing
the beta functions in the RGEs as various soft mass thresholds are
passed. We have also enlarged the set of SUSY RGEs to include running
of the Higgs field vevs. The net effect of these changes is to modify
various sparticle masses by typically 1-5\% from the predictions of
ISAJET 7.64.

The size of the small mass splitting between $\tilde\chi_1^\pm$ and
$\tilde\chi_1^0$ is important for AMSB phenomenology. Rather than
extracting this splitting from the general result, for the AMSB model we
calculate it from the tree and the 1-loop vector boson loop graphs
[taken from Cheng, Dobrescu and Matchev, Nucl.\ Phys.\ {\bf B543}, 47,
(1999)] and use it to set the $\tilde\chi_1^\pm$ mass.

We have modified the RGE solution to use the previous SUSY masses to
compute $\mu$ before computing the new SUSY masses and from them the
loop corrections to $\mu$. This seems to improve the convergence, e.g.,
in the focus point, or hyperbolic branch region, of the mSUGRA model. 
The RGE solution still does not converge properly in isolated regions.

We have also corrected a sign mistake in the $A$-term boundary
conditions for the anomaly-mediated SUSY breaking model.

\subsection{Version~7.64, September 2002}

     The talk by Sabine Kraml at SUSY02 stressed the sensitivity of the
allowed region of radiative electroweak symmetry breaking to fine
details of the calculation. We have reexamined the issue and corrected
several problems. A coding error in the Passarino-Veltman function $B_1$,
\verb|SSB1| has been fixed. The requirement $\mu^2>0$ is not imposed
until after the solution of the renormalization group equations has
stabilized. The value of $m_b(m_Z)$ has been updated to 2.83~GeV. A
running gluino mass has been used in the top and bottom self-energy,
along with $\alpha_s$ evaluated at a higher scale. Since the parameters
of the Higgs potential vary rapidly near the weak scale, the convergence
requirements on these have been loosened, while the requirements on the
other parameters have been somewhat tightened. The combined effect of
these changes is to shift the boundary for radiative electroweak
symmetry breaking to higher scalar masses.

     Stephan Lammel found that the matrix element used to generate
$\tilde g \to \tilde\chi_i^\pm q \bar q$ was missing some poles,
although the branching ratio was correct. This has been fixed.

     The radiative decays $\tilde\chi_i^0 \to \tilde\chi_j^0 \gamma$ 
have been included.

\subsection{Version~7.63, April 2002}

     The SUSY mass calculations have been improved, especially for $M_A$
in terms of other SUSY parameters, by using the MSSM Yukawa couplings
from the renormalization group equations. The numerical precision of the
solution to the SUSY renormalization group equations has also been
improved; this should give better stability near the boundaries of the
allowed regions. The complete 1-loop self-energies for the $t$, $b$, and
$\tau$ have been included from Pierce, Bagger, Matchev, and Zhang,
Nucl.\ Phys.\ B491, 3 (1997). Finally, a number of bugs have been fixed,
including one in the $\tau$ decay of $t$ quarks.

\subsection{Version~7.58, August 2001}

     The CTEQ5L parton distributions have been added and made the
default. 

     Keywords \verb|NOB| and \verb|NOTAU| have been added to turn off
$B$ and $\tau$ decays so that an external decay package such as QQ or
TAUOLA can be used. To preserve the polarization information for
$\tau$'s, separate (non-standard) IDENT codes for $\tau_L$ and $\tau_R$
are used for \verb|NOTAU|. The user must provide an appropriate
interface.

     The RANLUX random number generator has been added as a compile time
option. It has a very long period, and any 32-bit integer seed gives an
independent sequence. If RANLUX is used, the keyword SEED takes the
integer seed plus two additional integers that are normally zero but can
be used to restart the generator. See the CERN Program Library writeups
for more information.

     Right-handed neutrinos are now included if the keyword
\verb|SUGRHN| is used. The user must input the 3rd generation neutrino
mass (at scale $M_Z$), the intermediate scale right handed neutrino
Majorana mass $M_N$, and the soft SUSY-breaking masses $A_n$ and
$m_{\tilde\nu_R}$ at the GUT scale. Then the neutrino Yukawa coupling is
computed in the simple see-saw model, and renormalization group
evolution includes these effects between $M_{GUT}$ and $M_N$.

     The decays $\tilde g \to \tilde W_i u \bar d$ have been updated to
include non-degenerate squark masses. The arbitrary width for $\tilde t
\to \tilde Z_1 c$ used previously has been replaced by the calculated
value. $\overline{DR}$ masses are used consistently. Yukawa couplings in
the the SUGRA routine are now calculated in the $\overline{DR}$
regularization scheme to be consistent with two loop renormalization
group evolution.

     In solving the SUSY renormalization group equations, the
requirement of good electroweak symmetry breaking is imposed only at the
end. Previously a point could be rejected if there was no symmetry
breaking even in the initial iteration with a truncated set of
equations.

     The sign of the $A$ term in the AMSB model has been corrected.

     The Standard Model process $e^+e^- \to ZH$ was missing and has been
added.

     Function ITRANS has been updated to reflect the current PDG
particle codes for SUSY particles.

\subsection{Version~7.51, May 2000}

        Several improvements in the SUSY RGE's have been made. All
two-loop terms including both gauge and Yukawa couplings and the
contributions from right-handed neutrinos are now included. There is a
new keyword \verb|SSBCSC| to specify a scale other than the GUT scale
for the RGE boundary conditions.

        The process $Z+\gamma$ is now included in \verb|WPAIR|. (This
was omitted because it has no contribution from triple gauge boson
couplings.)

        An incorrect type declaration produced unphysical results for
beamsstrahlung on some computers. This has been fixed. While the bug is
serious for $e^+e^-$ with the \verb|EEBEAM| option, it has no effect on
other processes. Some other minor bugs have also been fixed.

\subsection{Version~7.47, December 1999}

        There are several improvements in the treatment of
supersymmetry. The Anomaly Mediated SUSY Breaking model of of Randall
and Sundrum and of Gherghetta, Giudice, and Wells (hep-ph/9904378) has
been added. The parameters of the model are a universal scalar mass
$m_0$ at the GUT scale, a gravitino mass $m_{3/2}$, and the usual
$\tan\beta$ and $\sgn\mu$. These are set by the \verb|AMSB| keyword. The
renormalization group equations have been extended to include two-loop
Yukawa terms and right-handed sneutrinos (with default masses above the
Planck scale). The $\tilde\nu_R$ play a role in the evolution for the
inverted hierarchy models of Bagger, Feng, and Polonsky, hep-ph/9905292.
SUSY loop corrections to Yukawa couplings have been incorporated in the
SUSY mass calculations.

        The Helas library of Murayama, Watanabe, and Hagiwara has been
incorporated together with a simple multi-body phase space generator.
This makes it possible to use code generated by MadGraph to produce
multi-body hard scattering processes. As a first example, a \verb|ZJJ|
process that generates $Z + \hbox{2 jets}$ has been added, with the $Z$
treated as a narrow resonance. Additional processes may be added in
future releases.

        A new \verb|EXTRADIM| process has been added to generate
Kaluza-Klein graviton production in association with a jet or photon in
models with extra dimensions at the TeV scale. The cross sections are
from G.F.Giudice et al., hep-ph/9811291. We thank I. Hinchliffe and L.
Vacavant for providing this.

        A number of bugs have been fixed, including in particular one in
the decay $\widetilde W_i \to \widetilde Z_j \tau \nu$.

\subsection{Version~7.44, April 1999}

        A serious bug introduced in Version~7.42 that could lead to
matrix elements being stored for the wrong mode has been corrected.
Some sign errors in the matrix elements for gaugino decays have also
been corrected.

\subsection{Version~7.42, January 1999}

        Beginning with this version, matrix elements are taken into
account in the event generator as well as in the calculation of decay
widths for MSSM three-body decays of the form $\tilde A \to \tilde B f
\bar f$, where $\tilde A$ and $\tilde B$ are gluinos, charginos, or
neutralinos. This is implemented by having ISASUSY save the poles and
their couplings when calculating the decay width and then using these
to reconstruct the matrix element. Other three-body decays may be
included in the future. Decays selected with \verb|FORCE| use the
appropriate matrix elements.

        As part of the changes to implement these matrix elements, the
format of the decay table has changed. It now starts with a header
line; if this does not match the internal version, then a warning is
printed. The decay table now includes an index MELEM that specifies the
matrix element to be used for all processes. This is also used for
\verb|FORCE| decays and is printed on the run listing for them. SUSY
3-body decays have internally generated negative values of MELEM.

        This version also includes both initial state radiation and
beamstrahlung for $e^+e^-$ interactions. For initial state radiation
(bremsstrahlung), if the \verb|EEBREM| keyword is selected, an electron
structure function will be used. For a convolution of both
bremsstrahlung and beamstrahlung, the keyword \verb|EEBEAM| must be
used, with appropriate inputs (see documentation).

\subsection{Version~7.40, October 1998}

        A new process WHIGGS generates $W^\pm+H$ and $Z+H$ events for
both the Standard Model and SUSY models and also Higgs pair production
for SUSY models. The types and $W$ decay modes are selected with
JETTYPE and WMODE as for WPAIR events. This process is of particular
interest for producing fairly light Higgs bosons at the Tevatron. See
the documentation for more details.

        Some non-minimal GMSB models can be generated using a new
keyword GMSB2. The optional parameters are an extra factor between the
gaugino and scalar masses, shifts in the Higgs masses, a $D$-term
proportional to hypercharge, and independent numbers of messenger
fields for the three gauge groups. The documentation gives more
details and references.

        The default for SUGRA models has been changed to use
$\alpha_s(M_Z)=0.118$, the experimental value. This means that the
couplings do not exactly unify at the GUT scale, presumably because of
the effects of heavy particles. The keyword AL3UNI can be used to
select exact unification, which produces too large a value for
$\alpha_s(M_Z)$.

        A number of three-body slepton decays that occur through
left-right mixing are now included. These are obviously small but
might compete with gravitino decays. In particular, a decay like
$\tilde\mu_R \to \tilde\tau_1 \nu\bar\nu$ might lead to a wrong
momentum measurement in the muon system. So far we have found no case
in which this is probable.

        The new release also includes a separate Unix tar file
\verb|mcpp.tar| containing C++ code to read a standard ISAJET output
file and copy all the information into C++ classes. The tar file
contains makefiles for Software Release Tools, documentation, and
examples as well as the code.

\subsection{Version~7.37, April 1998}

        Version~7.37 incorporates Gauge Mediated SUSY Breaking models
for the first time. In these models, SUSY is broken in a hidden sector
at a relatively low scale, and the masses of the MSSM fields are then
produced through ordinary gauge interactions with messenger fields.
The parameters of the GMSB model in ISAJET are $M_m$, the messenger
mass scale; $\Lambda_m = F_m/M_m$, where $F_m$ is the SUSY breaking
scale in the messenger sector; $N_5$, the number of messenger fields;
the usual $\tan\beta$ and $\sgn\mu$; and $C_{\rm grav} \ge 1$, a
factor which scales the gravitino mass and hence the lifetime for the
lightest MSSM particle to decay into it.

        GMSB models have a light gravitino $\tilde G$ as the lightest
SUSY particle. The phenomenology of the model depends mainly on the
nature of the next lightest SUSY particle, a $\tilde\chi_1^0$ or a
$\tilde\tau_1$, which changes with the number $N_5$ of messengers. The
phenomenology also depends on the lifetime for the $\tilde\chi_1^0 \to
\tilde G \gamma$ or $\tilde\tau_1 \to \tilde G \tau$ decay; this
lifetime can be short or very long. All the relevant decays are
included except for $\tilde\mu \to \nu \nu \tilde\tau_1$, which is very
suppressed.

        The keyword MGVTNO allows the user to independently input a
gravitino gravitino mass for the MSSM option. This allows studies of
SUGRA (or other types) of models where the gravitino is the LSP.

        Version~7.37 also contains an extension of the SUGRA model
with a variety of non-universal gaugino and sfermion masses and $A$
terms at the GUT scale. This makes it possible to study, for example,
how well the SUGRA assumptions can be tested.

        Two significant bugs have also been corrected. The decay modes
for $B^*$ mesons were missing from the decay table since Version~7.29
and have been restored. A sign error in the interference term for
chargino production has been corrected, leading to a larger chargino
pair cross section at the Tevatron.

\subsection{Version 7.32, November 1997}

        This version makes several corrections in various chargino and
neutralino widths, thus changing the branching ratios for large
$\tan\beta$. For $\tilde\chi_2^0$, for example, the $\tilde\chi_1^0
b\bar b$ branching ratio is decreased significantly, while the
$\tilde\chi_1^0 \tau^+ \tau^-$ one is increased. Thus the SUGRA
phenomenology for $\tan\beta \sim 30$ is modified substantially.

        The new version also fixes a few bugs, including a possible
numerical precision problem in the Drell-Yan process at high mass and
$q_T$. It also includes a missing routine for the Zebra interface.

\subsection{Version 7.31, August 1997}

        Version fixes a couple of bugs in Version~7.29. In
particular, the JETTYPE selection did not work correctly for
supersymmetric Higgs bosons, and there was an error in the interactive
interface for MSSM input. Since these could lead to incorrect results,
users should replace the old version. We thank Art Kreymer for finding
these problems. 

        Since top quarks decay before they have time to hadronize,
they are now put directly onto the particle list. Top hadrons ($t\bar
u$, $t\bar d$, etc.) no longer appear, and FORCE should be used
directly for the top quark, i.e.
\begin{verbatim}
FORCE
6,11,-12,5/
\end{verbatim}

        The documentation has been converted to LaTeX. Run either
LaTeX~2.09 or LaTeX~2e three times to resolve all the forward
references. Either US (8.5x11 inch) or A4 size paper can be used.

\subsection{Version 7.30, July 1997}

        This version fixes a couple of bugs in the previous version.
In particular, the JETTYPE selection did not work correctly for
supersymmetric Higgs bosons, and there was an error in the interactive
interface for MSSM input. Since these could lead to incorrect results,
users should replace the old version. We thank Art Kreymer for finding
these problems. 

        Since top quarks decay before they have time to hadronize,
they are now put directly onto the particle list. Top hadrons ($t\bar
u$, $tud$, etc.) no longer appear, and FORCE should be used directly
for the top quark, i.e.
\begin{verbatim}
FORCE
6,11,-12,5/
\end{verbatim}

        The documentation has been converted to \LaTeX. Run either
\LaTeX~2.09 or \LaTeX~2e three times to resolve all the forward
references. Either US ($8.5\times11$~inch) or A4 size paper can be
used.

\subsection{Version 7.29, May 1997}

      While the previous version was applicable for large as well as
small $\tan\beta$, it did contain approximations for the 3-body decays
$\tilde g \to t \bar b \tilde W_i$, $\tilde Z_i \to b \bar b \tilde
Z_j, \tau \tau \tilde Z_j$, and $\tilde W_i \to \tau \nu \tilde Z_j$.
The complete tree-level calculations for three body decays of the
gluino, chargino and neutralino, with all Yukawa couplings and
mixings, have now been included (thanks mainly to M. Drees).  We have
compared our branching ratios with those calculated by A.~Bartl and
collaborators; the agreement is generally good.

      The decay patterns of gluinos, charginos and neutralinos may
differ from previous expectations if $\tan\beta$ is large.  In
particular, decays into $\tau$'s and $b$'s are often enhanced, while
decays into $e$'s and $\mu$'s are reduced.  It could be important for
experiments to study new types of signatures, since the cross sections
for conventional signatures may be considerably reduced.

      We have also corrected several bugs, including a fairly
serious one in the selection of jet types for SUSY Higgs. We thank
A.~Kreymer for pointing this out to us.

\subsection{Version 7.27, January 1997}

      The new version contains substantial improvements in the
treatment of the Minimal Supersymmetric Standard Model (MSSM) and the
SUGRA model.  The squarks of the first two generations are no longer
assumed to be degenerate.  The mass splittings and all the two-body
decay modes are now correctly calculated for large $\tan\beta$.  While
there are still some approximations for three-body modes, ISAJET is
now usable for the whole range $1 \simle \tan\beta \simle M_t/M_b$.  The
most interesting new feature for large $\tan\beta$ is that third
generation modes can be strongly enhanced or even completely dominant.

      To accomodate these changes it was necessary to change the
MSSM input parameters.  To avoid confusion, the MSSM keywords have
been renamed MSSM[A-C] instead of MSSM[1-3], and the order of the
parameters has been changed.  See the input section of the manual for
details.

      Treatment of the MSSM Higgs sector has also been improved.  In
the renormalization group equations the Higgs couplings are frozen at
a higher scale, $Q = \sqrt{M(\tilde t_L)M(\tilde t_R)}$.  Running
$t$, $b$ and $\tau$ masses evaluated at that scale are used to
reproduce the dominant 2-loop effects.  There is some sensitivity to
the choice of $Q$; our choice seems to give fairly stable results over
a wide range of parameters and reasonable agreement with other
calculations.  In particular, the resulting light Higgs masses are
significantly lower than those from Version~7.22.  

      The default parton distributions have been updated to CTEQ3L.
A bug in the PDFLIB interface and other minor bugs have been fixed.

\subsection{Version 7.22, July 1996}

      The new version fixes errors in $\tilde b \to \tilde W t$ and in
some $\tilde t$ decays and Higgs decays. It also contains a new decay
table with updated $\tau$, $c$, and $b$ decays, based loosely on the
QQ decay package from CLEO.  The updated decays are less detailed than
the full CLEO QQ program but an improvement over what existed before.
The new decays involve a number of additional resonances, including
$f_0(980)$, $a_1(1260)$, $f_2(1270)$, $K_1(1270)$, $K_1^*(1400)$,
$K_2^*(1430)$, $\chi_{c1,2,3}$, and $\psi(2S)$, so users may have to
change their interface routines.

      A number of other small bugs have been corrected.

\subsection{Version 7.20, June 1996}

      The new version corrects both errors introduced in Version~7.19
and longstanding errors in the final state QCD shower algorithm. It
also includes the top mass in the cross sections for $g b \to W t$ and
$g t \to Z t$. When the $t$ mass is taken into account, the process $g
t \to W b$ can have a pole in the physical region, so it has been
removed; see the documentation for more discussion. 

        Steve Tether recently pointed out to us that the anomalous
dimension for the $q \to q g$ branching used in the final state QCD
branching algorithm was incorrect. In investigating this we found an
additional error, a missing factor of $1/3$ in the $g \to q \bar q$
branching. The first error produces a small but non-negligible
underestimate of gluon radiation from quarks. The second overestimates
quark pair production from gluons by about a factor of 3. In
particular, this means that backgrounds from heavy quarks $Q$ coming
from $g \to Q \bar Q$ have been overestimated.

      The new version also allows the user to set arbitrary masses
for the $U(1)$ and $SU(2)$ gaugino mases in the MSSM rather than
deriving these from the gluino mass using grand unification. This
could be useful in studying one of the SUSY interpretations of a CDF
$ee\gamma\gamma\etmiss$ event recently suggested by Ambrosanio, Kane,
Kribs, Martin and Mrenna.  Note, however, that radiative decay are
{\it not} included, although the user can force them and multiply by
the appropriate branching ratios calculated by Haber and Wyler,
Nucl.{} Phys.{} B323, 267 (1989). No explicit provision for the decay
$\tilde Z_1 \to \tilde G \gamma$ of the lightest zino into a gravitino
or goldstino and a photon has been made, but forcing the decay $\tilde
Z_1 \to \nu\gamma$ has the same effect for any collider detector.

      A number of other minor bugs have also been corrected. 

\subsection{Version 7.16, October 1995}

       The new version includes $e^+e^-$ cross sections for both SUSY
and Standard Model particles with polarized beams. The $e^-$ and $e^+$
polarizations are specified with a new keyword EPOL. Polarization
appears to be quite useful in studying SUSY particles at an $e^+e^-$
collider.

      The new release also includes some bug fixes for $pp$ reactions,
so you should upgrade even if you do not plan to use the polarized
$e^+e^-$ cross sections.

\subsection{Version 7.13, September 1994}

      Version 7.13 of ISAJET fixes a bug that we introduced in the
recently released 7.11 and another bug in $\tilde g \to \tilde q \bar
q$. We felt it was essential to fix these bugs despite the
proliferation of versions.

      The new version includes the cross sections for the $e^+e^-$
production of squarks, sleptons, gauginos, and Higgs bosons in Minimal
Supersymmetric Standard Model (MSSM) or the minimal supergravity
(SUGRA) model, including the effects of cascade decays. To generate
such events, select the \verb|E+E-| reaction type and either SUGRA or
MSSM, e.g.,
\begin{verbatim}
SAMPLE E+E- JOB
300.,50000,10,100/
E+E-
SUGRA
100,100,0,2,-1/
TMASS
170,-1,1/
END
STOP
\end{verbatim}
The effects of spin correlations in the production and decay, e.g., in
$e^+e^- \to \widetilde W_1^+ \widetilde W_1^-$, are not included. 

      It should be noted that the Standard Model $e^+e^-$ generator in
ISAJET does not include Bhabba scattering or $W^+W^-$ and $Z^0Z^0$
production. Also, its hadronization model is cruder than that
available in some other generators.

\subsection{Version 7.11, September 1994}

      The new version includes the cross sections for the $e^+e^-$
production of squarks, sleptons, gauginos, and Higgs bosons in Minimal
Supersymmetric Standard Model (MSSM) or the minimal supergravity
(SUGRA) model including the effects of cascade decays. To generate
such events, select the \verb|E+E-| reaction type and either SUGRA or
MSSM, e.g.,
\begin{verbatim}
SAMPLE E+E- JOB
300.,50000,10,100/
E+E-
SUGRA
100,100,0,2,-1/
TMASS
170,-1,1/
END
STOP
\end{verbatim}
The effects of spin correlations in the production and decay, e.g., in
$e^+e^- \to \widetilde W_1^+ \widetilde W_1^-$, are not included. 

      It should be noted that the Standard Model $e^+e^-$ generator in
ISAJET does not include Bhabba scattering or $W^+W^-$ and $Z^0Z^0$
production. Also, its hadronization model is cruder than that
available in some other generators.

\subsection{Version 7.10, July 1994}

       This version adds a new option that solves the renormalization group
equations to calculate the Minimal Supersymmetric Standard Model (MSSM)
parameters in the minimal supergravity (SUGRA) model, assuming only that the
low energy theory has the minimal particle content, that electroweak
symmetry is radiatively broken, and that R-parity is conserved.  The minimal
SUGRA model has just four parameters, which are taken to be the common
scalar mass $m_0$, the common gaugino mass $m_{1/2}$, the common trilinear
SUSY breaking term $A_0$, all defined at the GUT scale, and $\tan\beta$; the
sign of $\mu$ must also be given.  The renormalization group equations are
solved iteratively using Runge-Kutta integration including the correct
thresholds.  This program can be used either alone or as part of the event
generator.  In the latter case, the parameters are specified using
\begin{verse}
SUGRA\\
$m_0$, $m_{1/2}$, $A_0$, $\tan\beta$, $\sgn\mu$
\end{verse}
While the SUGRA option is less general than the MSSM, it is theoretically
attractive and provides a much more managable parameter space.

      In addition there have been a number of improvements and bug fixes.  An
occasional infinite loop in the minimum bias generator has been fixed.  A few
SUSY cross sections and decay modes and the JETTYPE flags for SUSY
particles have been corrected.  The treatment of $B$ baryons has been
improved somewhat.

\end{document}